\documentclass[manuscript]{aastex}
\usepackage{graphicx,apjfonts,emulateapj5,psfig,onecolfloat5,times}

\slugcomment{Received 2008 September 16; accepted 2009 January 9; published
  2009 March 24. ApJS 181:557-571  doi:10.1088/0067-0049/181/2/557}

\shorttitle{GBT polarization observations of pulsars}
\shortauthors{Han et al.}

\begin{document}
\twocolumn[

\title{Polarization observations of 100 pulsars at 774~MHz by
the GREEN BANK TELESCOPE}

\author{J.~L. Han\altaffilmark{1},
      P.~B. Demorest\altaffilmark{2},
      W. van Straten\altaffilmark{3},
      A.~G. Lyne\altaffilmark{4} 
}
\altaffiltext{1}{National Astronomical Observatories, Chinese Academy of
        Sciences, Jia 20 DaTun Road, Beijing 100012, China}
\altaffiltext{2}{National Radio Astronomy Observatory, Charlottesville, VA
        22903, USA}
\altaffiltext{3}{Centre for Astrophysics and Supercomputing, Swinburne
        University of Technology, PO Box 218, Hawthorn, VIC 3122, Australia}
\altaffiltext{4}{University of Manchester, Jodrell Bank Observatory,
        Macclesfield, SK11, 9DL, UK}

\begin{abstract}
We report on polarimetric observations of 100 pulsars centered on 774~MHz,
made using the Green Bank Telescope, presenting their polarization
profiles and polarized flux densities and comparing
them with previous observations when possible. For 67 pulsars, these are the
first such measurements made.  Polarization profiles of 8 millisecond
pulsars in our sample show wide profiles and flat position-angle curves.
Strong linear polarization, sometimes approaching 100\% of the total
intensity, has been detected in all or a part of the average pulse profiles
of some pulsars. In general, circular polarization is very weak, although it
is observed to be extremely strong in the leading component of PSR
J1920+2650.  Sense reversal of circular polarization as a function of pulse
phase has been detected from both core and other components of more than 20
pulsars.  Any relationship between the spin-down luminosity and the
percentage of linear polarization is not evident in our data 
at this frequency.
\end{abstract}


]
\altaffiltext{1}{National Astronomical Observatories, Chinese Academy of
        Sciences, Jia 20 DaTun Road, Beijing 100012, China}
\altaffiltext{2}{National Radio Astronomy Observatory, Charlottesville, VA
        22903, USA}
\altaffiltext{3}{Centre for Astrophysics and Supercomputing, Swinburne
        University of Technology, PO Box 218, Hawthorn, VIC 3122, Australia}
\altaffiltext{4}{University of Manchester, Jodrell Bank Observatory,
        Macclesfield, SK11, 9DL, UK}

\section{Introduction}

More than 40 years after the discovery of the first pulsars, there is
currently no consensus on how electromagnetic radiation is generated in the
pulsar magnetosphere.  Soon after their discovery, it was observed that the
radio emission from pulsars is highly polarized \citep[]{ls68}, with the
linearly polarized fraction approaching 100\% in some or all profile
components.  Some pulsars also have prominent circular polarization
\citep[see a summary by][]{hmxq98}.  Observations of pulsar polarization
continue to provide new insights into the emission mechanism
\citep[e.g.][]{lm88,ran93}.

Pulse profiles tend to have many peaks, which are related to the brightness
distribution inside the pulsar radiation beam.  \citet{ran83,ran93}
suggested that pulsar emission beams consist of two distinct component
types: a central core surrounded by one or more hollow cones.  The cones in
such a beam tend to be shifted to earlier longitudes with respect to the
center of the core \citep{gg03}. We note that the conventional model with
one core plus two cones cannot explain more than five components:
e.g. greater than six components for PSR B0740-28 \citep{kra94}, nine
components for PSR B0329+54 \citep{gg01}, greater than 13 components for PSR
J0437$-$4715 \citep{nms+97} and PSR J2124$-$3358 \citep{mh04}.  A model with
a multiconal beam has been proposed for some pulsars \citep{gg01}.  A
pulsar emission beam filled with ``randomly'' distributed emission patches
\citep{lm88,hm01} may explain more complicated multicomponent profiles.

Pulsar emission is thought to come from two possible regions, one near the
magnetic poles and the other near the light cylinder
\citep[e.g.][]{rs75,chr86a,zh00}.  It is also possible that the radio or
high-energy emission is radiated by particles flowing from the poles all the
way to the light-cylinder \citep[e.g.][]{gg01,gg03,qlw+04}.
Backward-flowing particles may also contribute to the observed radio
emission \citep{dzg05}.  Different components may originate at different
altitudes \citep{kj07} and/or from different magnetic field lines, and the
core component is probably emitted at a lower altitude \citep{gil91}.

At a given rotational phase, the polarization angle (P.A.) of the observed
linear polarization is thought to be related to the plane of curved magnetic
field lines tangential to the line of sight (LOS) in the neutron star
magnetosphere. In the ideal case, the P.A. curve of the on-pulse phases
follows the ``S''-shape of the ``rotating vector model'' \citep{rc69a}.
This model was confirmed by the earliest observations of some pulsars, which
led to the long-term belief that the radio emission is generated only near
the magnetic poles.

Radio emission generated at higher altitudes \citep{man96,jw06}, even near
the light cylinder, should produce a flatter P.A. curve, and suffer less from
propagation effects in the magnetosphere. To date, flat P.A. curves for just
one component together with the ``S''-shaped P.A. for other components have
been observed for many pulsars, which may indicate that different components
originate at different emission altitudes.  That is to say, even the
emission at a single frequency may come from different regions in the
neutron star magnetosphere.

Millisecond pulsars have much smaller light cylinders than normal pulsars,
so that the emission region is limited to a magnetosphere that is 6 or 7
orders of magnitude smaller in volume. The total intensity profiles of
millisecond pulsars are qualitatively similar to those of normal pulsars,
though with a tendency to cover a wider region of pulse phase
\citep{kxl+98}. In contrast, the polarization profiles of millisecond pulsars
\citep{xkj+98,stc99,mh04,ovhb04} are typically much more complex than those
of normal pulsars.  Polarization observations of millisecond pulsars are
relatively scarce, and most are presently available at only one radio
frequency (see references above).

Single pulse polarization observations show that orthogonal modes of
polarization can be emitted at a given pulse phase \citep{scr+84,ms00}.  In
the integrated polarization profile, transitions between orthogonal modes
typically stand out as jumps in the P.A. curve; in general, orthogonally polarized
modes are elliptically polarized \citep{crb78,mck04}.  Furthermore,
histograms of the orientation of the polarization vector provide evidence of
nonorthogonal modes of emission at a given pulse longitude \citep{es04}.

We have observed 100 pulsars with the Green Bank Telescope (GBT) in order to
measure their polarization properties.  We also determined the rotation
measures (RMs) of these pulsars.  The RMs will be presented in a second
paper \citep{hvd+08}, and, together with an additional 377 RMs, will be used
in an analysis of the large-scale Galactic magnetic field. In this paper, we
present the polarization profiles and polarized flux densities obtained
for these pulsars, beginning with a description of the observations in
Section 2.  In Section 3, the polarization profiles are grouped according to
similar characteristics. In Section 4, the results are summarized, and the
dependence of polarization percentage on the rotational energy loss rate
($\dot{E}$) is discussed.

\section{Observations}

Pulsar polarization observations were carried out with the NRAO
GBT\footnote{The National Radio Astronomy Observatory is a facility of the
National Science Foundation operated under cooperative agreement by
Associated Universities, Inc.} between 2007 November~17 and 26, using the
800 MHz prime focus receiver.  This dual-polarization receiver amplifies the
signals from two linear receptors, $X$ and $Y$, using cooled field-effect
transistors. The choice of central observation frequency, 774~MHz, was
primarily based on the absence of significant radio frequency
interference. The amplified RF signals were mixed to an intermediate
frequency of 400~MHz then fed into the Green Bank Astronomy Signal Processor
(GASP) pulsar backend \citep{dem07,fsb+04}.  In the GASP system, 128~MHz
total bandwidth was sampled in each polarization and then divided into 32
4-MHz subbands in hardware using a digital polyphase filter bank.  The
4-MHz subbands were distributed to a 16-node computer cluster for
additional real-time software signal processing.  In this case, the signal
was divided further in frequency for a final spectral resolution of
0.25~MHz. The $X$ and $Y$ signals in each channel were detected and
cross-multiplied to form the coherency products $XX$, $YY$, $X^*Y$, and
$XY^*$.  These values were later converted to Stokes parameters during the
calibration process via a simple linear transformation, as described by
\citet{van04c}.

The coherency products in each channel were averaged or folded (also in
real time) into 1024 pulse-phase bins modulo the predicted apparent
pulse period of each source. Pulsar ephemerides were obtained from the
ATNF Pulsar Catalog \citep{mhth05}, with the exception of 25 sources
where new timing data from Jodrell Bank were used instead.  All
ephemerides were processed with the pulsar analysis program
TEMPO\footnote{see http://www.atnf.csiro.au/research/pulsar/tempo/}. The
folded profiles were saved in 30~s subintegrations using the standard
PSRFITS data format \citep{hvm04}.

The computation-limited real-time processing allowed us to record a
total bandwidth of 96~MHz (384 channels) covering the range from 726 to
822~MHz, resulting in an  effective central observation frequency
of 774~MHz. Due to aliasing effects and strong instrumental
polarization, the edge channels within each 4~MHz subband were flagged and ignored
in our subsequent analysis.  There are also two receiver resonances in
this band, one at 796.6~MHz with a 2.1~MHz bandwidth, and the other at
817~MHz with a 3.3~MHz bandwidth. Channels in these regions were also
ignored, resulting in a total (discontinuous) usable bandwidth of
$\sim$80~MHz.

\begin{figure}
\centering
\mbox{\psfig{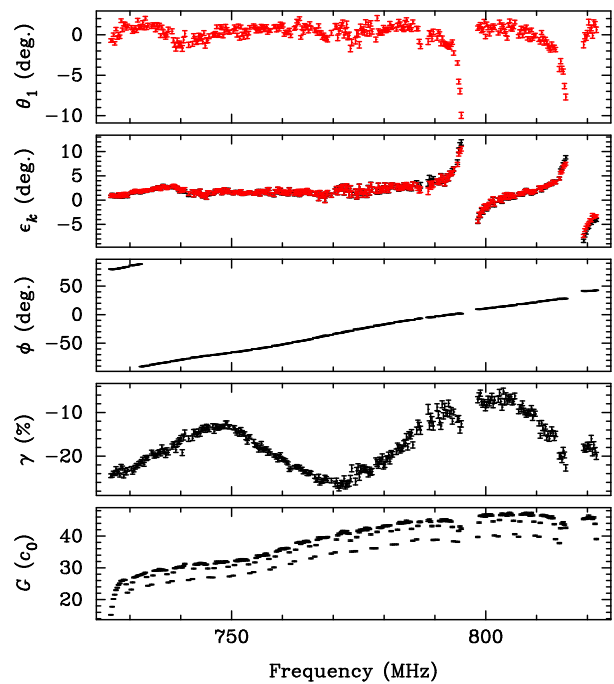}}
\caption{Best-fit instrumental parameters as a function of observing
frequency. From top to bottom are plotted the nonorthogonality,
$\theta_1$, and ellipticities, $\epsilon_k$, of the feed receptors,
the differential phase, $\phi$, the differential gain ratio, $\gamma$,
and the absolute gain, $G$.  The receiver resonances at 796.6~MHz and
817~MHz are clearly seen as discontinuities in receptor ellipticity
and orthogonality estimates.}
\label{calibrator}
\end{figure}

Before observing any pulsars, the overall system gain was calibrated by
observing the flux calibrators 3C286 and 3C295.  A low-level (4~K)
calibration signal (cal) produced by a noise diode that is coupled to the $X$
and $Y$ receptors before the primary amplifier was switched on and off at
25~Hz.  This nominally linearly polarized cal signal was recorded for 4*30~s 
(i.e. 4 separate recordings of 30~s; the same notation is used below) by
GASP while pointing at each flux calibrator.  The levels of all signals
were adjusted using attenuators to preserve linearity of the system.  The
telescope was then pointed at a position 2\degr away from each flux calibrator
and the pulsed cal signal was recorded again.  These two data sets were used to
determine the flux density of the cal signal relative to 3C286 and 3C295,
whose flux densities at 800~MHz were calculated according to \citet{bgpw77}.

The cal was reobserved prior to observing each pulsar.  The telescope was
pointed at a position $15'$ offset from the pulsar ($15'$ is the FWHM
telescope beam size at 800~MHz), and the pulsed cal was recorded for 2*30~s.
The pulsar was then observed with the cal signal turned off, and
observed the pulsar as described above for 8*30~s. The GBT prime focus
receivers are located on a focus rotation mount on a retractable boom.  We
took advantage of this feature by rotating the whole receiver by $90\degr$,
and re-observing the pulsar for another 8*30~s.  The addition of the two
data sets reduces the error in Stokes $Q$ that arises from the experimental
uncertainty in the differential gain calibration.

Data reduction and polarimetric calibration were performed using
PSRCHIVE \citep{hvm04}. First, we plotted the data in each
subintegration and frequency channel and manually removed any
obvious interference. Second, we obtained the receiver cross-coupling
parameters from observations of PSR B1929+10 at several different hour
angles \citep{van04c};
the best-fit solution is shown in Figure~\ref{calibrator}.
Third, as described by \citet{ovhb04}, we calibrated the pulsar data using a
combination of the instrumental polarization solution, the flux density
calibration information, and the cal data recorded prior to each pulsar
observation.  From residual variation versus hour angle in the
calibrated polarization of B1929+10, we infer that the systematic
error in the calibration procedure is at most $\sim$5\% of the total
polarized flux density.  For most of our targets this is less than the
uncertainty due to noise.

Finally the two sets of calibrated data recorded with different feed angles
were averaged to form the final calibrated polarization profiles as a
function of frequency.  From these data, the RM of each
pulsar was obtained, and the polarization data
were corrected for Faraday rotation and averaged to form a single
polarization profile, integrated over the band.

\begin{table*}
\begin{footnotesize}
\caption{Polarization parameters of 100 pulsars observed at 774~MHz by GBT}
\label{tab1}
\tabcolsep 1mm
\begin{tabular}{llrrrrrrrrrcll}
\hline
PSR Jname & PSR Bname & Period   & DM  & $\log\dot{E}$ & $S$  & $L$ &  $V$ & $|V|$ & $\sigma$ & W50   &
Fig. & Notes & Prev.polarization obs.\\ \cline{6-10}
 & & \multicolumn{1}{c}{(s)} &  (pc cm$^3$) & (erg~s$^{-1}$) &\multicolumn{5}{c}{(mJy)} &
($^\circ$) & & & Ref.:Freq.(MHz) \\
\hline
J1900--2600&B1857--26&0.6122&  37.99& 31.55&  78.64& 32.81&$ -0.62$&10.21&0.11& 26.5&2 & cal+cp & See text \\
J1903+0135&B1900+01  &0.7293& 245.17& 32.61&  40.86&  0.87&$  5.16$& 5.15&0.06&  3.8&2 & cal    & See text \\
J1932+1059&B1929+10  &0.2265&   3.18& 33.59&  61.62& 47.39&$ -4.36$& 4.84&0.06&  9.0&2 & cal+hiL& See text \\
J1935+1616&B1933+16  &0.3587& 158.52& 33.71& 266.29& 51.10&$  0.16$&39.18&0.14&  6.5&2 & cal+cp & See text \\[1mm] \hline 
J0014+4746&B0011+47  &1.2407&  30.85& 31.07&  19.87&  5.17&$ -0.67$& 0.75&0.10& 33.4&4 & hiL    & GL98:610,1418,1642\\
J0055+5117&B0052+51  &2.1152&  44.12& 31.60&   3.66&  1.62&$  0.18$& 0.20&0.11& 10.8&7 & db     & GL98:408,610,925,1408,1642 \\
J0117+5914&B0114+58  &0.1014&  49.42& 35.34&   1.75&  0.25&$ -0.06$& 0.08&0.08& 13.6&9 &        & GL98:408,610,1408\\
J0156+3949&B0153+39  &1.8116&  60.00& 30.00&   0.96&  0.38&$  0.01$& 0.07&0.10& 28.6&9 &        & GL98:606  \\
J0215+6218&	     &0.5489&  84.00& 32.20&  13.08&  2.91&$  0.84$& 0.94&0.13& 26.4&8 & ort  &   \\[1mm]
J0415+6954&B0410+69  &0.3907&  27.46& 31.71&   1.47&  0.13&$  0.00$& 0.07&0.10&  4.9&9 &        & GL98:408,610,1408 \\
J0849+8028&B0841+80  &1.6022&  34.66& 30.63&   1.09&  0.38&$  0.05$& 0.11&0.08& 10.1&9 &        & GL98:606 \\
J0921+6254&B0917+63  &1.5680&  13.16& 31.57&   1.58&  0.47&$ -0.06$& 0.05&0.10&  8.0&7 & db     & GL98:234,410,606,1408\\
J1012+5307&	     &0.0053&   9.02& 33.67&   3.05&  1.78&$  0.02$& 0.17&0.07& 47.6&3 & MSP+hiL& S+99:610;X+98:1410\\
J1300+1240&B1257+12  &0.0062&  10.17& 34.27&   1.34&  0.33&$  0.05$& 0.19&0.08& 39.9&3 & MSP+hiL+cp& X+98:1410 \\[1mm]
J1321+8323&B1322+83  &0.6700&  13.31& 31.87&   0.77&  0.63&$ -0.07$& 0.07&0.08& 10.9&4 & hiL    & GL98:410,606,925,1408\\
J1453+1902&	     &0.0058&  14.05& 33.37&   0.54&  0.25&$ -0.00$& 0.03&0.07& 21.6&3 & MSP    &   \\
J1503+2111&	     &3.3140&  11.75& 29.18&   0.70&  0.10&$  0.01$& 0.04&0.12&  4.1&9 &        &   \\
J1518+4904&	     &0.0409&  11.61& 31.20&   8.36&  0.74&$ -1.21$& 1.26&0.09& 38.5&3 & MSP+ort& X+98:1410;S+99:1414\\
J1549+2113&	     &1.2625&  24.80& 31.23&   0.85&  0.08&$ -0.06$& 0.07&0.09&  3.4&9 &        &   \\[1mm]
J1627+1419&	     &0.4909&  33.80& 32.12&   4.03&  0.39&$  0.17$& 0.20&0.07& 18.3&9 &        &   \\
J1640+2224&	     &0.0032&  18.43& 33.55&   2.40&  0.29&$  0.10$& 0.16&0.04& 20.2&3 & MSP    & X+98:1410 \\
J1649+2533&	     &1.0153&  35.50& 31.32&   3.89&  1.30&$ -0.07$& 0.22&0.12&  9.0&9 &        &   \\
J1652+2651&	     &0.9158&  40.80& 31.53&  10.79&  4.60&$ -0.60$& 0.77&0.10& 12.6&6 & cp     &   \\
J1720+2150&	     &1.6157&  41.10& 30.84&   0.76&  0.28&$  0.03$& 0.09&0.09& 13.6&6 & cp+db? &   \\[1mm]
J1741+2758&	     &1.3607&  29.30& 31.46&   1.03&  0.16&$ -0.08$& 0.11&0.09&  8.6&9 &        &   \\
J1746+2245&	     &3.4650&  52.00& 30.67&   1.50&  0.38&$  0.14$& 0.20&0.11&  1.5&7 & db?    &   \\
J1746+2540&	     &1.0581&  51.50& 31.54&   1.75&  0.73&$  0.11$& 0.15&0.12&  7.6&6 & cp     &   \\
J1754+5201&B1753+52  &2.3914&  35.35& 30.66&   4.35&  1.71&$ -0.42$& 0.52&0.10&  2.7&6 & cp     & GL98:410,606,1408\\
J1758+3030&	     &0.9473&  34.90& 31.52&   4.15&  0.52&$  0.06$& 0.12&0.10&  8.9&9 &        &   \\[1mm]
J1819+1305&	     &1.0604&  64.90& 31.08&   3.99&  1.13&$ -0.04$& 0.32&0.12& 21.8&6 & cp     &   \\
J1821+1715&	     &1.3667&  60.47& 31.13&   6.78&  1.46&$ -0.13$& 0.61&0.14& 11.0&6 & cp     &   \\
J1828+1359&	     &0.7416&  56.00& 31.85&   0.69&  0.23&$  0.21$& 0.21&0.10&  6.5&8 & ort    &   \\
J1838+1650&	     &1.9020&  33.90& 31.19&   1.89&  0.49&$ -0.21$& 0.21&0.10&  2.3&7 & db     &   \\
J1845+0623&	     &1.4217& 113.00& 30.88&   1.41&  0.27&$ -0.07$& 0.09&0.14&  3.6&9 &        &   \\[1mm]
J1848+0604&	     &2.2186& 242.70& 31.13&   1.40&  0.41&$ -0.03$& 0.13&0.15&  5.0&6 & cp     &   \\
J1849+2423&	     &0.2756&  62.24& 32.46&   3.21&  1.05&$  0.16$& 0.21&0.10& 19.4&4 & hiL    &   \\
J1853+1303&	     &0.0041&  30.57& 33.71&   1.66&  0.42&$ -0.01$& 0.20&0.10& 54.8&3 & MSP+cp &   \\
J1859+1526&	     &0.9340&  97.45& 32.28&   0.42&  0.10&$ -0.08$& 0.09&0.08&  ---&9 &        &   \\
J1900+0227&	     &0.3743& 201.10& 33.63&   0.93&  0.19&$ -0.12$& 0.17&0.18&  ---&9 &        &   \\[1mm]
J1901--0312&	     &0.3557& 106.40& 33.30&   0.18&  0.09&$ -0.06$& 0.06&0.11&  ---&4 & hiL?   &   \\
J1901+0124&	     &0.3188& 314.40& 33.60&   0.29&  0.16&$  0.07$& 0.07&0.16& 13.2&4 & hiL?   &   \\
J1902-0340&	     &1.5247& 114.00& 31.35&   0.56&  0.10&$  0.02$& 0.07&0.10& 12.2&9 &        &  \\ 
J1902+0723&	     &0.4878& 105.00& 31.85&   1.45&  0.26&$  0.27$& 0.25&0.13& 12.7&9 &        &  \\ 
J1903+0601&	     &0.3741& 388.00& 34.16&   1.03&  0.25&$ -0.02$& 0.06&0.16&  ---&9 &        &  \\ [1mm]
J1903+2225&	     &0.6512& 109.20& 31.80&   1.02&  0.21&$ -0.05$& 0.07&0.10&  6.7&9 &        &   \\
J1906+0414&	     &1.0434& 349.00& 32.60&   0.44&  0.11&$  0.04$& 0.04&0.16&  7.1&9 &        &   \\
J1906+0649&	     &1.2866& 249.00& 30.45&   0.98&  0.37&$  0.09$& 0.12&0.21&  ---&9 &        &   \\
J1906+1854&	     &1.0191& 156.72& 30.88&   2.47&  0.20&$  0.05$& 0.18&0.10& 23.8&9 &        &   \\
J1907+0345&	     &0.2402& 311.70& 34.37&   0.00&  0.00&$  0.00$& 0.00&0.14&  7.4&4 & hiL?   &   \\[1mm]
J1909+0616&	     &0.7560& 352.00& 33.27&   1.90&  0.29&$  0.26$& 0.29&0.07& 18.4&9 &        &   \\
J1909+1450&	     &0.9961& 119.50& 30.94&   0.32&  0.15&$  0.01$& 0.01&0.14&  ---&9 &        &   \\
J1909+1859&	     &0.5425&  64.48& 31.38&   1.32&  0.47&$  0.11$& 0.20&0.13&  8.3&9 &        &   \\
J1911+1758&	     &0.4604&  48.97& 30.72&   0.36&  0.12&$  0.01$& 0.03&0.10&  9.3&9 &        &   \\
J1912+1036&B1910+10  &0.4093& 147.00& 33.96&   0.21&  0.11&$ -0.03$& 0.04&0.12&  ---&4 & hiL?   &   \\[1mm]
J1912+2525&	     &0.6220&  37.84& 31.57&   0.74&  0.08&$ -0.07$& 0.07&0.09&  4.5&9 &        &   \\
J1915+0738&	     &1.5427&  39.00& 31.55&   1.47&  0.77&$ -0.31$& 0.32&0.15&  1.9&9 &        &   \\
J1915+0752&	     &2.0583& 105.30& 29.80&   1.45&  0.39&$ -0.08$& 0.13&0.16&  4.3&9 &        &   \\
J1915+0838&	     &0.3428& 358.00& 33.19&   0.85&  0.23&$  0.10$& 0.10&0.13& 16.7&9 &        &  \\ 
J1920+2650&B1918+26  &0.7855&  27.62& 30.45&   5.85&  1.47&$  1.32$& 1.55&0.13&  6.5&5 & hiCP+cp+nom& W+99:1400;GL98:410,610,1642\\
\hline
\end{tabular}
\end{footnotesize}
\end{table*} \addtocounter{table}{-1}
\begin{table*}
\begin{footnotesize}
\caption{Parameters of 100 pulsars observed at 774~MHz by GBT --  continued}
\tabcolsep 1mm
\begin{tabular}{llrrrrrrrrrcll}
\hline
PSR Jname & PSR Bname & Period   & DM  & $\log\dot{E}$ & $S$  & $L$ &  $V$ & $|V|$ & $\sigma$ & W50   &
Fig. & Notes & Prev.polarization obs.\\ \cline{6-10}
 & & \multicolumn{1}{c}{(s)} &  (pc cm$^3$) & (erg~s$^{-1}$) &\multicolumn{5}{c}{(mJy)} &
($^\circ$) & & & Ref.:Freq.(MHz) \\
\hline
J1921+2003&B1919+20  &0.7607& 101.00& 30.65&   1.49&  0.32&$ -0.27$& 0.28&0.10&  2.7&7 & db     & W+04:430\\
J1924+2040&B1922+20  &0.2378& 213.00& 33.79&   3.95&  0.43&$ -0.12$& 0.16&0.11& 16.5&9 &        & GL98:610\\
J1926+1434&B1924+14  &1.3249& 211.41& 30.57&   4.04&  0.99&$ -0.32$& 0.41&0.14&  5.2&7 & db     & RB81:430;GL98:606,1408\\
J1927+1856&B1925+188 &0.2983&  99.00& 33.52&   1.02&  0.15&$ -0.05$& 0.13&0.08& 18.3&9 &        &  \\
J1927+2234&B1925+22  &1.4311& 180.00& 31.02&   1.99&  0.29&$  0.07$& 0.14&0.12&  9.4&9 &        & W+99:1400\\[1mm]
J1930+1316&B1927+13  &0.7600& 207.30& 32.52&   0.69&  0.15&$ -0.02$& 0.03&0.08&  3.2&9 &        & W+04:430;W+99:1400\\
J1931+1536&B1929+15  &0.3144& 140.00& 33.80&   0.50&  0.11&$ -0.01$& 0.05&0.12& 12.4&9 &        & W+04:430\\
J1933+1304&B1930+13  &0.9283& 177.90& 31.20&   1.48&  0.40&$ -0.03$& 0.16&0.12& 10.2&7 & db     & W+99:1400\\
J1933+2421&B1931+24  &0.8137& 106.03& 32.77&   3.59&  1.34&$ -0.44$& 0.49&0.09&  9.7&6 & cp  & GL98:410,606\\
J1936+1536&B1933+15  &0.9673& 165.00& 32.25&   0.79&  0.23&$  0.17$& 0.19&0.10&  7.9&9 &        & W+99:1400\\[1mm]
J1937+2544&B1935+25  &0.2010&  53.22& 33.50&   7.69&  2.49&$ -0.13$& 0.26&0.11& 25.4&7 & db+hiL & GL98:606,925,1408,1642; \\
          &      &       &       &      &	 &       &       &     &    &     &      &        & W+99:1400;J+07:691,3100\\
J1939+2449&B1937+24  &0.6453& 142.88& 33.43&   5.77&  1.95&$  0.34$& 0.46&0.12& 10.9&4 & hiL+ort&   \\
J1941+1026&	     &0.9054& 138.91& 31.72&   0.75&  0.15&$ -0.06$& 0.08&0.13&  4.7&9 &        &  \\	
J1944+1755&B1942+17  &1.9969& 175.00& 30.56&   2.25&  0.88&$ -0.11$& 0.15&0.17&  3.4&7 & db+hiL & W+04:430\\
J1945+1834&B1943+18  &1.0687& 217.70& 30.89&   0.29&  0.02&$  0.02$& 0.05&0.12&  8.6&9 &        &   \\[1mm]
J1946+2611&	     &0.4351& 165.00& 34.02&   2.24&  0.91&$ -0.22$& 0.20&0.12& 12.1&9 &        &  \\
J1951+1123&	     &5.0941&  31.29& 29.96&   0.85&  0.12&$ -0.00$& 0.02&0.11&  1.8&9 &        &  \\
J1952+1410&B1949+14  &0.2750&  31.46& 32.39&   0.97&  0.20&$ -0.13$& 0.14&0.09&  7.2&9 &        &  \\
J1955+2908&B1953+29  &0.0061& 104.58& 33.71&   2.89&  0.80&$ -0.02$& 0.49&0.07&100.9&3 & MSP    & X+98:1410;TS90:1400\\
J1957+2831&	     &0.3077& 138.99& 33.62&   2.76&  0.72&$ -0.07$& 0.21&0.09& 12.0&4 & hiL+cp+ort&  \\[1mm]
J2002+3217&B2000+32  &0.6968& 142.21& 34.09&   4.65&  1.56&$ -0.35$& 0.37&0.13&  5.1&9 &        & W+99:1400;HR08:430;\\
          &      &       &       &      &	 &       &       &     &    &     &      &        & GL98:408,610,925,1408,1642 \\
J2008+2513&	     &0.5892&  60.55& 33.02&   3.10&  0.92&$ -0.17$& 0.19&0.11&  7.0&4 & hiL+ort&  \\
J2017+2043&	     &0.5371&  61.50& 32.40&   1.31&  0.31&$ -0.00$& 0.13&0.09& 11.3&6 & cp+db  & \\
J2023+5037&B2022+50  &0.3726&  33.02& 33.28&   7.74&  1.65&$ -0.26$& 0.36&0.10&  4.0&4 & hiL+ort& GL98:410,610,925,1408,1642\\
J2029+3744&B2027+37  &1.2168& 190.66& 32.43&   3.59&  0.64&$ -0.25$& 0.26&0.11&  5.3&9 &        & W+99:1400;GL98:410,610,1408,1642\\[1mm]
J2037+3621&B2035+36  &0.6187&  93.56& 32.88&   9.36&  2.69&$ -0.17$& 0.22&0.13& 10.1&8 & nort& GL98:6061408,1642;W+99:1400\\
J2038+5319&B2036+53  &1.4246& 160.10& 31.11&   1.73&  0.51&$ -0.06$& 0.21&0.10&  5.3&6 & cp     & GL98:606,1408\\
J2044+4614&	     &1.3927& 315.40& 30.96&   5.86&  1.96&$  0.06$& 0.44&0.15&  6.4&7 & db     &  \\
J2046+5708&B2045+56  &0.4767& 101.81& 33.61&   1.17&  0.55&$ -0.08$& 0.09&0.09&  9.5&9 &        & GL98:606\\
J2139+2242&	     &1.0835&  44.10& 31.64&  34.16& 14.76&$  0.79$& 1.50&0.10& 10.4&6 & cp     &  \\[1mm]
J2151+2315&	     &0.5935&  23.60& 32.13&   0.39&  0.16&$  0.00$& 0.13&0.03&192.5&4 & hiL?   &  \\
J2155+2813&	     &1.6090&  77.40& 30.94&   0.94&  0.10&$ -0.03$& 0.07&0.09&  3.6&9 &        &  \\
J2156+2618&	     &0.4981&  48.80& 30.66&   0.42&  0.07&$  0.03$& 0.05&0.09&  7.1&9 &        &  \\
J2205+1444&	     &0.9380&  36.72& 31.63&   0.84&  0.39&$  0.09$& 0.09&0.07& 18.1&7 & db     &  \\
J2215+1538&	     &0.3742&  29.26& 33.25&  16.55&  5.03&$  0.10$& 0.52&0.08&  5.0&6 & cp     &  \\[1mm]
J2234+2114&	     &1.3587&  35.08& 30.54&   4.30&  1.02&$  0.34$& 0.38&0.09& 21.4&9 &        &  \\
J2235+1506&	     &0.0598&  18.09& 31.46&   1.77&  0.21&$ -0.00$& 0.11&0.11& 16.6&3 & MSP    &  \\
J2253+1516&	     &0.7922&  29.18& 30.72&   5.65&  0.93&$ -0.27$& 0.48&0.17&  3.1&7 & db     &  \\
J2302+6028&	     &1.2064& 156.70& 31.66&   9.75&  2.78&$  0.83$& 0.98&0.11&  4.7&9 &        &  \\
J2305+4707&B2303+46  &1.0664&  62.06& 31.27&   1.18&  0.07&$  0.11$& 0.12&0.09&  8.2&9 &        & GL98:408,606,1642\\[1mm]
J2307+2225&          &0.5358&   7.08& 30.35&   3.65&  1.79&$  0.19$& 0.20&0.15&  5.6&9 &        &  \\
\hline
\end{tabular}
\\ 
\parbox{170mm}{ \noindent 1) The pulsar spin period (in seconds),
dispersion measure (DM, in pc~cm$^{-3}$) and spin-down luminosity
($\dot{E}$, in erg~s$^{-1}$) were taken from the pulsar catalog
\citep{mhth05}. The newly observed average flux density $S$ at 774 MHz,
average linearly polarized flux density $L$, average circularly polarized
flux density $V$, average absolute circularly polarized flux density $|V|$,
as well as the uncertainty of these flux densities, are all in mJy. The full
pulse width measured at half peak flux is in degrees of rotational phase.
\\
\noindent 2) Notes: MSP: millisecond pulsar; hiL: high
linear polarization; hiCP: high circular polarization; cp: circular
polarization with sense reversal; db: double components; ort: orthogonal
modes; nort:non-orthogonal modes.  \\ 3). References for previous
polarization observations: GL98: \citet{gl98}; HR08: \citet{hr08}; J+07: \citet{jkk+07}; RB81:
\citet{rb81}; S+99: \citet{stc99}; TS90: \citet{ts90}; W+99: \citet{wcl+99};
W+04: \citet{wck+04}; X+98: \citet{xkj+98}.}
\end{footnotesize}
\end{table*}

\begin{figure}
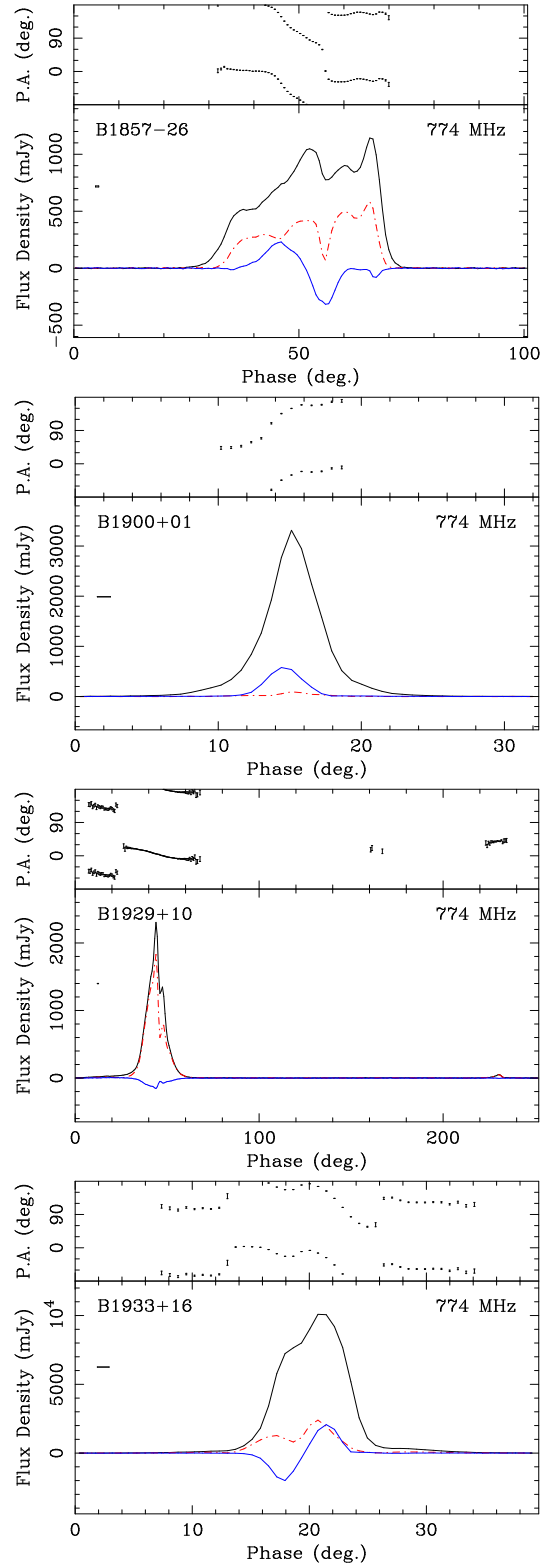
   
\centering
\begin{tabular}{l}
\mbox{\psfig{file=fig2_1.ps,angle=270,width=70mm,height=51mm}}\\
\mbox{\psfig{file=fig2_2.ps,angle=270,width=70mm,height=51mm}}\\
\mbox{\psfig{file=fig2_3.ps,angle=270,width=70mm,height=51mm}}\\
\mbox{\psfig{file=fig2_4.ps,angle=270,width=70mm,height=51mm}}\\
\end{tabular}%
\caption{Polarization profiles of four strong pulsars observed for calibration
purposes. In the lower part of each plot, the upper thick line is the total
intensity (Stokes parameter $I$), the dash-dot-dashed line is the linearly
polarized intensity $L$, and the thin line is the circularly polarized
intensity $V=I_L-I_R$, where $I_L$ and $I_R$ are the left and right
circularly polarized intensities, respectively.  The error box plotted on
the left below the pulsar name has a width equal to the pulse-phase
instrumental broadening (including dispersion smearing) and a height equal
to 4 times the baseline rms noise (i.e. $\pm 2\sigma$). In the upper
part, the P.A. angle of the linearly polarized emission is plotted with
$2\sigma$ error bars
only for those phases where the linear polarization is greater than $2\sigma$.}
\label{cal}
\end{figure}%

\section{Polarization profiles}

We observed 100 pulsars with the GBT centered on 774~MHz, including four
calibration pulsars.  The targets were selected from the set of pulsars in
the northern sky with unknown or uncertain rotation measures.  For 67 of
these pulsars, our observations are the first published polarization
measurements. Where possible in this section, the polarization profiles have
been grouped and presented according to their main polarization
properties. Although each profile is plotted only once, some will be
discussed in more than one subsection. We also compare our profiles with
previous polarization observations when available.

In Table~\ref{tab1}, we list for each source: pulsar J2000 source name,
B1950 source name where applicable, spin period (in seconds),
dispersion measure (DM, in pc~cm$^{-3}$), and spin-down luminosity
($\dot{E}$, in erg~s$^{-1}$).  These values were obtained from the
ATNF pulsar catalog \citep{mhth05}.  They are followed by our measured
parameters at 774 MHz: the observed average flux density $S$ (in
mJy), average linearly polarized flux density $L$ (in mJy),
average circularly polarized flux density $V$ (in mJy), average
absolute circularly polarized flux density $|V|$ (in mJy), as well
as the uncertainty of these flux densities, followed by the full pulse
width measured at half peak flux density (in degrees). We also give the
number of the figure in which the polarization profile is presented, and
what previous polarization observations are available.

\subsection{Calibration Pulsars}

Figure~\ref{cal} shows polarization profiles of the four pulsars
 that we
observed for calibration purposes, PSRs B1857-26 (J1900$-$2600), B1900+01
(J1903+0135), B1929+10 (J1932+1059), and B1933+16 (J1935+1616); these appear
consistent with those presented in previous publications (see references
below).

The bright pulsar PSR B1929+10, a standard polarization
calibrator, was observed several times at many parallactic angles. From
these observations, we derived the instrumental polarization parameters
\citep{van04c}. Consistent with previous
observations \citep[e.g.,][]{scw+84,gl98,stc99}, the main pulse of PSR
B1929+10 shows more than three components, i.e., two peaks plus unresolved
leading shoulders, all highly polarized. The right-hand circular
polarization is weak and nulls between the main peak and trailing component.
The inter-pulse is almost 100\% polarized.

PSR B1900+01 is a bright pulsar with a large DM that was observed to check
the order of frequency channels. It has strong left-hand circular
polarization but extremely weak linear polarization. The polarization
profile is consistent with that given by GL98.

PSRs B1933+16 and B1857$-$26 were also observed
to validate the phase convention of the instrument, which affects the
handedness of circular polarization as well as the position angle.
The polarization profile of PSR B1933+16 at 774~MHz has two barely
resolvable components with reversed circular polarization, similar to those
at 925~MHz \citep[see][hereafter GL98]{gl98} and 1400~MHz
\citep{rsw89,wcl+99}. The P.A. values of this pulsar at 774~MHz also compare
well with those at 1.4~GHz given by \citet{jhv+05} and are consistent
with the small RM ($=-1.9$~rad~m$^{-2}$) of this pulsar.
Our observations of PSR B1857$-$26 at 774~MHz have better
sensitivity than previously published data at 610~MHz, 800~MHz and 925~MHz
\citep[][GL98]{vdhm97,mhq98}.  All profiles at these frequencies show
consistent linear and circular polarization.

\begin{figure*}  
\centering
\begin{tabular}{ll}
\mbox{\psfig{file=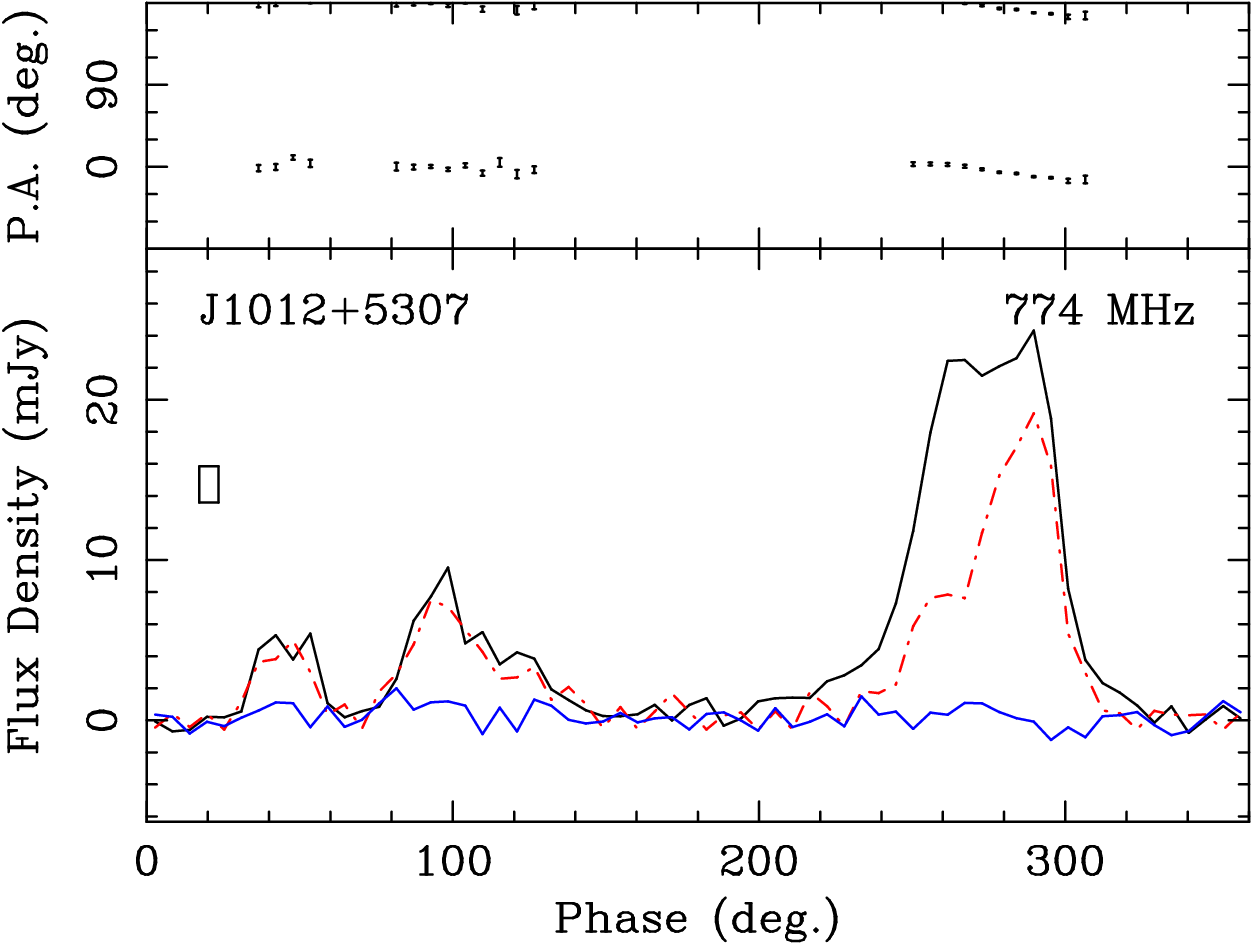,angle=270,width=70mm,height=46mm}} &
\mbox{\psfig{file=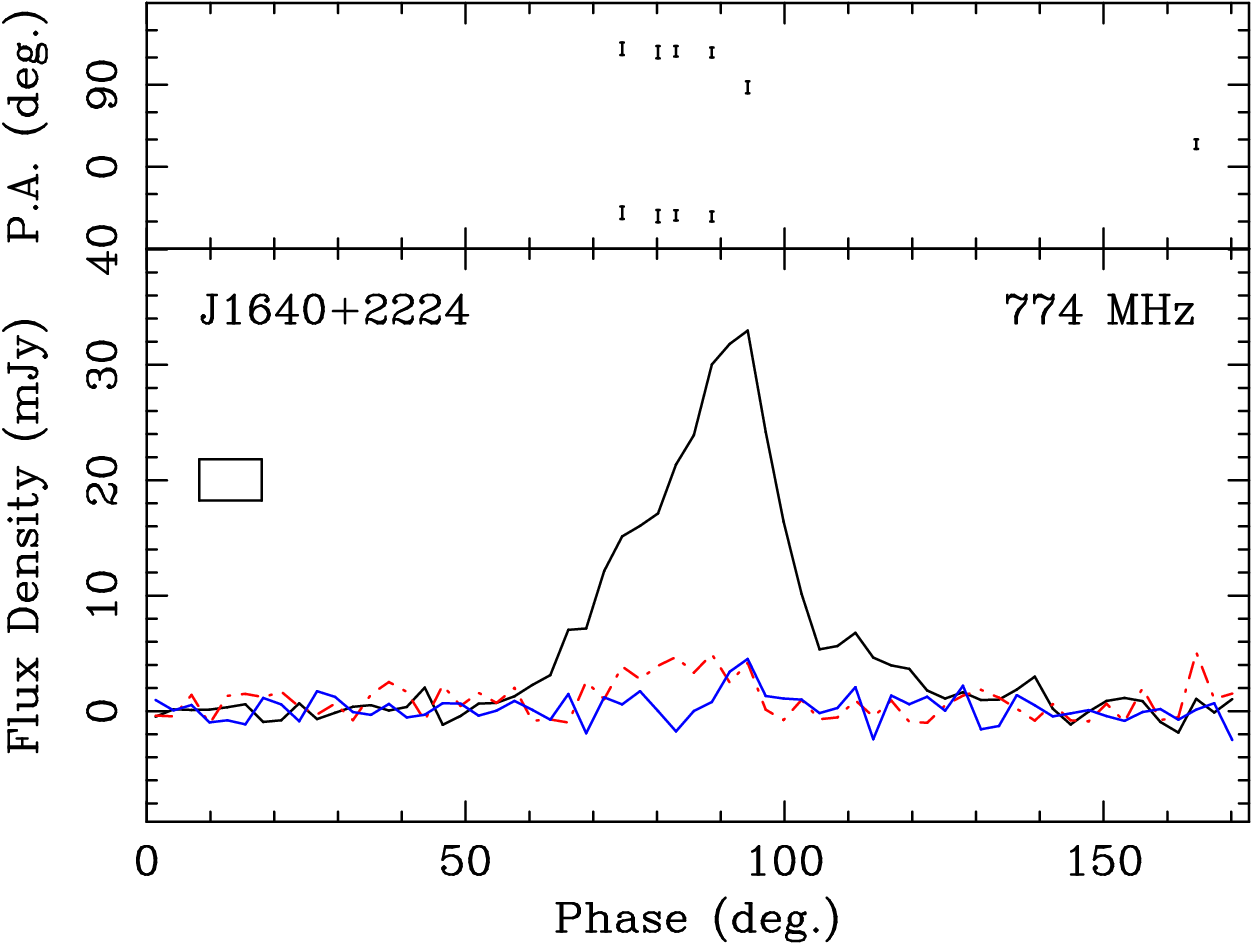,angle=270,width=70mm,height=46mm}}\\
\mbox{\psfig{file=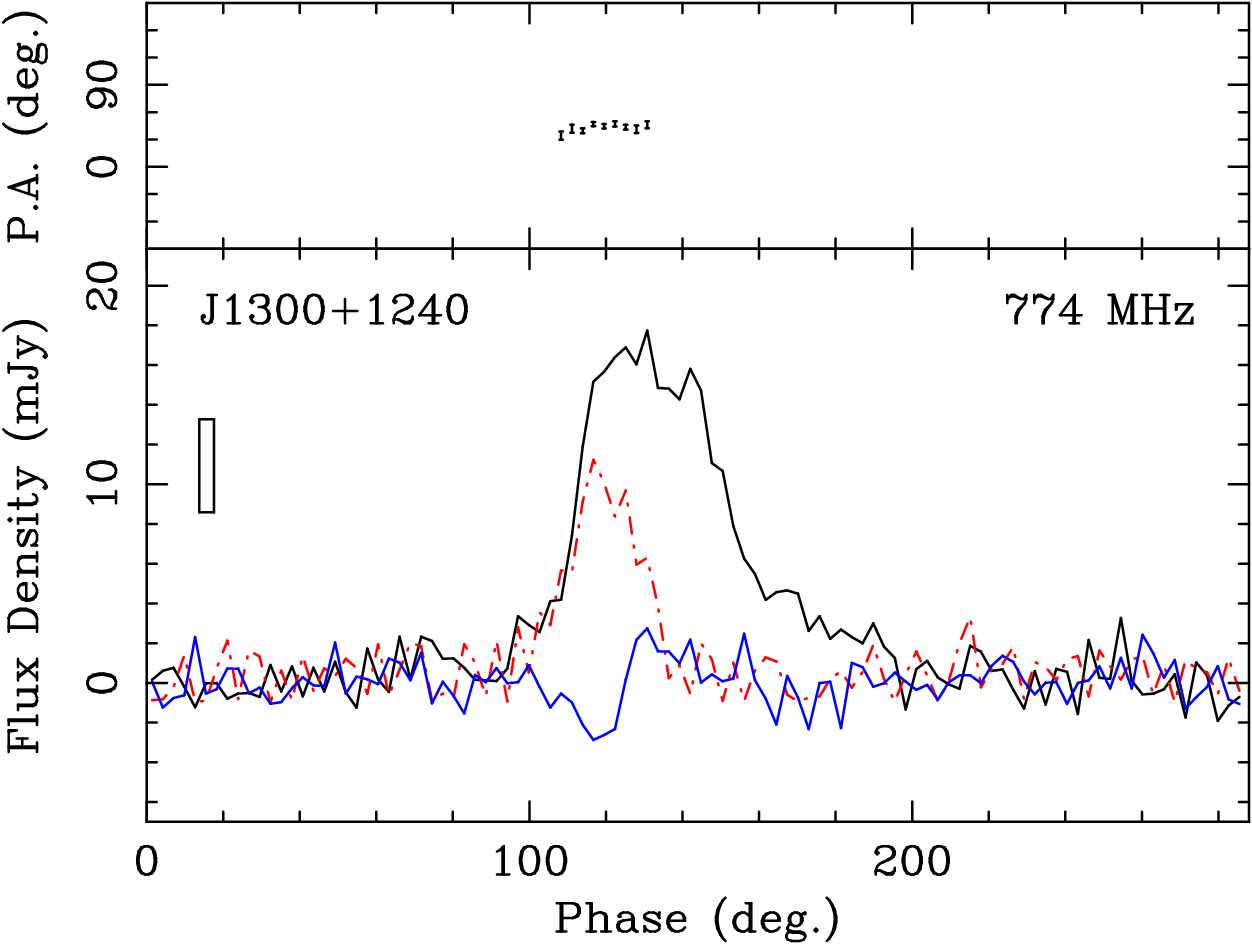,angle=270,width=70mm,height=46mm}} &
\mbox{\psfig{file=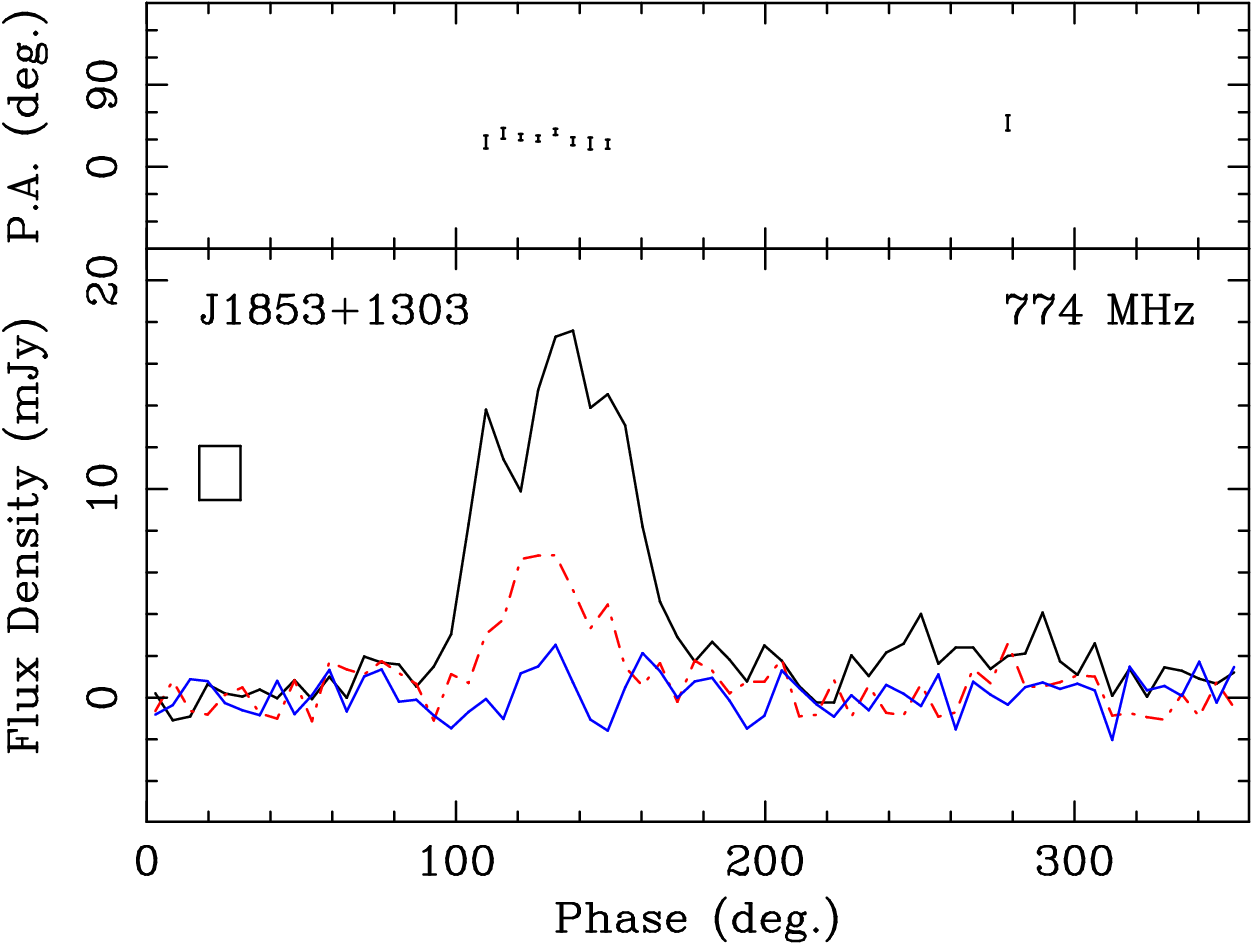,angle=270,width=70mm,height=46mm}}\\
\mbox{\psfig{file=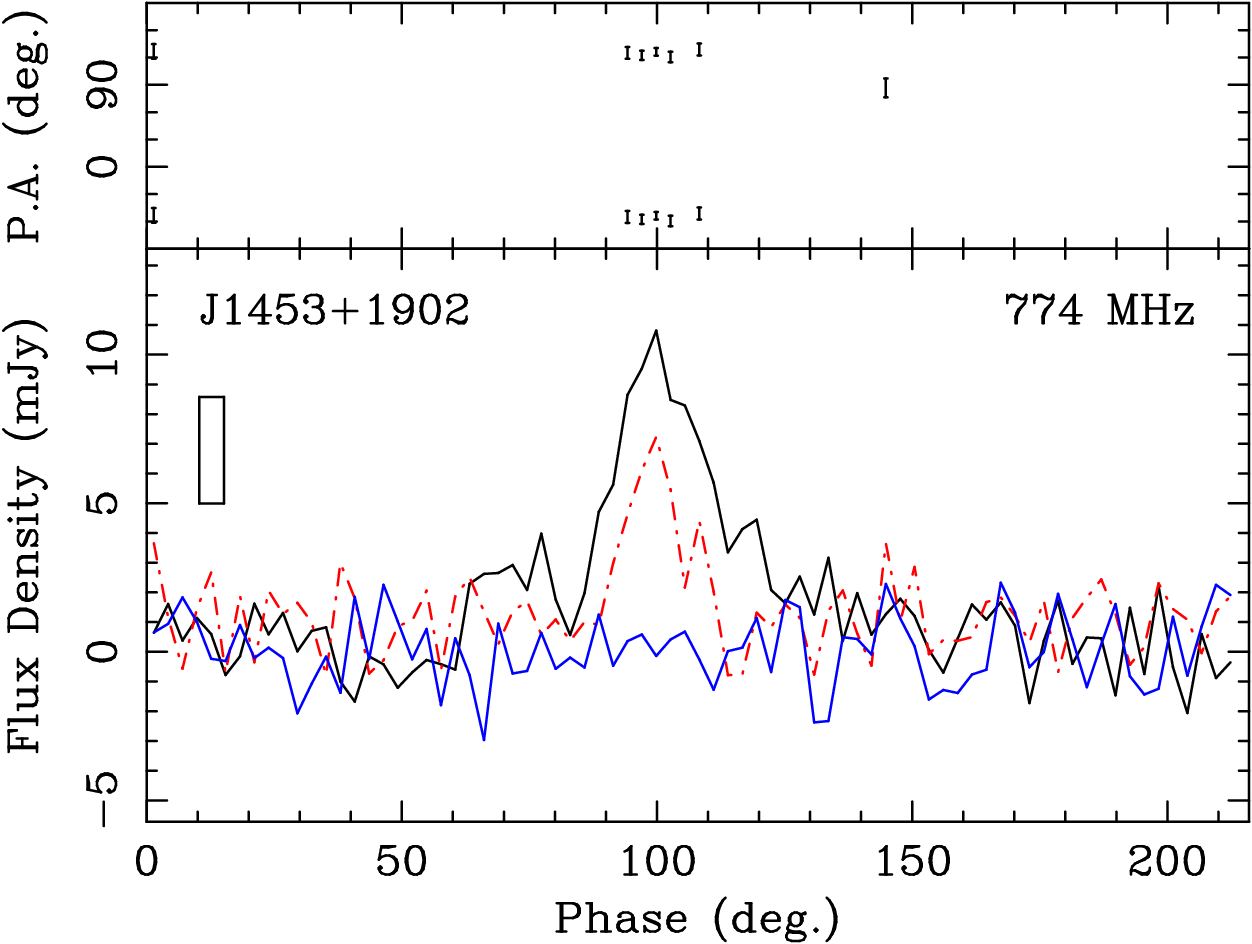,angle=270,width=70mm,height=46mm}} &
\mbox{\psfig{file=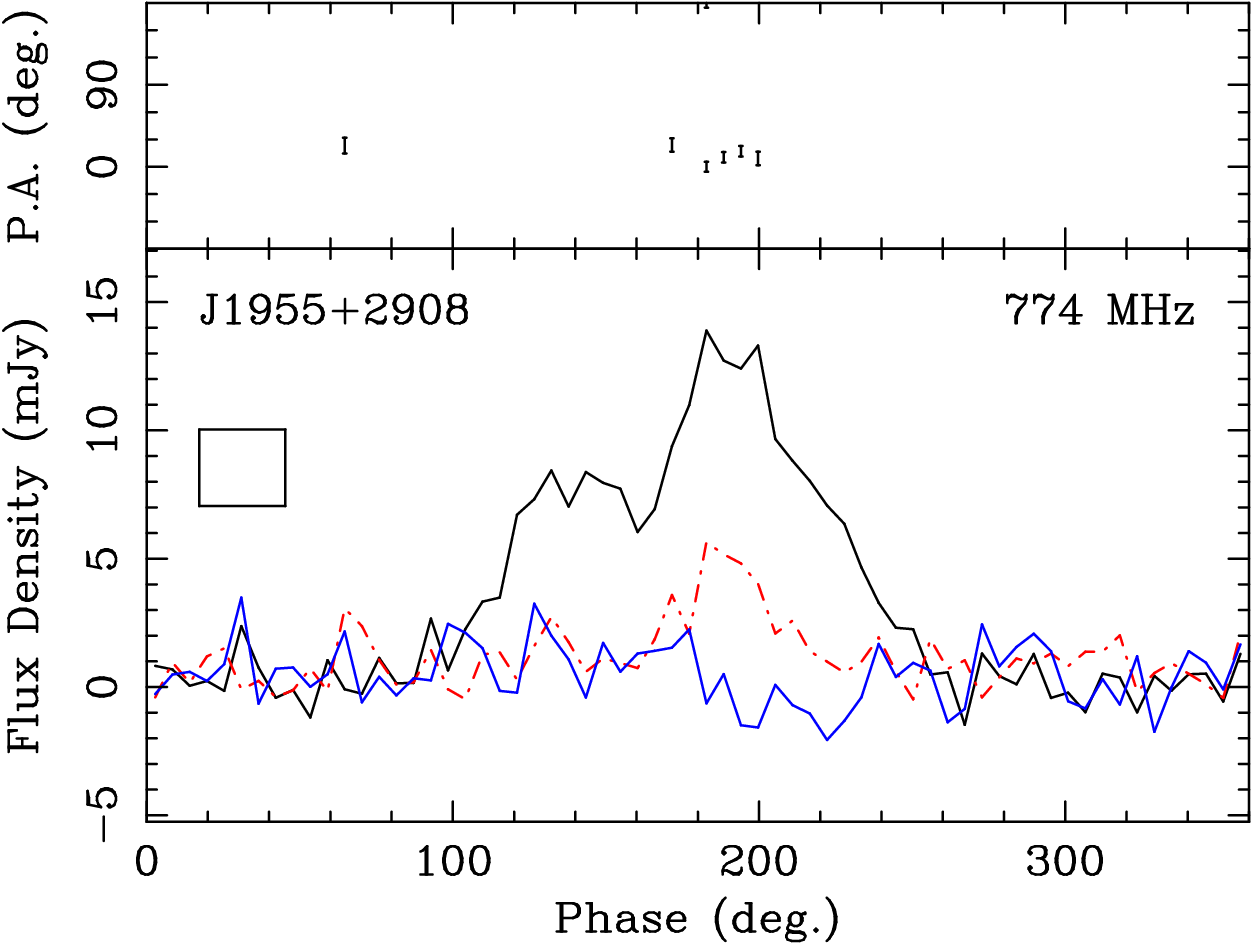,angle=270,width=70mm,height=46mm}}\\
\mbox{\psfig{file=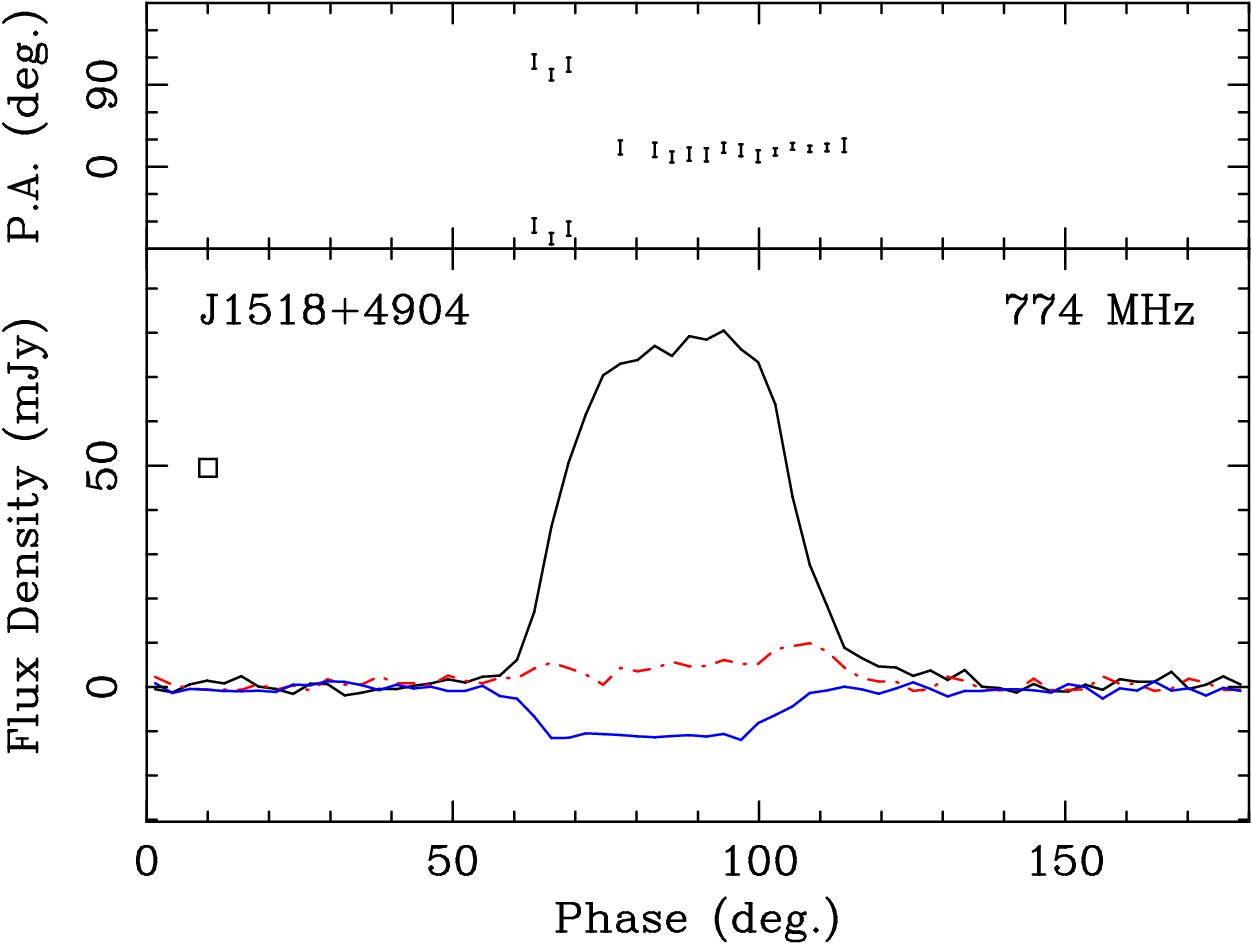,angle=270,width=70mm,height=46mm}} &
\mbox{\psfig{file=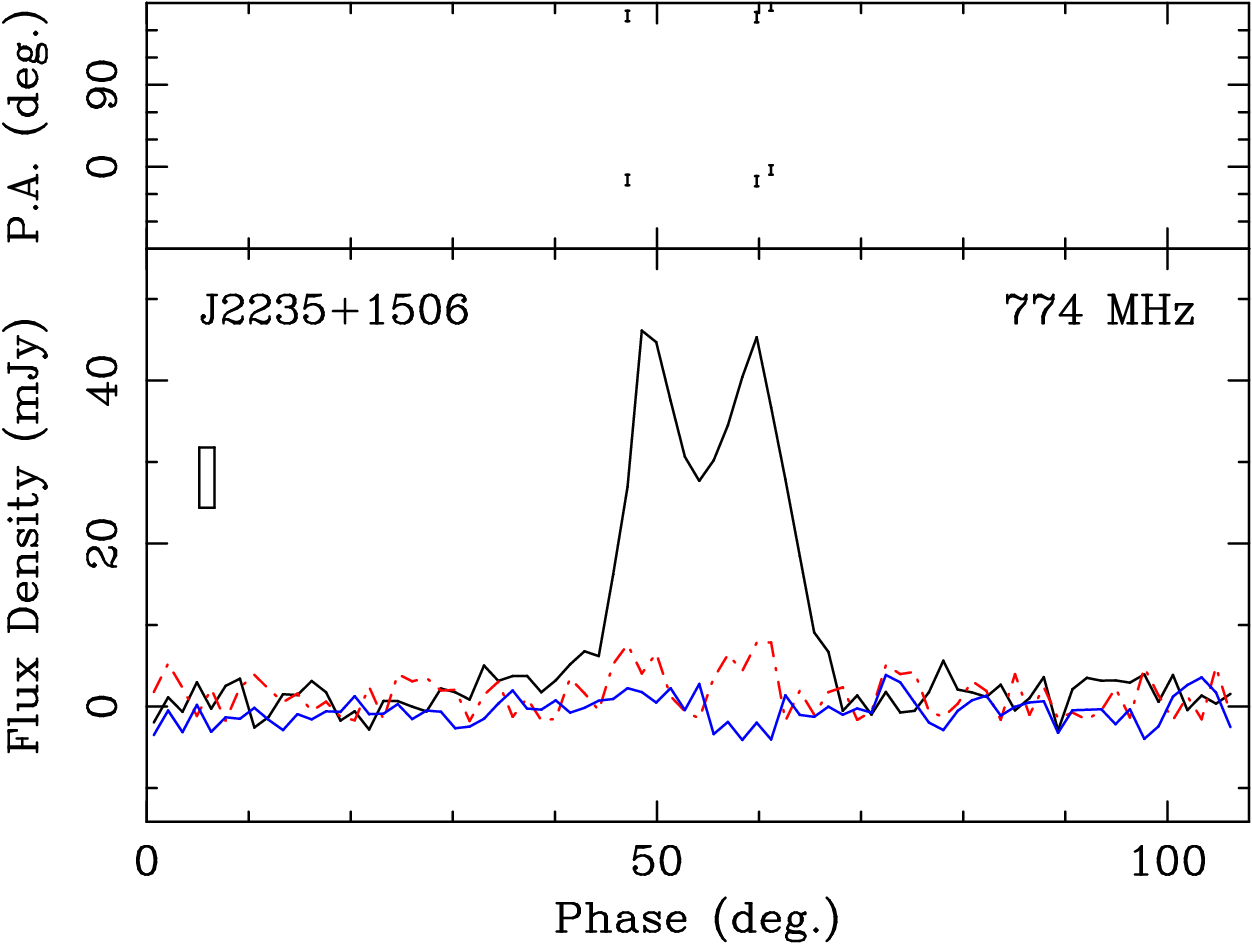,angle=270,width=70mm,height=46mm}}\\
\end{tabular}%
\caption{Polarization profiles of 8 millisecond pulsars. See keys in
Figure~\ref{cal}}
\label{msp}
\end{figure*}

\subsection{Millisecond pulsars}

The eight millisecond pulsars that we observed are shown in Figure~\ref{msp}.

PSR J1012+5307 has emission at almost all longitudes and appears to have
three main regions of emission.  Two and a half of these are 100\%
linearly polarized, confirming the observations at 610~MHz by
\citet{stc99} and at 1.41~GHz by \citet{xkj+98}. Note that the PA
convention in \citet{stc99} is different from other pulsar polarization
observations. We did not detect much circular polarization at 774~MHz,
in contrast with \citet{stc99} and \citet{xkj+98}.

PSR B1257+12 (J1300+1240) has a 100\% linearly polarized leading edge,
confirming the measurements at 1.41 GHz by \citet{xkj+98}. However, our 
observations indicate a reversal from right to left circular polarization
in the leading half
of the profile, and the circular polarization at 1.4~GHz in \citet{xkj+98}
is very different from our data \citep[see comments at the end of][]{mh04}.

PSR J1518+4904 has right-hand circular polarization at 774~MHz for
almost the full profile, rather than just the first half of the profile, 
as observed at 1.41~GHz by \citet{xkj+98} and \citet{stc99}.

Observations of PSR J1640+2224 show weak linear polarization and very weak
circular polarization at 774~MHz, in contrast to the high linear and
circular polarization at 1.41~GHz reported  by \citet{xkj+98}.

At 774 MHz, PSR B1953+29 (J1955+2908) shows observable linear polarization
in the stronger trailing component. Circular polarization is almost
undetectable anywhere in the profile, which is consistent with the result at
1.41~GHz reported by \citet{ts90} but in contrast to the strong
right circular polarization of the leading component observed at the same
frequency by \citet{xkj+98}.

PSRs J1453+1902, J1853+1303, and J2235+1506 have never been observed before.
Our GBT 774 MHz data show strong linear polarization of the former two and
weak polarization of the latter one, though the total signal-to-noise ratio 
(S/N) is not very high.

We see from our observations that the profiles of millisecond pulsars
are very wide, with their $W_{50}$ values among the highest (see
Table~\ref{tab1} and Figure~\ref{msp}). Second, the P.A. curves of 
all millisecond pulsars are rather flat where measurable (see 
Figure~\ref{msp}).

\begin{figure*}
\centering
\begin{tabular}{lll}
\mbox{\psfig{file=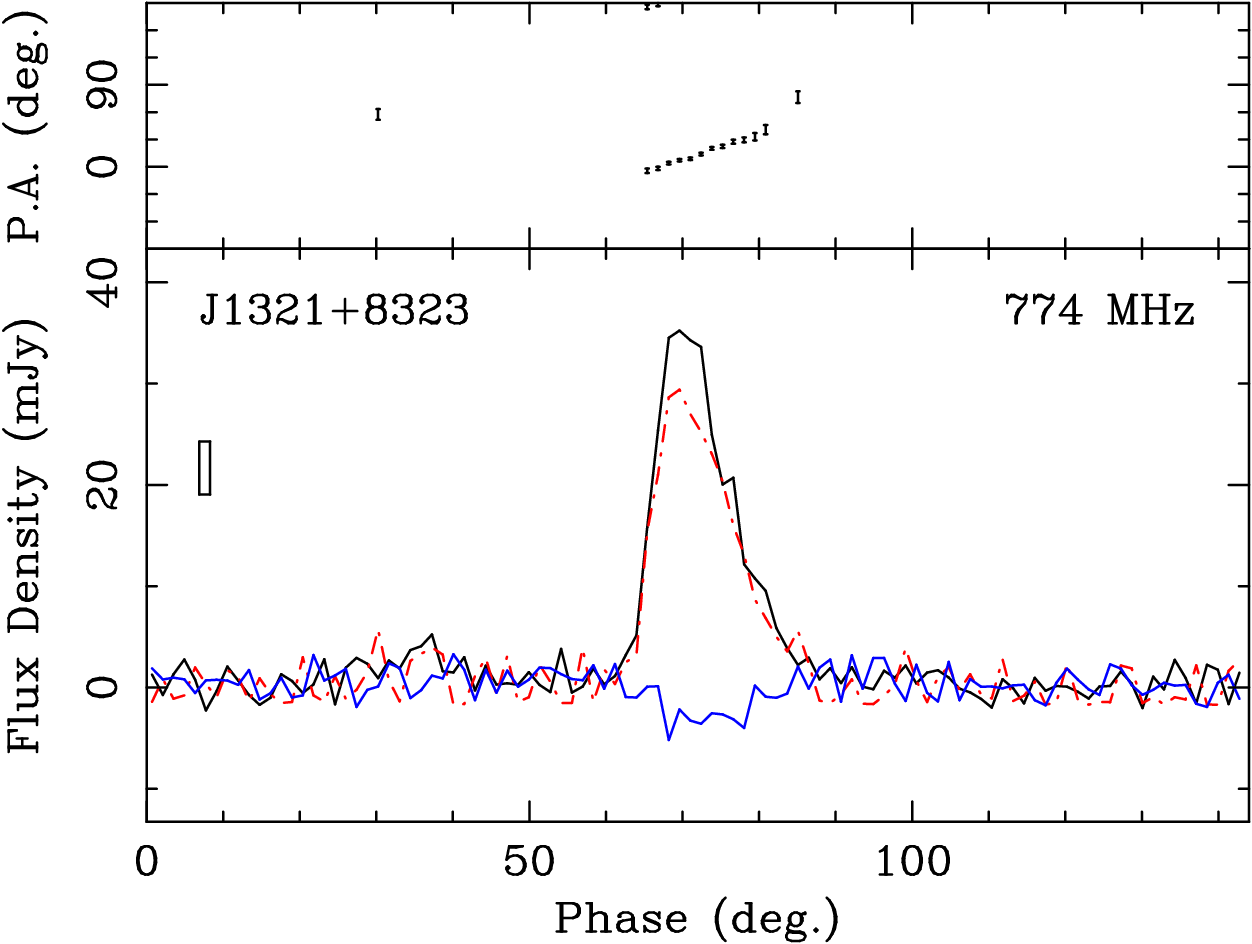,angle=270,width=55mm}}&
\mbox{\psfig{file=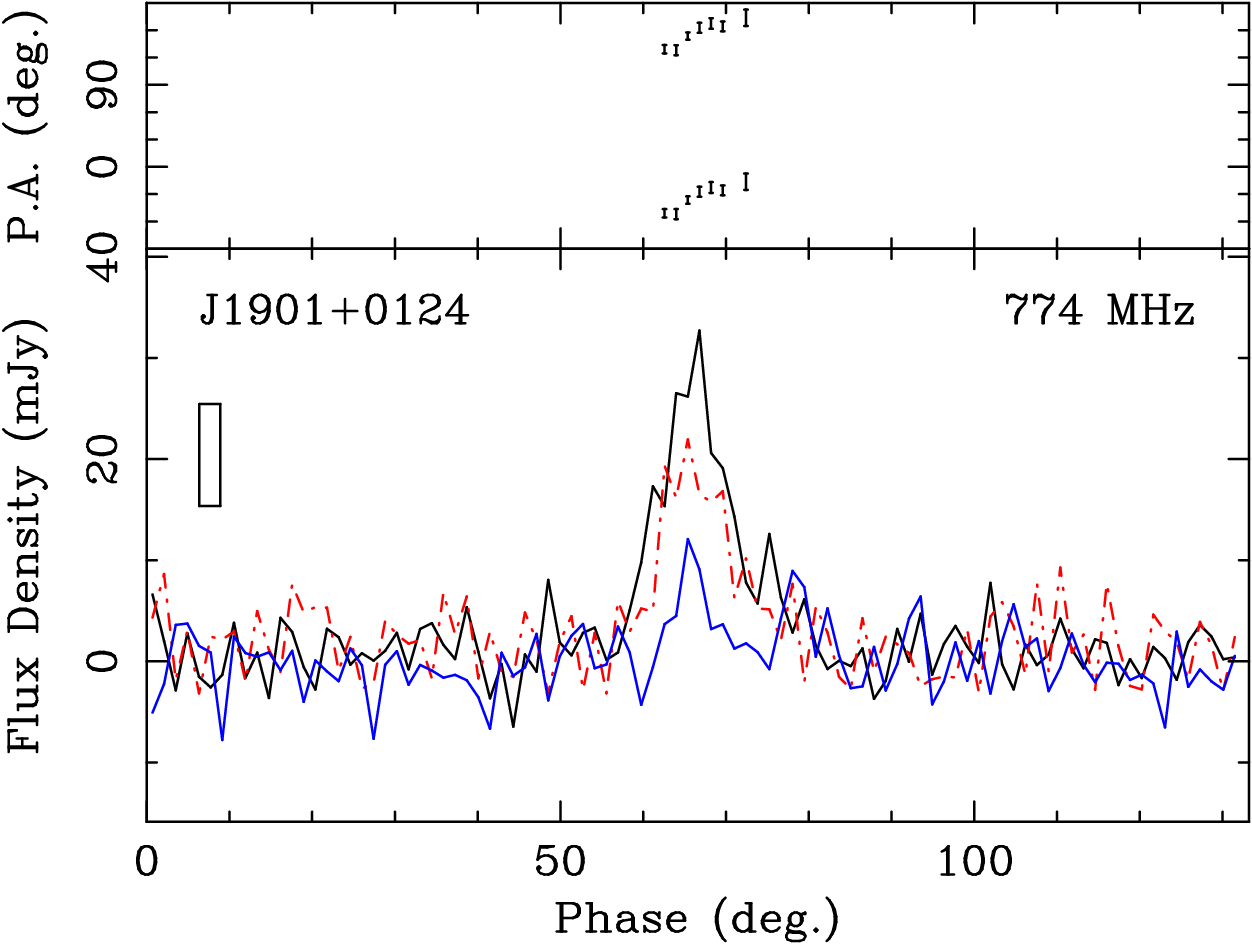,angle=270,width=55mm}}&
\mbox{\psfig{file=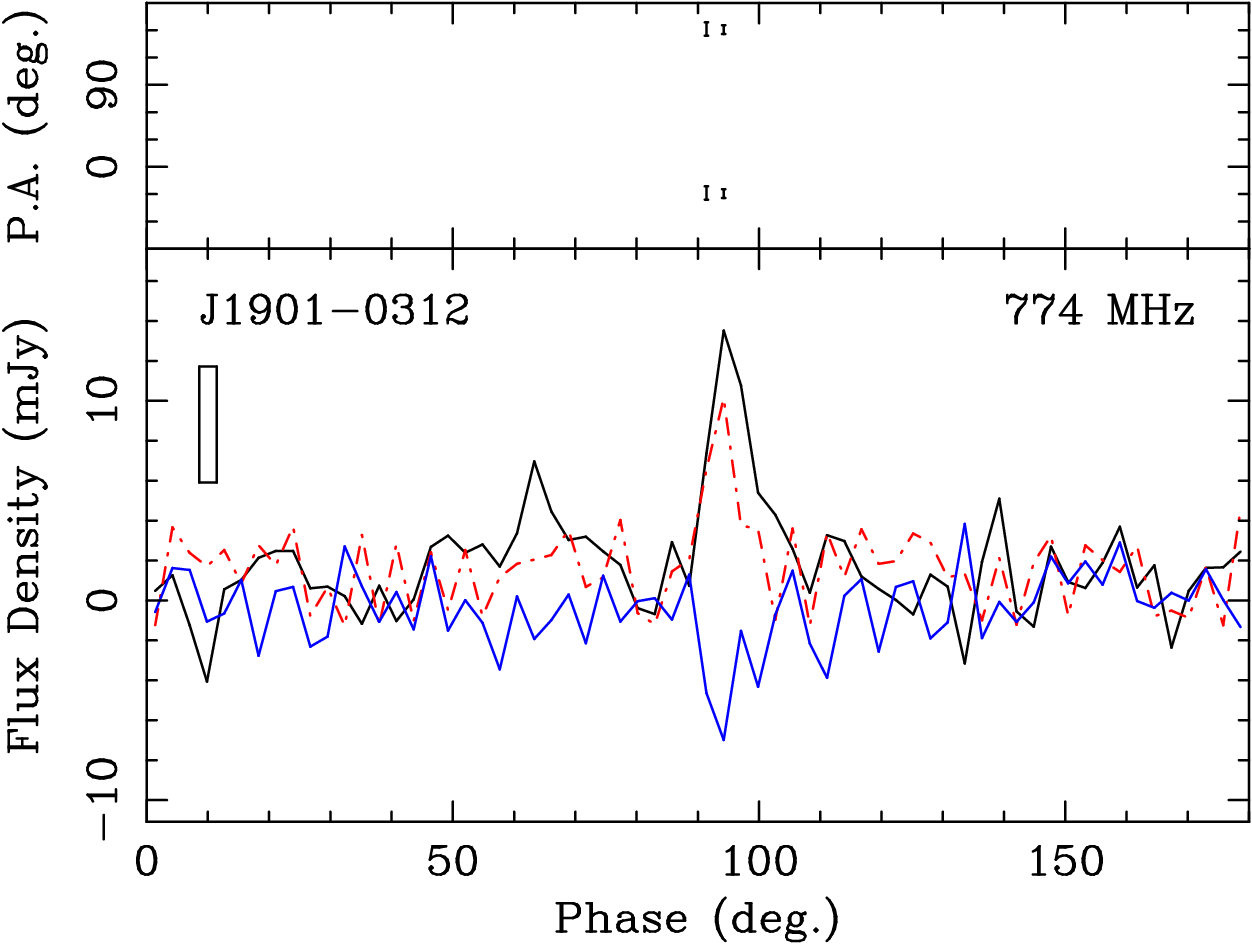,angle=270,width=55mm}}\\
\mbox{\psfig{file=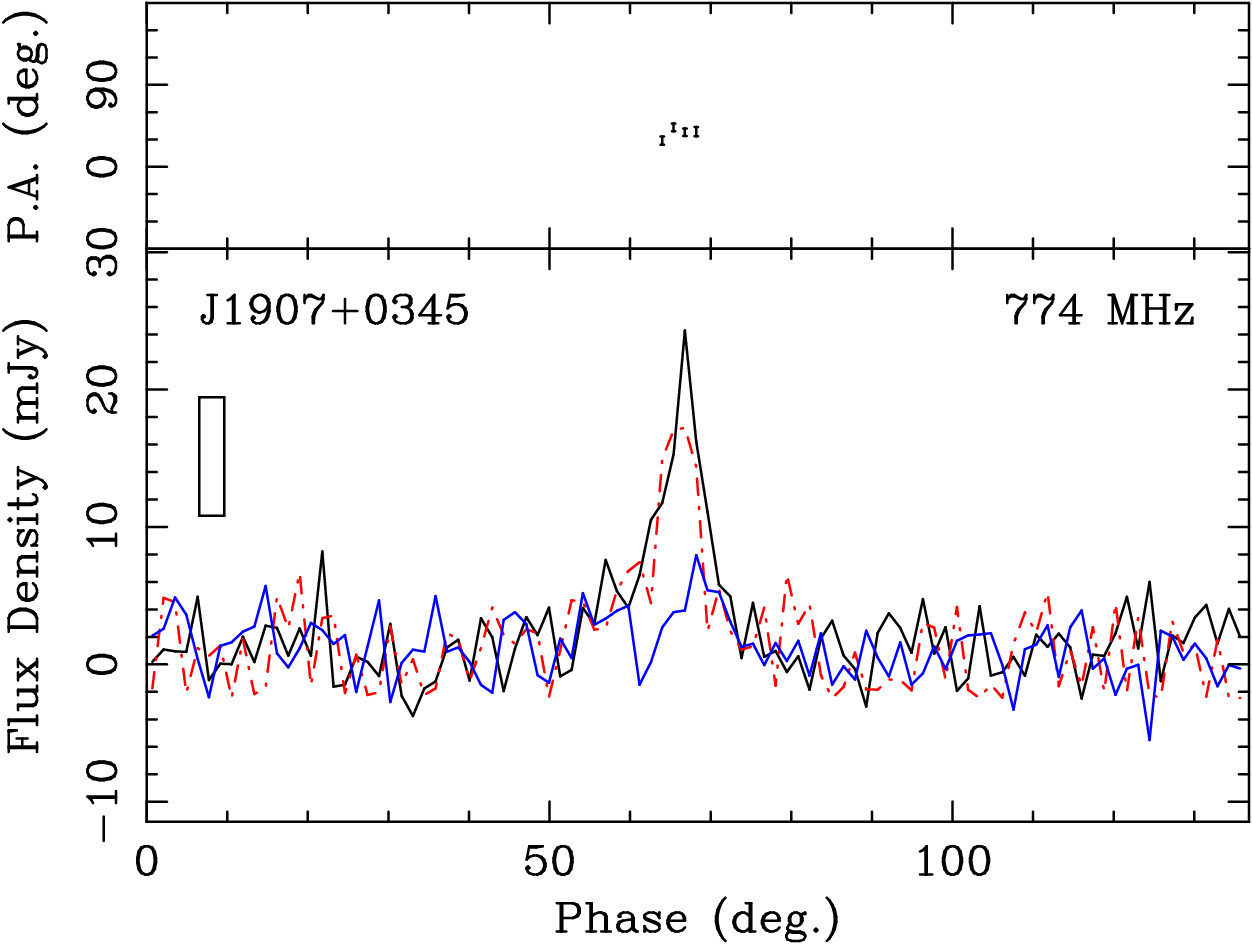,angle=270,width=55mm}}&
\mbox{\psfig{file=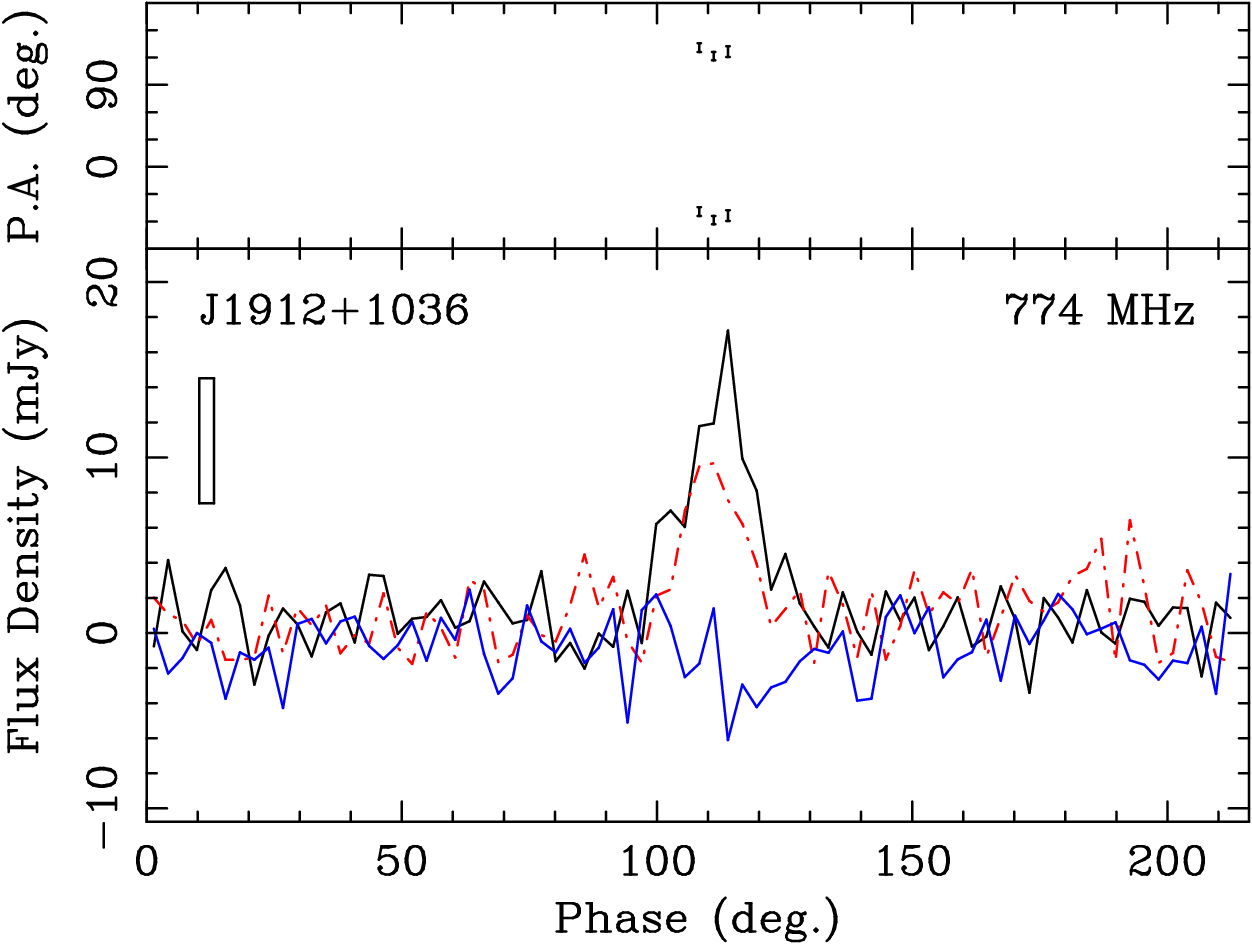,angle=270,width=55mm}}&
\mbox{\psfig{file=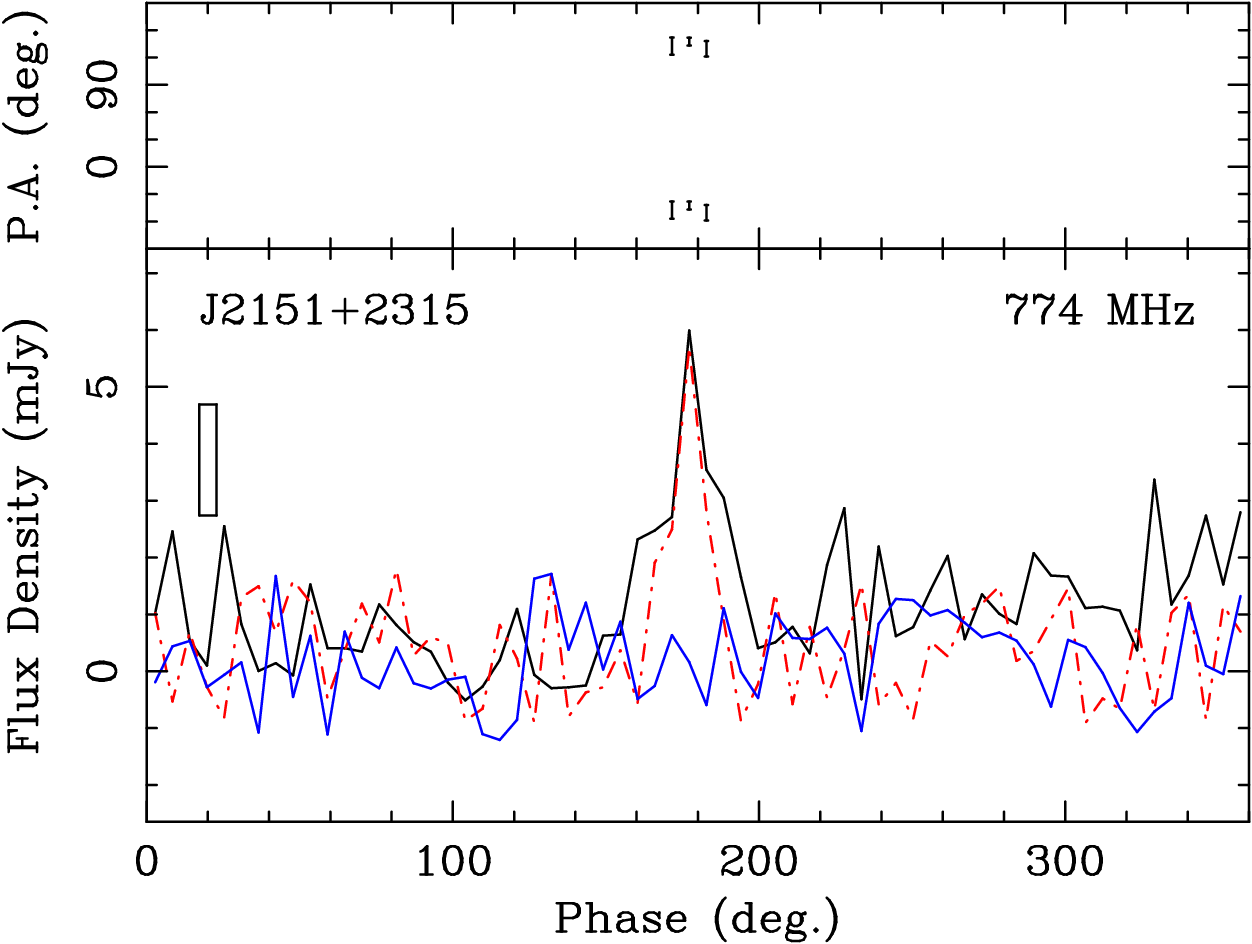,angle=270,width=55mm}}\\
\mbox{\psfig{file=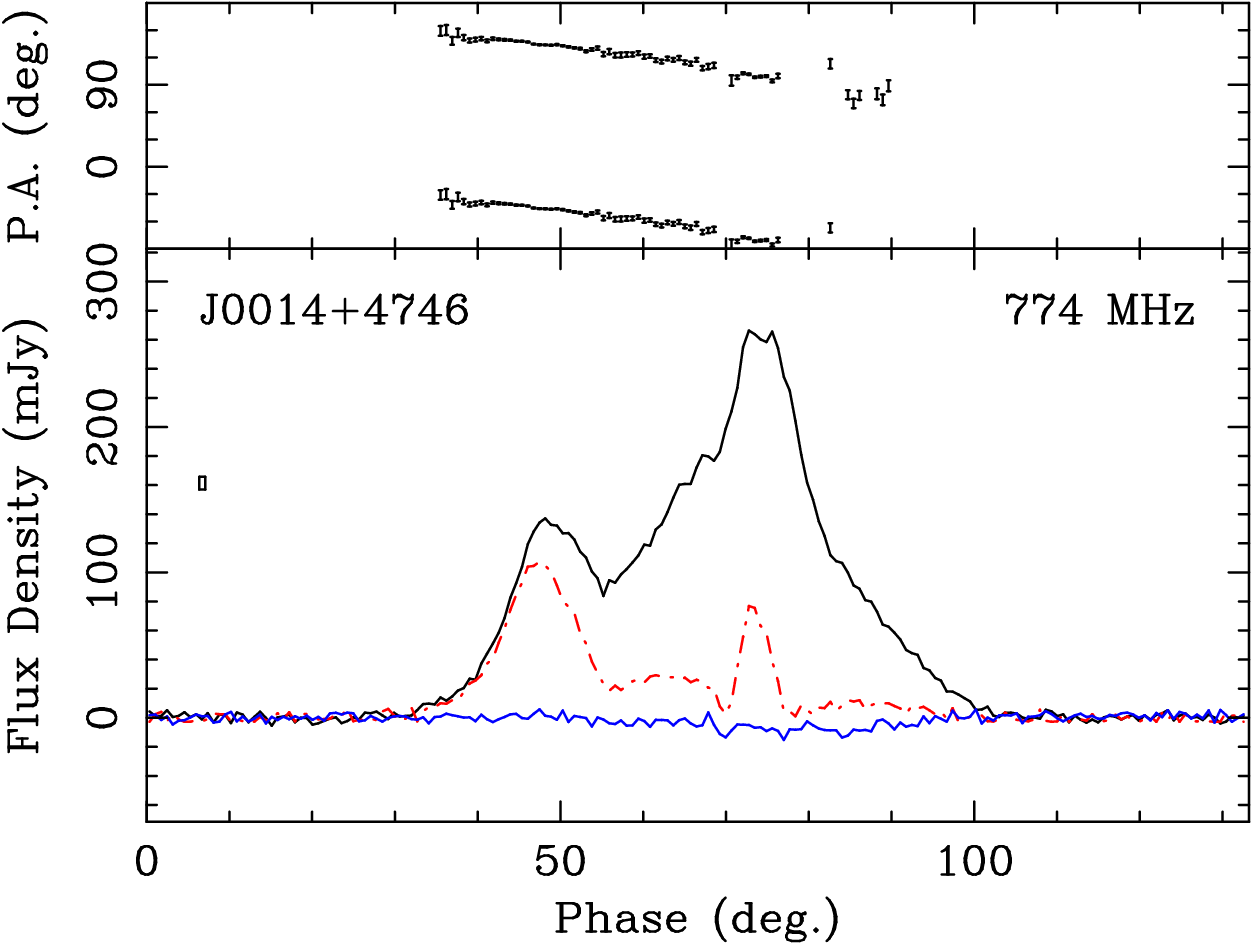,angle=270,width=55mm}}&
\mbox{\psfig{file=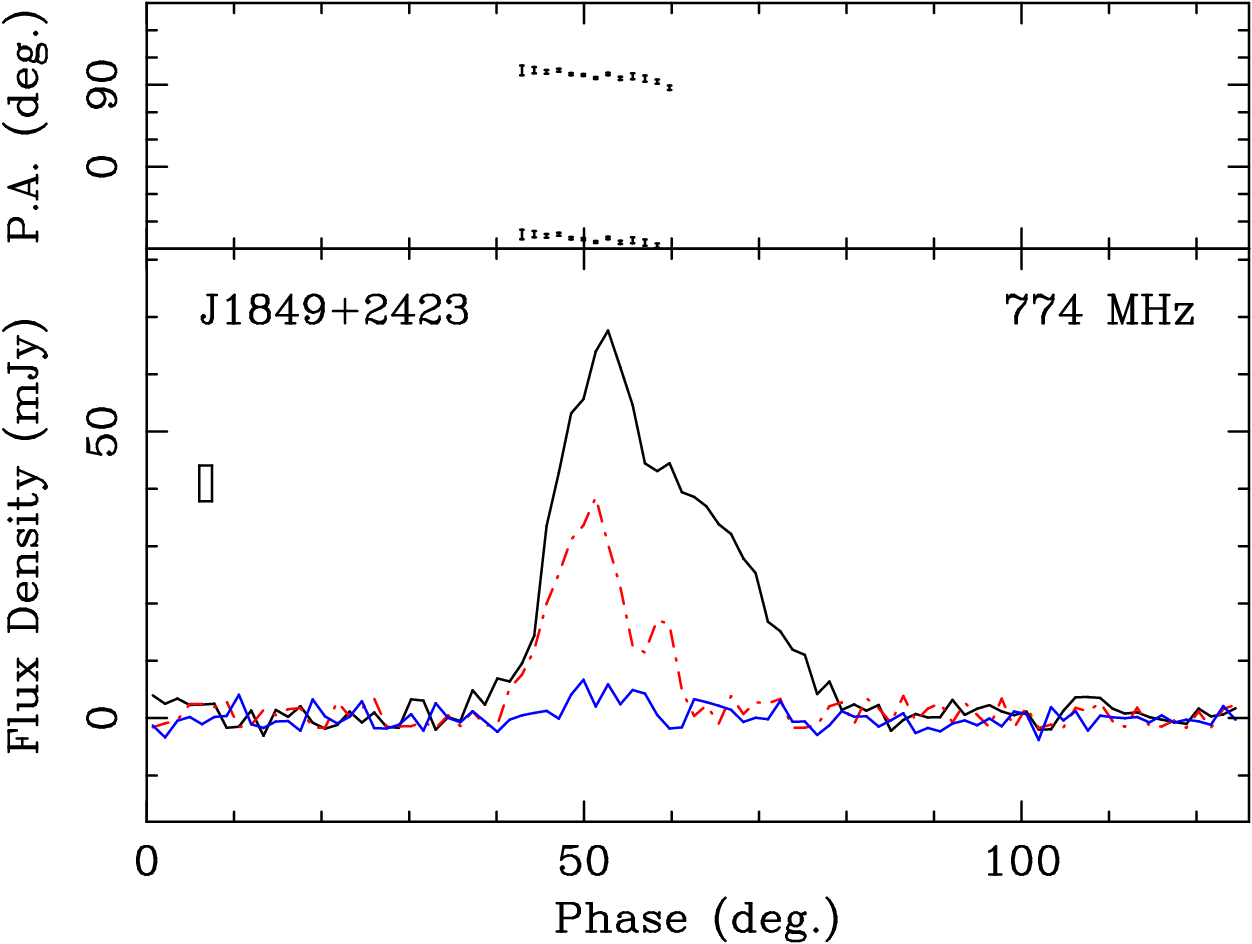,angle=270,width=55mm}}&
\mbox{\psfig{file=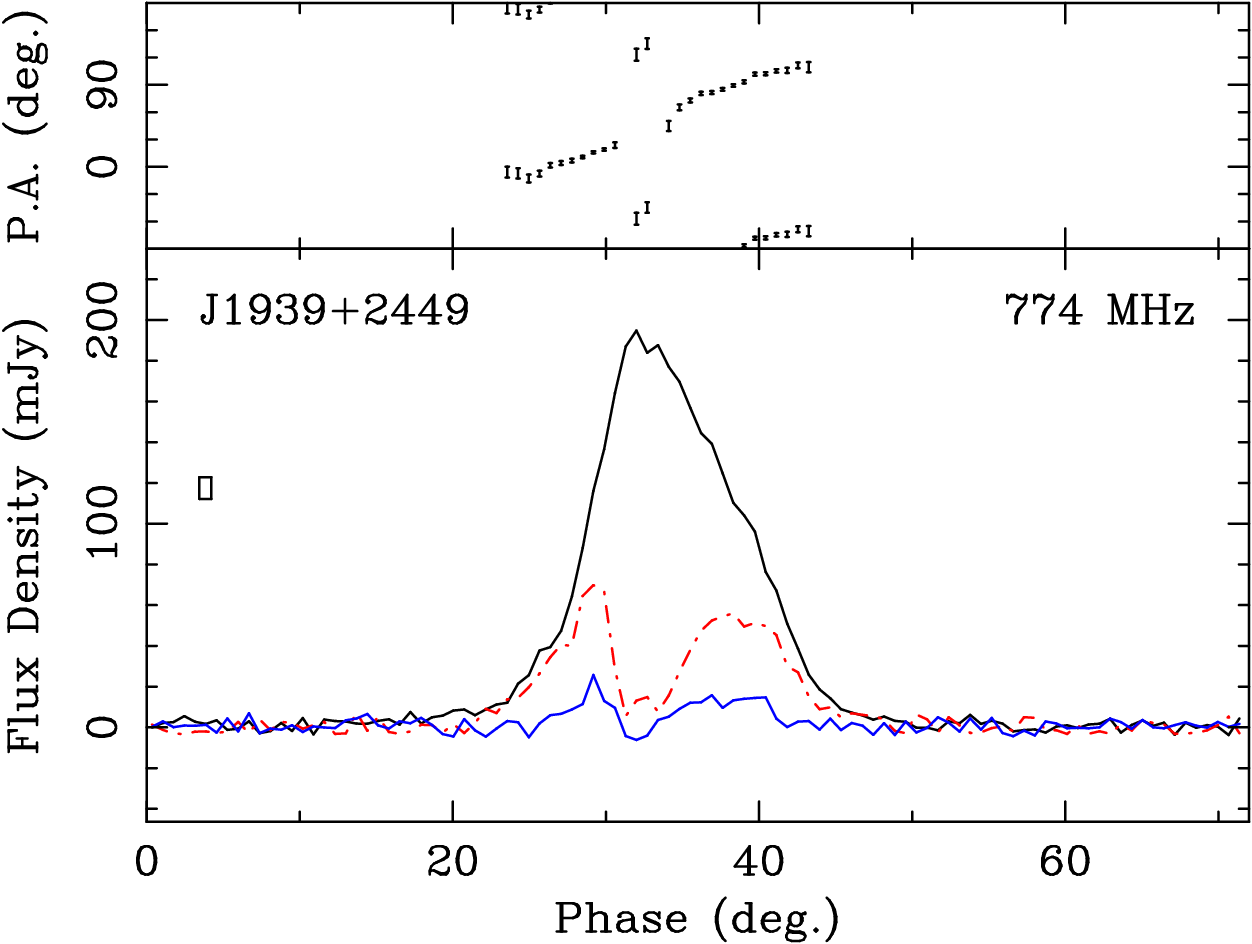,angle=270,width=55mm}}\\
\mbox{\psfig{file=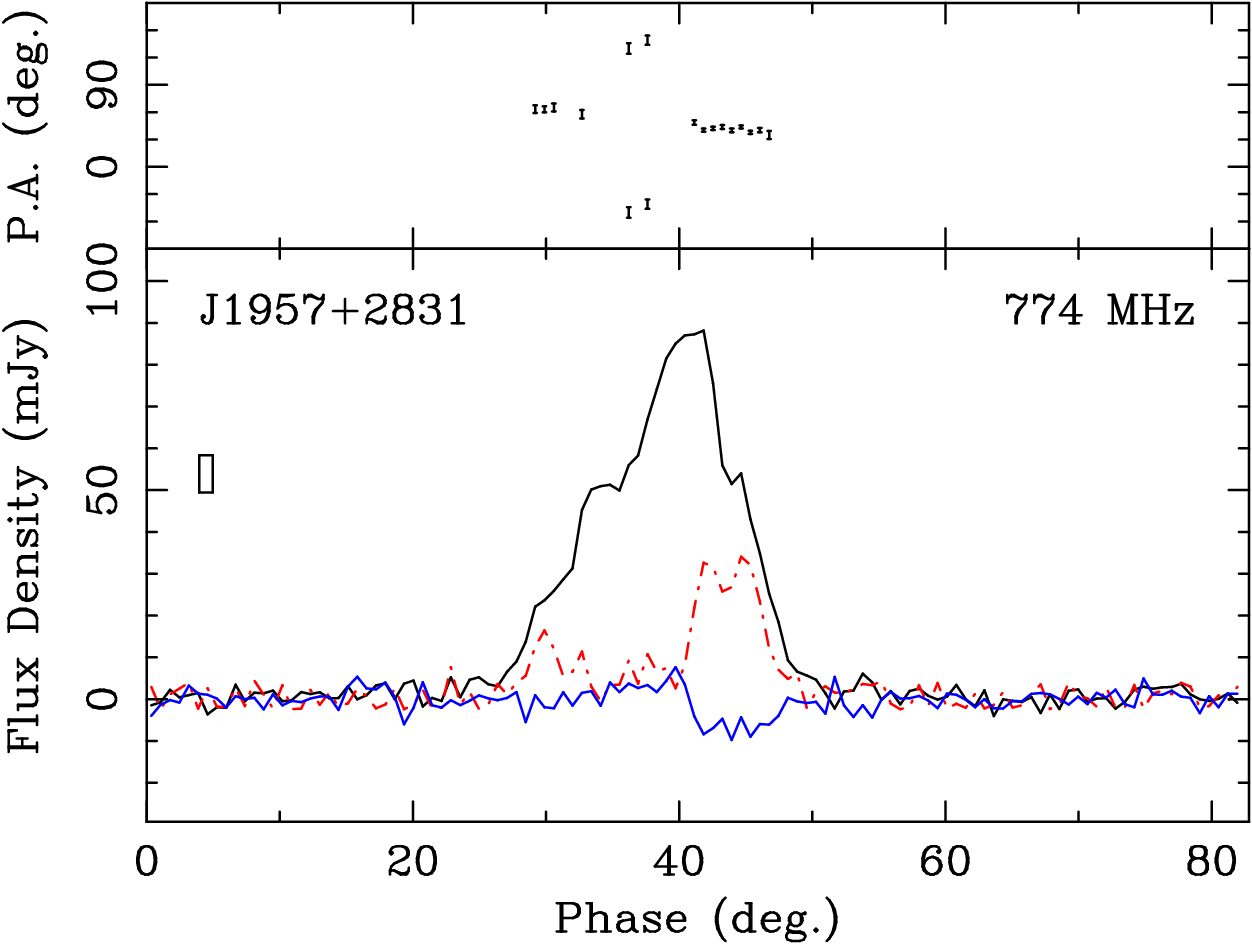,angle=270,width=55mm}}&
\mbox{\psfig{file=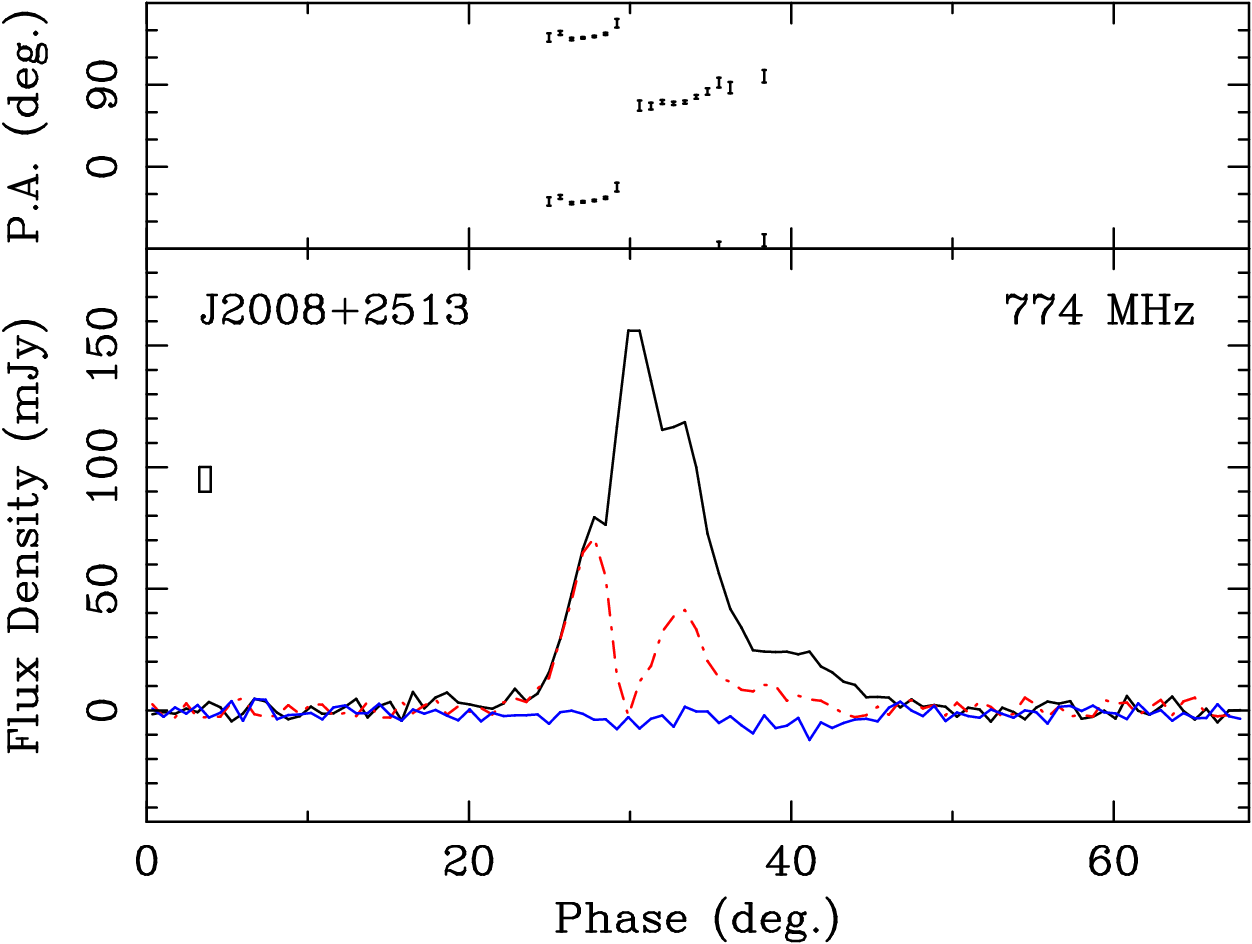,angle=270,width=55mm}}&
\mbox{\psfig{file=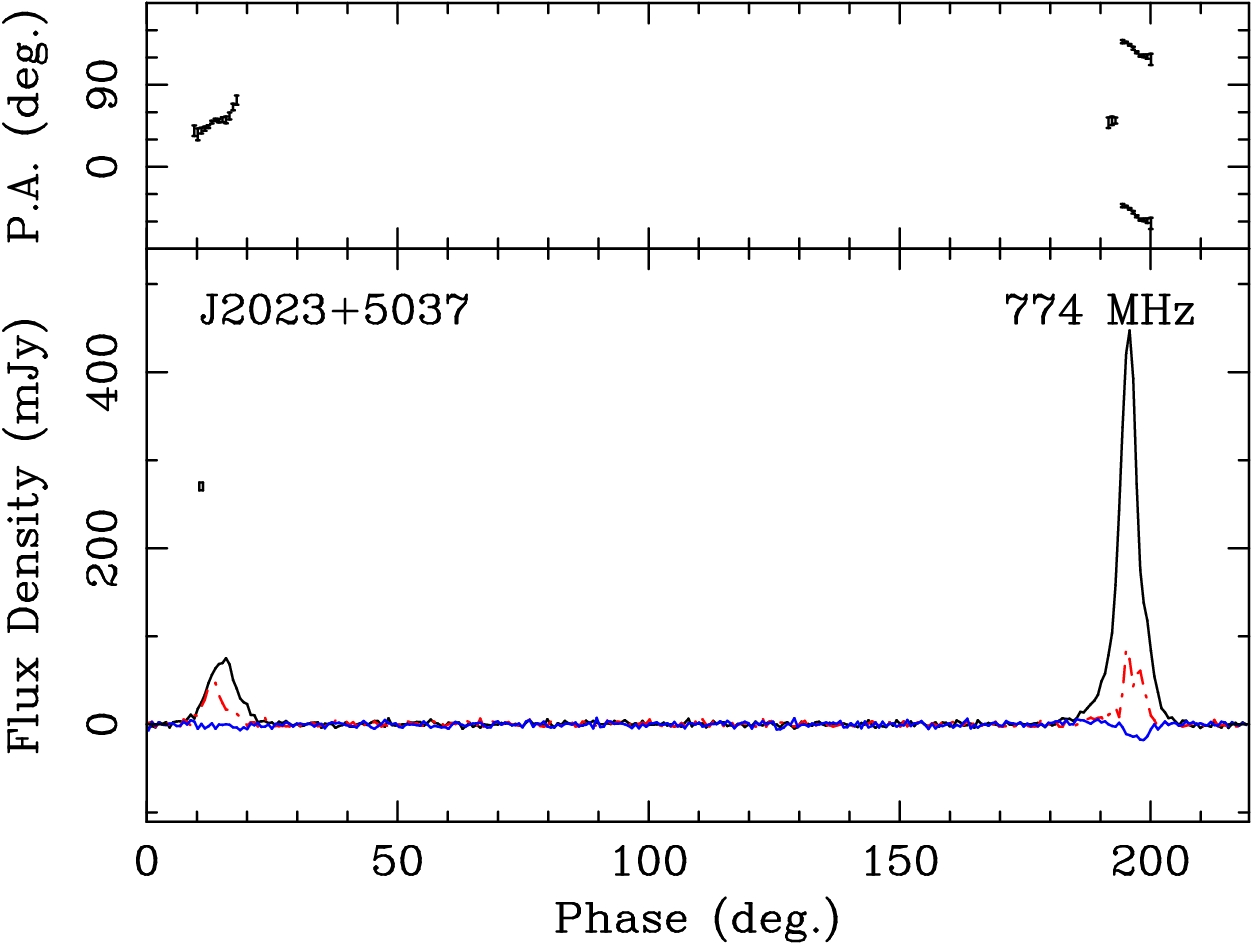,angle=270,width=55mm}}\\
\end{tabular}%
\caption{Top six pulsars have nearly 100\% polarization (linear plus
circular) across the entire pulse profile, and the bottom six pulsars have
nearly 100\% linear polarization for parts of their profiles. See keys in
Figure~\ref{cal}. Other pulsars belonging to this category are PSR B1929+10 in
Figure~\ref{cal}, PSRs J1012+5307 and B1257+12 (J1300+1240) in Figure~\ref{msp},
and PSRs B1935+25 (J1937+2544) and B1942+17 (J1944+1755) in Figure~\ref{dbp}.}
\label{hil}
\end{figure*}%
\subsection{Profiles with Highly Linearly Polarized Emission}

Many pulsars in our data set show highly linearly polarized emission 
(indicated by ``hiL'' in Table~\ref{tab1}) for all or part of their
profiles (Figure~\ref{hil}).

Five or six pulsars, PSRs B1322+83 (J1321+8323), J1901+0124, J1901$-$0312,
J1907+0345, B1910+10 (J1912+1036), and perhaps also PSR J2151+2315 appear to
have a single dominant component with up to 100\% linear polarization.
Of these pulsars, only PSR B1322+83 (J1321+8323) was previously observed;
our GBT observation at 774~MHz is consistent with those of GL98 (at 410,
606, 925, and 1408 MHz). The linear polarization percentage drops from
100\% at 410 MHz to 50\% at 1408 MHz, and the whole profile is right-hand
circularly polarized. The precursor is clearly visible at 606 MHz but only
marginally detected around the phase of $35\degr$ at 774~MHz 
in our Figure \ref{hil}.

A number of pulsars exhibit nearly 100\% linear polarization at 774~MHz in
only the leading or trailing edge of the pulse profile; for example, the
bottom six pulsars in Figure~\ref{hil}, PSR B1929+10 in Figure~\ref{cal}, two
millisecond pulsars, PSRs J1012+5307 and B1257+12 (J1300+1240) in
Figure~\ref{msp}, and two double component pulsars, PSRs B1935+25 (J1937+2544)
and B1942+17 (J1944+1755) in Figure~\ref{dbp}.
PSR B0011+47 (J0014+4746) was previously observed by GL98 at 404, 610, 925,
1414, and 1642~MHz. Our GBT observation at 774~MHz has high S/N and for the
first time clearly shows the 100\% linearly polarized leading edge.
The interpulse pulsar PSR B2022+50 (J2023+5037) has been observed by GL98
at 410, 610, 925, 1408, and 1642 MHz. Our GBT observation has the best S/N,
and clearly shows both the orthogonal mode of the main pulse and the 100\%
linear polarization of the leading edge of the interpulse.
PSR B1935+25 (J1937+2544) in Figure \ref{dbp} shows 100\% linear polarization
in the trailing edge of the leading component; another double component
pulsar, PSR B1942+17 (J1944+1755) (Figure~\ref{dbp}) exhibits highly linearly
polarized emission in the bridge.

\begin{figure}[htb]
\centering
\mbox{\psfig{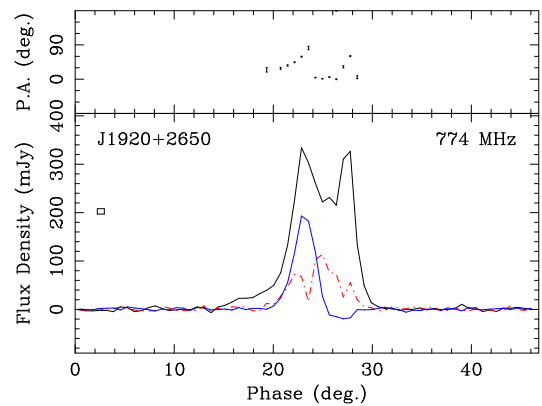}}
\caption{PSR B1918+26 (J1920+2650) shows extremely strong circular polarization. 
See keys in Figure~\ref{cal}.}
\label{hicp}
\end{figure}
\subsection{Highly Circular Polarized Component}

The average pulse profiles of most pulsars exhibit relatively weak circular
polarization. Therefore, in our sample, PSR B1918+26 (J1920+2650, see
Figure~\ref{hicp}) is outstanding for the strong left-hand circular
polarization observed in its first major component, with fractional
circular polarization reaching 64\%.  Observations at 1.4~GHz
\citep{wcl+99} and at 610~MHz (GL98) also show extraordinarily high
circular polarization in this component. Similar characteristics have been
previously observed in very few other pulsars, notably PSRs B1702$-$19
\citep{lm88}, J1603$-$7202 \citep{mh04} and J1907+0918 \citep{lx00}. This
is only the fourth case of fractional circular polarization exceeding $50\%$
in any component of a mean pulse profile.

\begin{figure*}
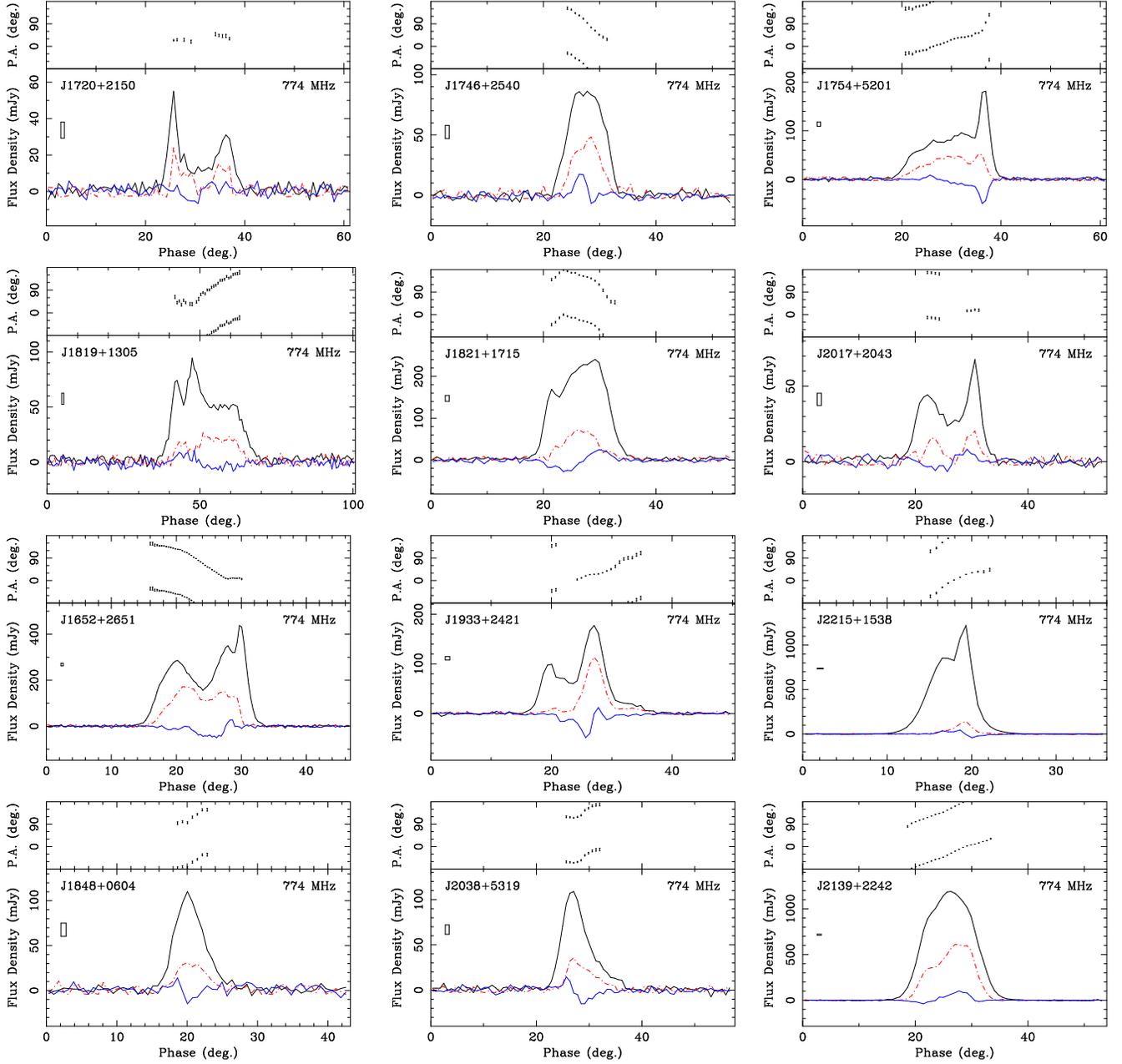

\centering
\begin{tabular}{lll}
\mbox{\psfig{file=fig6_1.ps,angle=270,width=55mm}}&
\mbox{\psfig{file=fig6_2.ps,angle=270,width=55mm}}&
\mbox{\psfig{file=fig6_3.ps,angle=270,width=55mm}}\\
\mbox{\psfig{file=fig6_4.ps,angle=270,width=57mm}}&
\mbox{\psfig{file=fig6_5.ps,angle=270,width=55mm}}&
\mbox{\psfig{file=fig6_6.ps,angle=270,width=55mm}}\\
\mbox{\psfig{file=fig6_7.ps,angle=270,width=55mm}}&
\mbox{\psfig{file=fig6_8.ps,angle=270,width=55mm}}&
\mbox{\psfig{file=fig6_9.ps,angle=270,width=55mm}}\\
\mbox{\psfig{file=fig6_10.ps,angle=270,width=55mm}}&
\mbox{\psfig{file=fig6_11.ps,angle=270,width=55mm}}&
\mbox{\psfig{file=fig6_12.ps,angle=270,width=55mm}}\\
\end{tabular}%
\caption{These pulsars, together with PSRs B1933+16 and B1857$-$26
(J1900$-$2600) in Figure~\ref{cal}, PSR B1257+12 (J1300+1240) in
Figure~\ref{msp}, PSR J1957+2831 in Figure~\ref{hil} and PSR B1918+26
(J1920+2650) in Figure~\ref{hicp}, show clear reversals of circular
polarization sense.  See keys in Figure~\ref{cal}.  The top six pulsars show the
sense reversal at the pulse center, the next three pulsars have a reversal in a
component at a later longitude, and the bottom three pulsars have a reversal at
an earlier longitude.}
\label{cpr}
\end{figure*}%
\subsection{Sense Reversal of Circular Polarization}

Many pulsars show sense reversal of circular polarization as a function
of pulse phase. Together with the steepness of the polarization angle
curve, this is an important feature for identifying the core
emission component \citep{ran83,rr90,ran90}. The core component is
located near the center of an emission beam, and sometimes has
an offset toward a later longitude due to a retardation effect
\citep{xqh97,gg03}.
Similar reversal of the sense of circular polarization is observed in many
of the pulsars in our sample, as shown in Figure~\ref{cpr}. 

The sense reversal of circular polarization of the first six pulsars in the
upper part of Figure~\ref{cpr} and PSR J1957+2831 in Figure~\ref{hil} is
probably associated with an either unresolved or marginally resolved core
component near the pulse center.
Three pulsars, PSRs J1652+2651, B1931+24 (J1933+2421), J2215+1538, and also
PSR B1919+26 (J1920+2650) in Figure~\ref{hicp}, show a sense reversal from a
component located at a later longitude than the pulsar center, which
probably is not core emission, except for B1931+24.
The sense reversals of another three pulsars, PSRs J1848+0604, B2036+53
(J2038+5319), and J2139+2242, as well as the millisecond pulsar PSR
J1300+1240, happen at an earlier longitude than the pulse center and do
not originate from core emission. The sense reversal in PSR B2036+53
(J2038+5319) is associated with the component peak, but two reversals of PSR
J2139+2242 are located between two or three unresolved components, very
similar to the case of the calibration pulsar PSR B1933+16.

%
\begin{figure*}
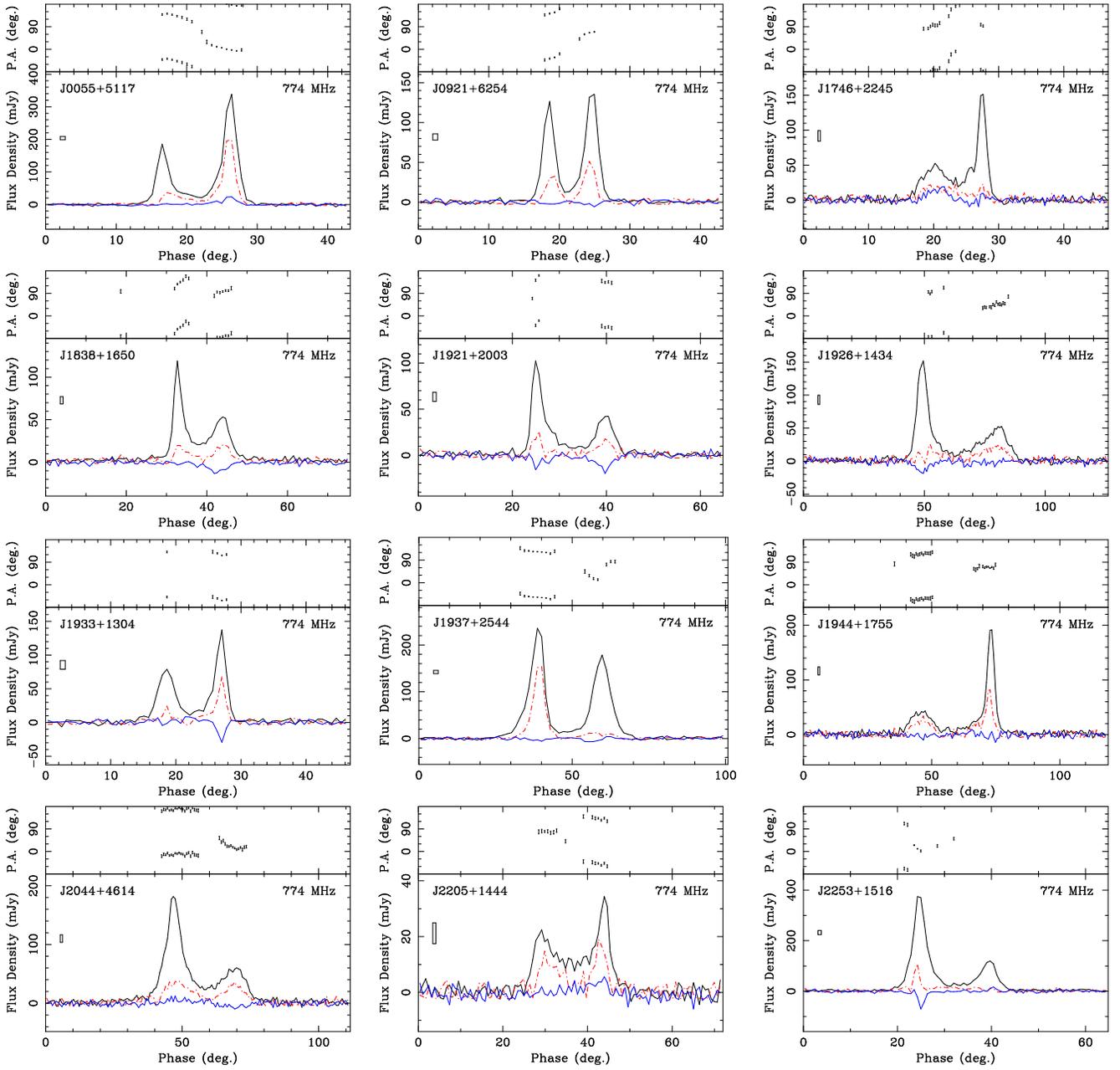

\begin{center}
\begin{tabular}{lll}
\mbox{\psfig{file=fig7_1.ps,angle=270,width=55mm}}& 
\mbox{\psfig{file=fig7_2.ps,angle=270,width=55mm}}&
\mbox{\psfig{file=fig7_3.ps,angle=270,width=55mm}}\\
\mbox{\psfig{file=fig7_4.ps,angle=270,width=55mm}}&
\mbox{\psfig{file=fig7_5.ps,angle=270,width=55mm}}&
\mbox{\psfig{file=fig7_6.ps,angle=270,width=55mm}}\\ 
\mbox{\psfig{file=fig7_7.ps,angle=270,width=55mm}}&
\mbox{\psfig{file=fig7_8.ps,angle=270,width=57mm}}&
\mbox{\psfig{file=fig7_9.ps,angle=270,width=55mm}}\\ 
\mbox{\psfig{file=fig7_10.ps,angle=270,width=55mm}}&
\mbox{\psfig{file=fig7_11.ps,angle=270,width=55mm}}&
\mbox{\psfig{file=fig7_12.ps,angle=270,width=55mm}}\\ 
\end{tabular}%
\caption{Polarization profiles of double-component pulsars at 774~MHz. 
PSR J2017+2043 in Figure~\ref{cpr} is a double-component pulsar as well.
Some of these pulsars are classical {\it conal-double} pulsars.}
\label{dbp}
\end{center}
\end{figure*}
\subsection{Double Component Pulsars}

Many pulsars show dominant double-peaked components (Figure \ref{dbp})
that, in some cases, originate from the (possibly single outer) conal
emission beam.  These are the classical {\it conal-double} pulsars which
are also characterized by the ``S'' shape of polarization angle (PA)
curves \citep{ran83}.  Strong correlation between the senses of PA
variation and the handedness of circular polarization has been found for
conal double pulsars by \citet{hmxq98}.  Decreasing P.A. curves are
associated with left hand (positive Stokes V) circular polarization, and
increasing P.A. curves with right-hand circular polarization. See
\citet{yh06} for updates. Below, we discuss whether the conal-double 
pulsars in our sample follow this correlation.

We have observed many double-component pulsars (see Figure~\ref{dbp}).
Not all double-component pulsars are conal-double in nature; for
example, most of the double-peak pulsars discussed by \citet{jw06} are
not conal-double pulsars. In our observations, PSRs B1935+25 (J1937+2544)
and J2253+1516 have double peaks but, with a fast P.A. sweep and even
orthogonal polarization modes, the second peak of PSR B1935+25 and the
first peak of PSR J2253+1516 (see Figure~\ref{dbp}) look more
like core components. The polarization observation at 691~MHz by
\citet{jkk+07} not only confirms our profile at 774~MHz, but also shows
up to three unresolved components in the second peak of PSR B1935+25.
The leading two separate and the trailing one disappears at 3100~MHz
\citep{jkk+07}. Therefore, PSRs B1935+25 and J2253+1516 likely
belong to the partial cone class \citep{lm88}.

The remaining ten pulsars in Figure~\ref{dbp} are possible conal-double
pulsars, with a few questionable cases such as the strange P.A. curves of
PSRs J1838+1650 and J2044+4614 and the very weak detection of circular
polarization from PSRs B1942+17 (J1944+1755) and J2044+4614.
We found that 
PSRs B0052+51 (J0055+5117) and J2205+1444 
have a decreasing P.A. and weak left-hand (positive) circular
polarization.
PSRs B0917+63 (J0921+6254), J1838+1650, B1924+14 (J1926+1434), and
B1942+17 (J1944+1755)
have an increasing P.A. and right-hand (negative) circular polarization. 
These six pulsars agree with the previously observed correlation between
the P.A. curve slope and circular polarization sense;
however, three other pulsars do not comply with the correlation. PSR
J1746+2245 shows an increasing P.A. curve but left-hand circular
polarization. PSRs B1919+20 (J1921+2003) and B1930+13 (J1933+1304) have
right-hand circular polarization, but their P.A. curves seem to decrease.

PSR  B0052+51 (J0055+5117) was included in the updated list of conal-double
pulsars in Table~4 of \citet{yh06} according to data by GL98. A possible
core component hinted by 408~MHz data (GL98) is unrecognizable in our
774~MHz profile.

PSR B0917+63 (J0921+6254) was observed by GL98 below 1.4~GHz. Our GBT
data have better S/N, and confirm that it is a conal double with 
clear increasing P.A. swing and weak right-hand circular polarization.

\begin{figure*}
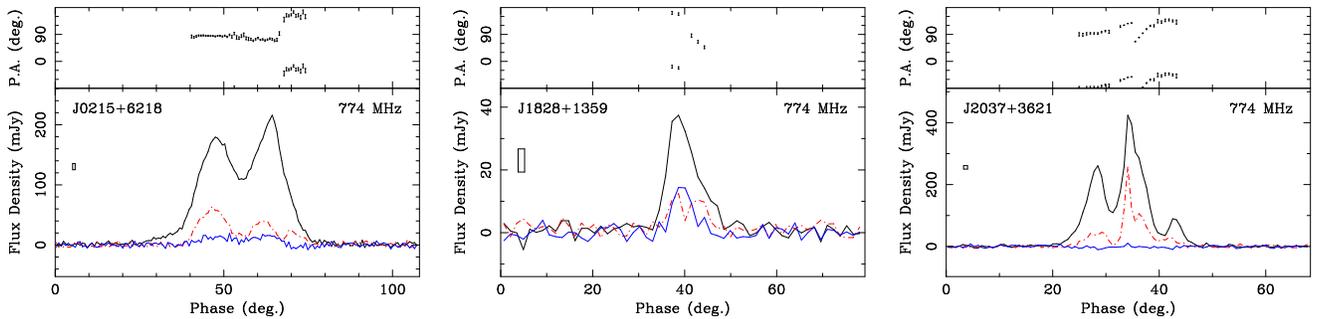

\begin{center}
\begin{tabular}{lll}
\mbox{\psfig{file=fig8_1.ps,angle=270,width=55mm}}&
\mbox{\psfig{file=fig8_2.ps,angle=270,width=55mm}}&
\mbox{\psfig{file=fig8_3.ps,angle=270,width=55mm}}\\ 
\end{tabular}%
\caption{These three pulsars show sudden mode jumps in polarization angle
curves, as discussed in the text. PSR B2035+36 
(J2037+3621) appears to have a nonorthogonal mode jump.}
\label{ort}
\end{center}
\end{figure*}
\subsection{Orthogonal Modes}

Radio pulsar emission at a given rotational phase may be an incoherent sum
of orthogonally polarized modes. These modes may be generated in one region,
one with P.A. parallel and one perpendicular to the magnetic field. It is also
possible that the modes come from different emission regions and hence
different PAs may appear at a given pulse longitude, producing
nonorthogonal modes \citep[e.g.,][]{xqh97}. The orthogonal polarization
modes or superposition of emission from two regions leads to complicated PA
curves in the integrated profiles.  The typical signature of this effect is
a sudden $\sim$90$\degr$ jump in P.A. at a certain pulse phase, accompanied by
a reduction in the degree of linear polarization.

Many of the targets that we observed show evidence for orthogonal modes in
their P.A. curves: PSRs B1929+10, B1933+16, B1857-26 (Figure~\ref{cal}),
J0215+6218 (Figure~\ref{ort}), J1518+4904 (Figure~\ref{msp}), J1828+1359
(Figure~\ref{ort}), B1937+24 (J1939+2449), J1957+2831, J2008+2513, B2022+50
(J2023+5037) (Figure~\ref{hil}), and J2253+1515 (Figure~\ref{dbp}).
We also note that the P.A. curves of PSRs B1918+26 (J1920+2650,
Figure~\ref{hicp}) and B2035+36 (J2037+3621, Figure~\ref{ort}) may result from
nonorthogonal emission modes.  Our observation of PSR B2035+36 (J2037+3621)
has a better S/N than those at 1.41~GHz by \citet{wcl+99} and 606, 925,
1408 and 1642~MHz by GL98.

\begin{figure*}
\begin{tabular}{rrrr}
\mbox{\psfig{file=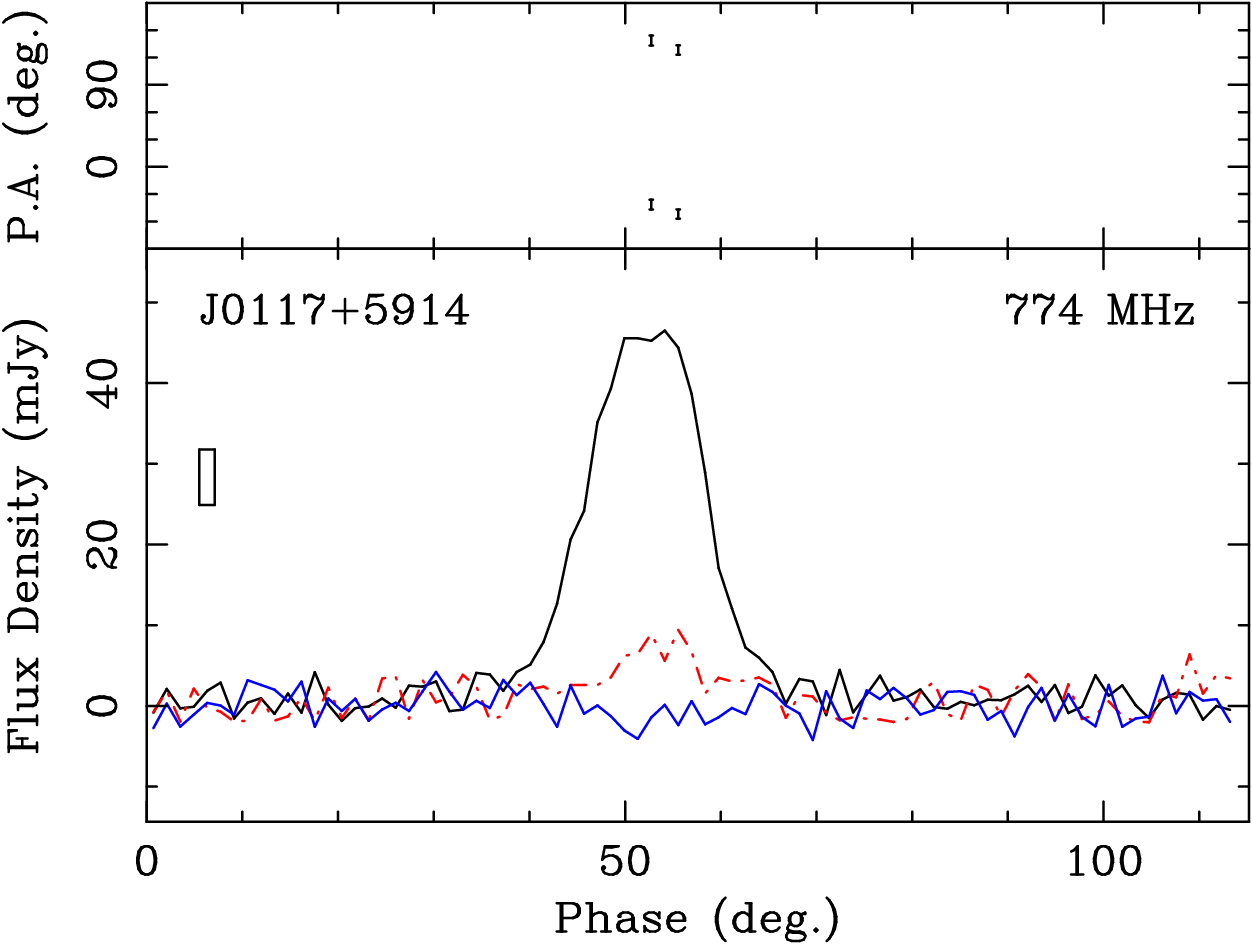,angle=270,width=55mm,height=39mm}}&	 
\mbox{\psfig{file=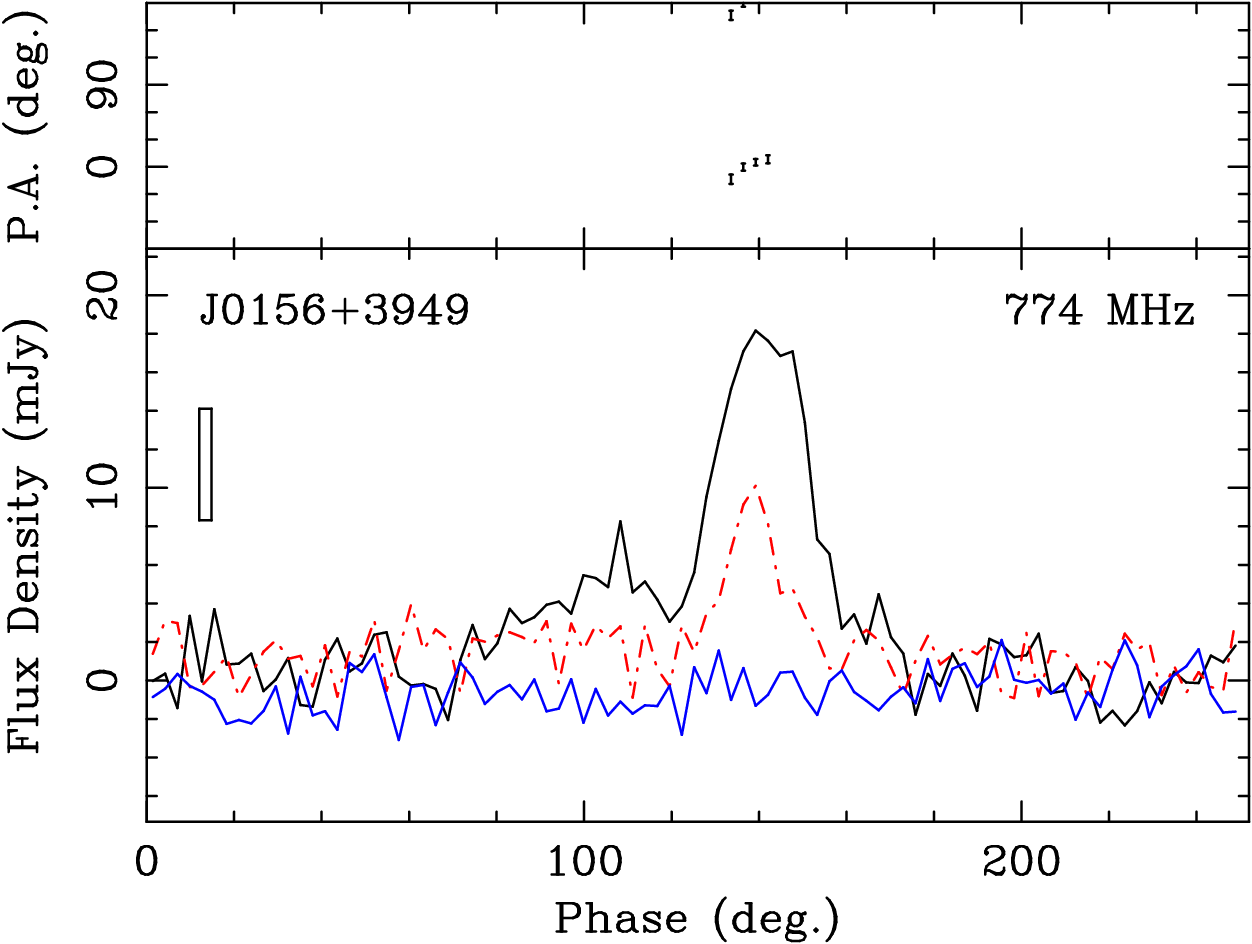,angle=270,width=55mm,height=39mm}}&	 
\mbox{\psfig{file=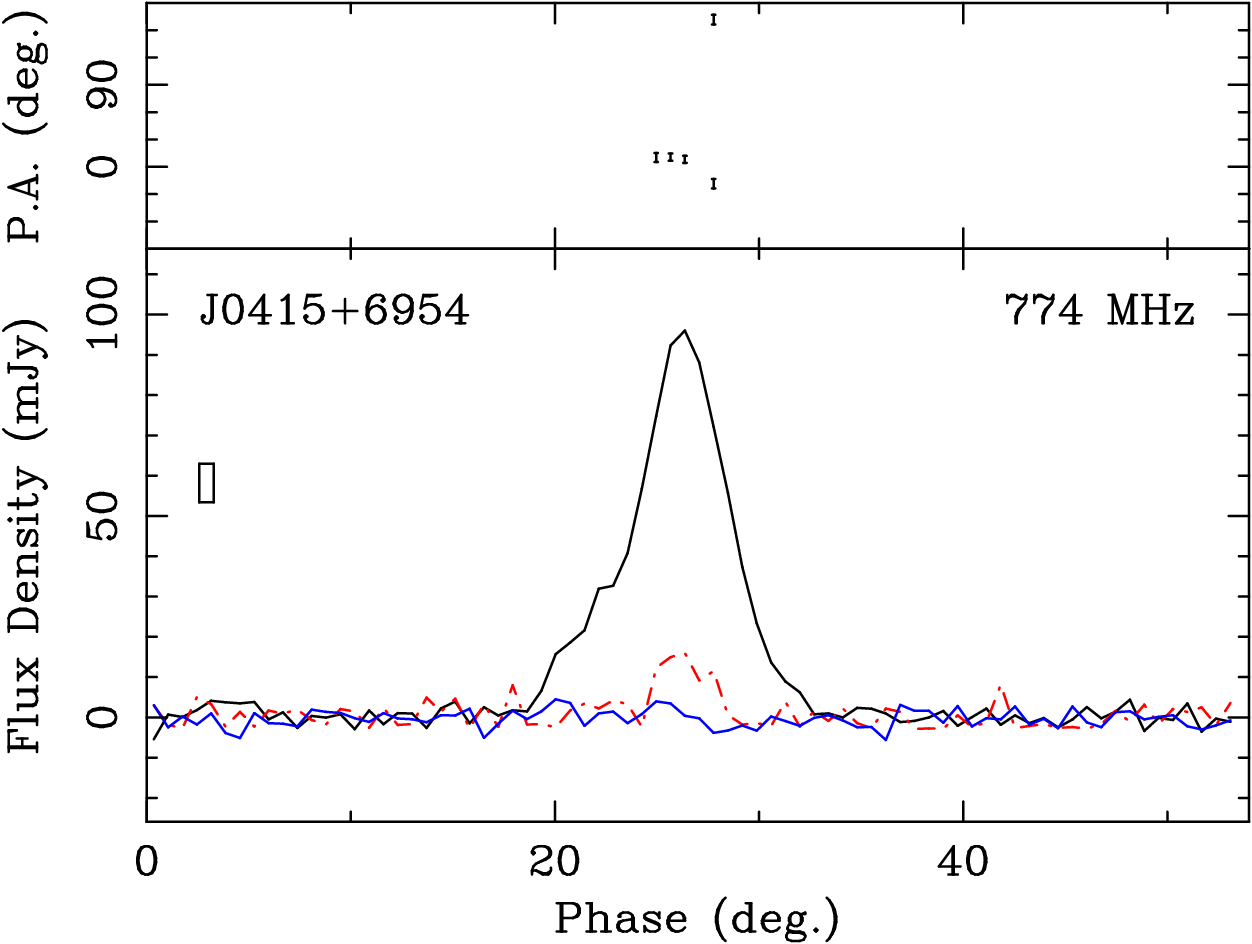,angle=270,width=55mm,height=39mm}}\\
\mbox{\psfig{file=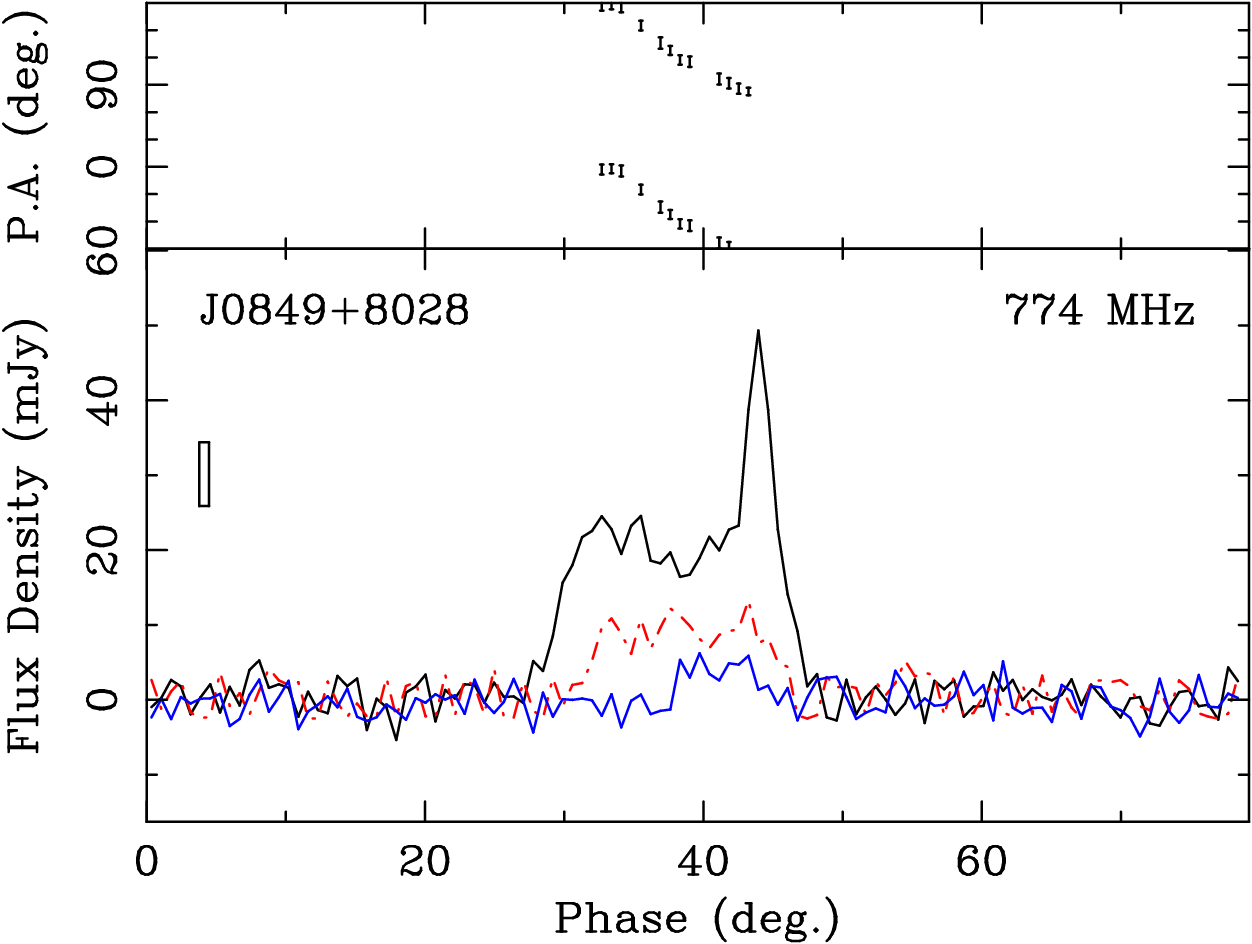,angle=270,width=55mm,height=39mm}}&	 
\mbox{\psfig{file=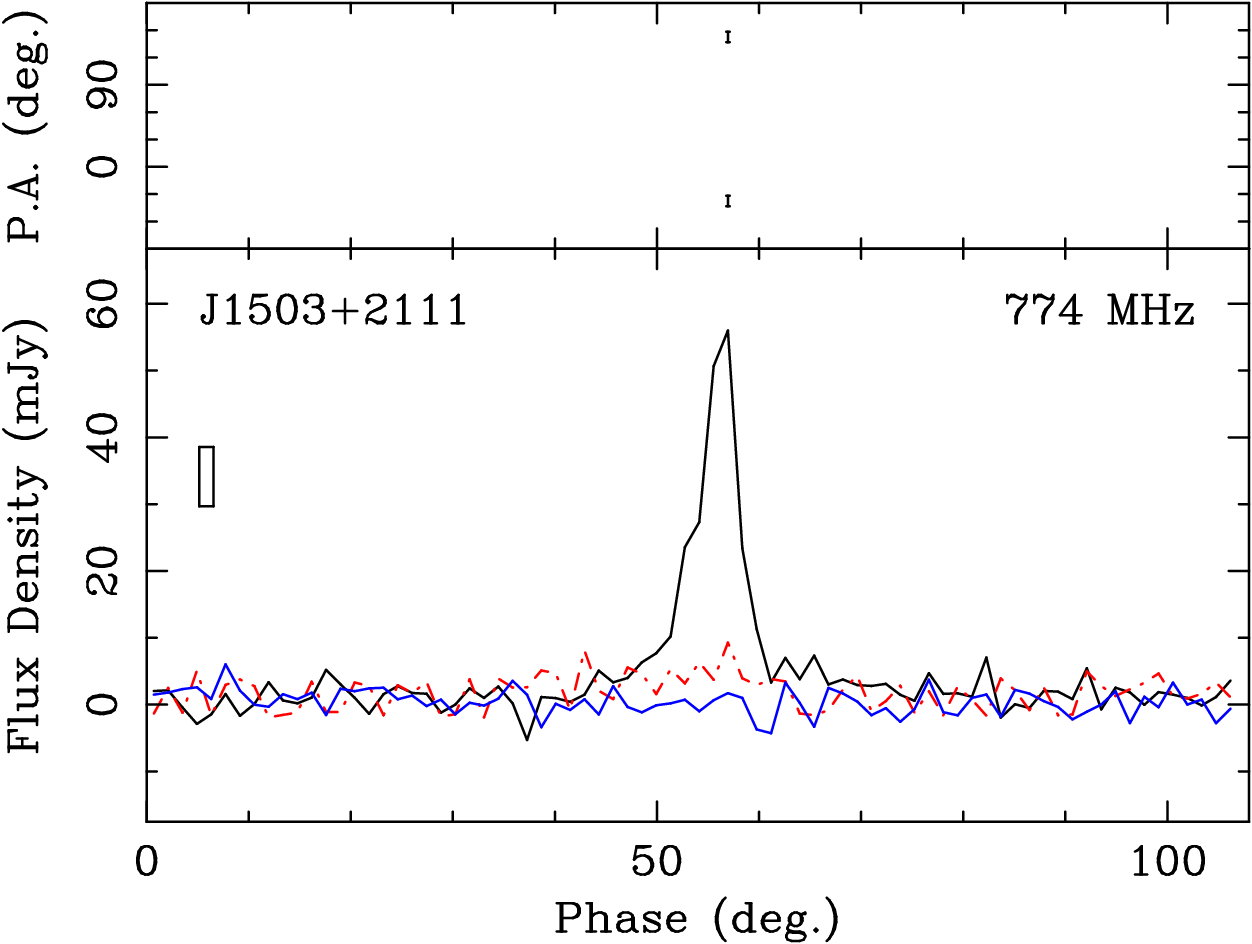,angle=270,width=55mm,height=39mm}}&
\mbox{\psfig{file=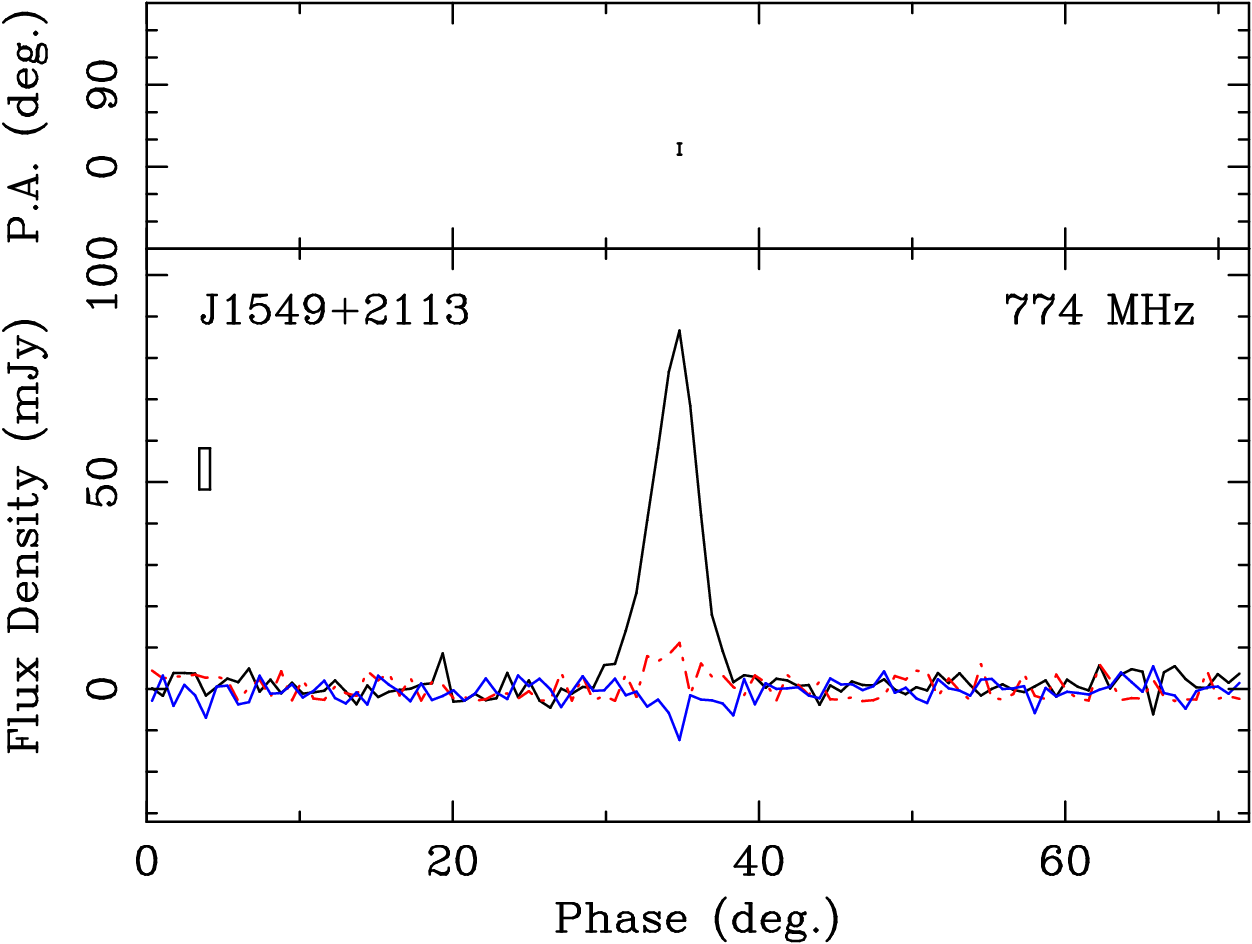,angle=270,width=55mm,height=39mm}}\\
\mbox{\psfig{file=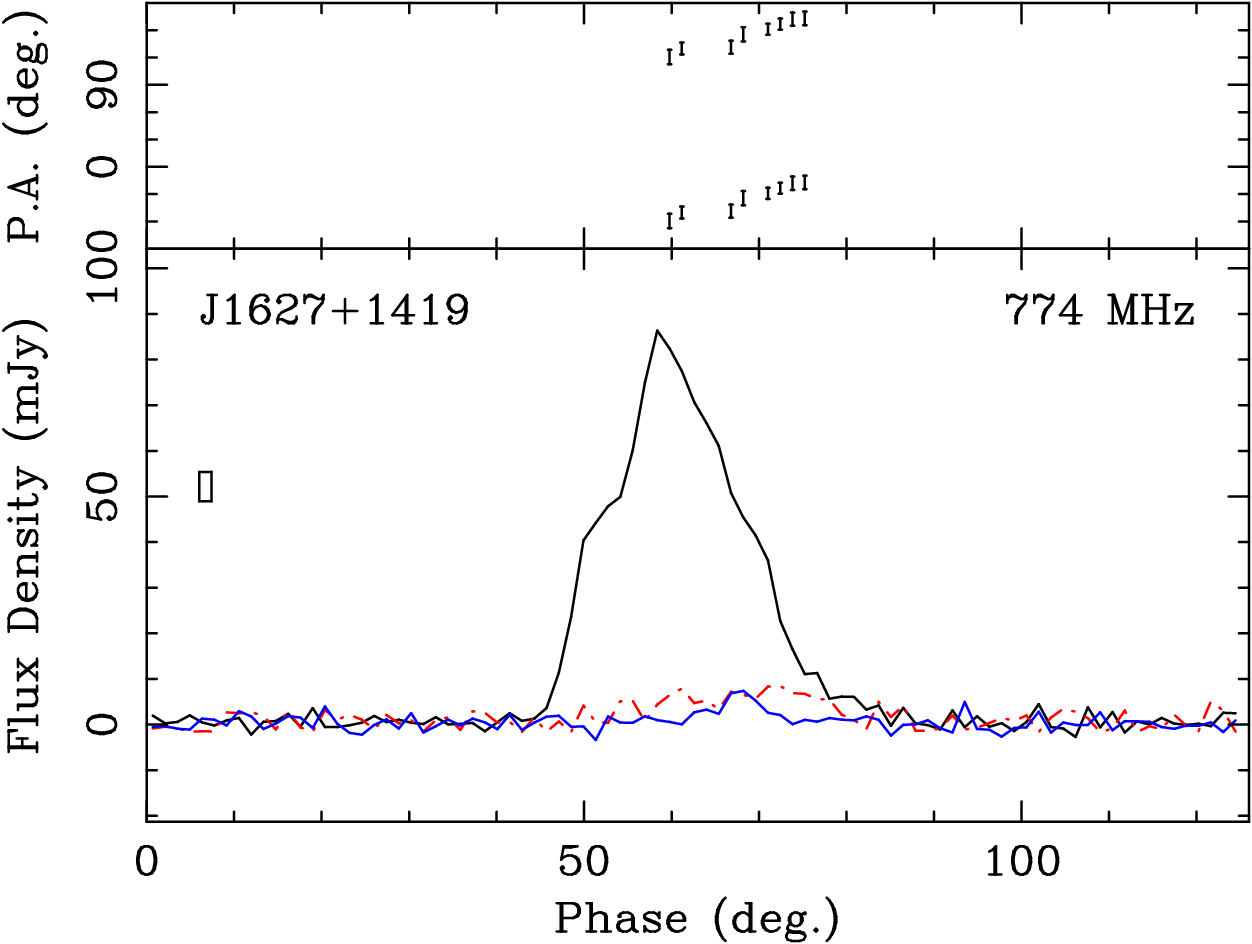,angle=270,width=55mm,height=39mm}}&	 
\mbox{\psfig{file=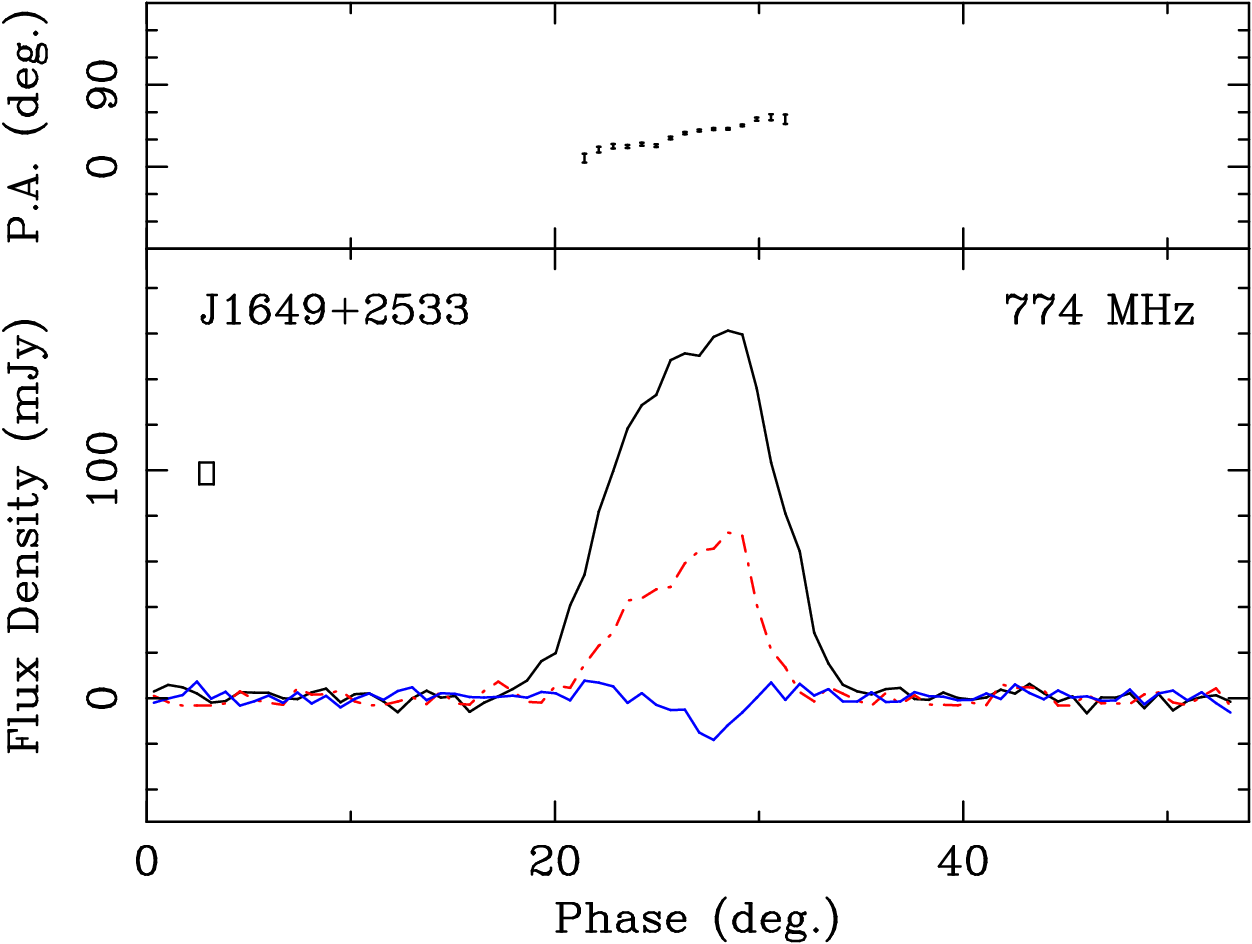,angle=270,width=55mm,height=39mm}}&
\mbox{\psfig{file=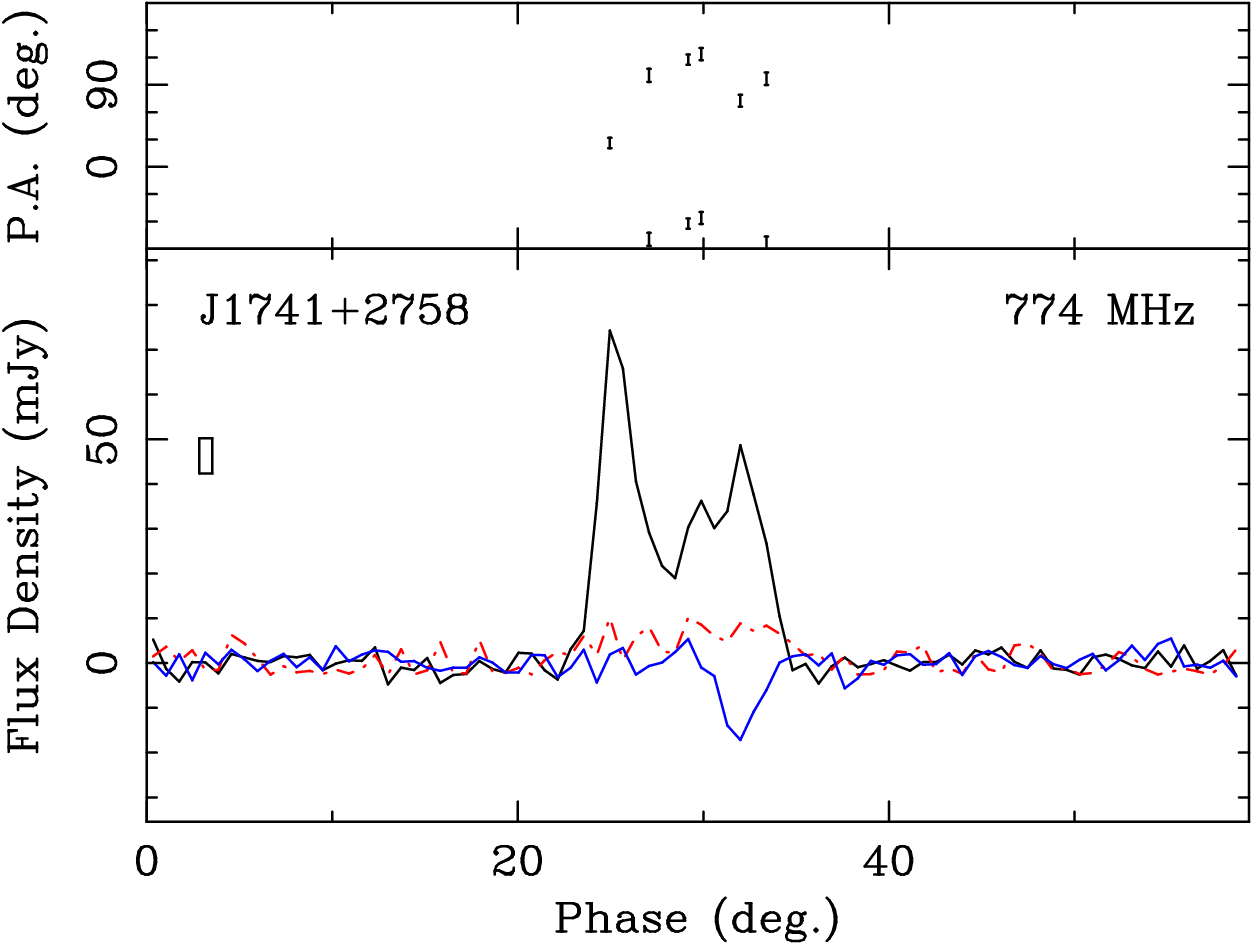,angle=270,width=55mm,height=39mm}}\\	 
\mbox{\psfig{file=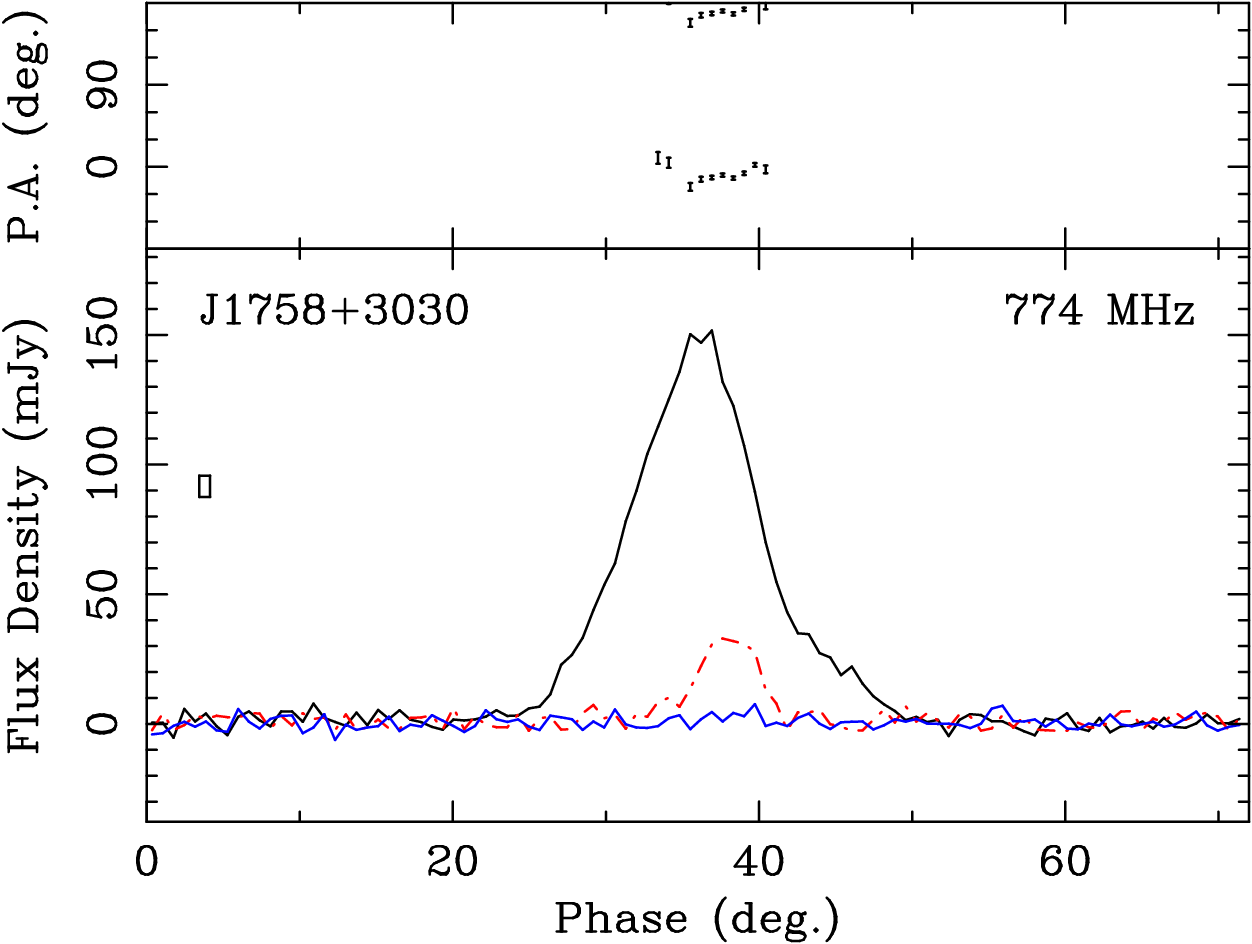,angle=270,width=55mm,height=39mm}}&
\mbox{\psfig{file=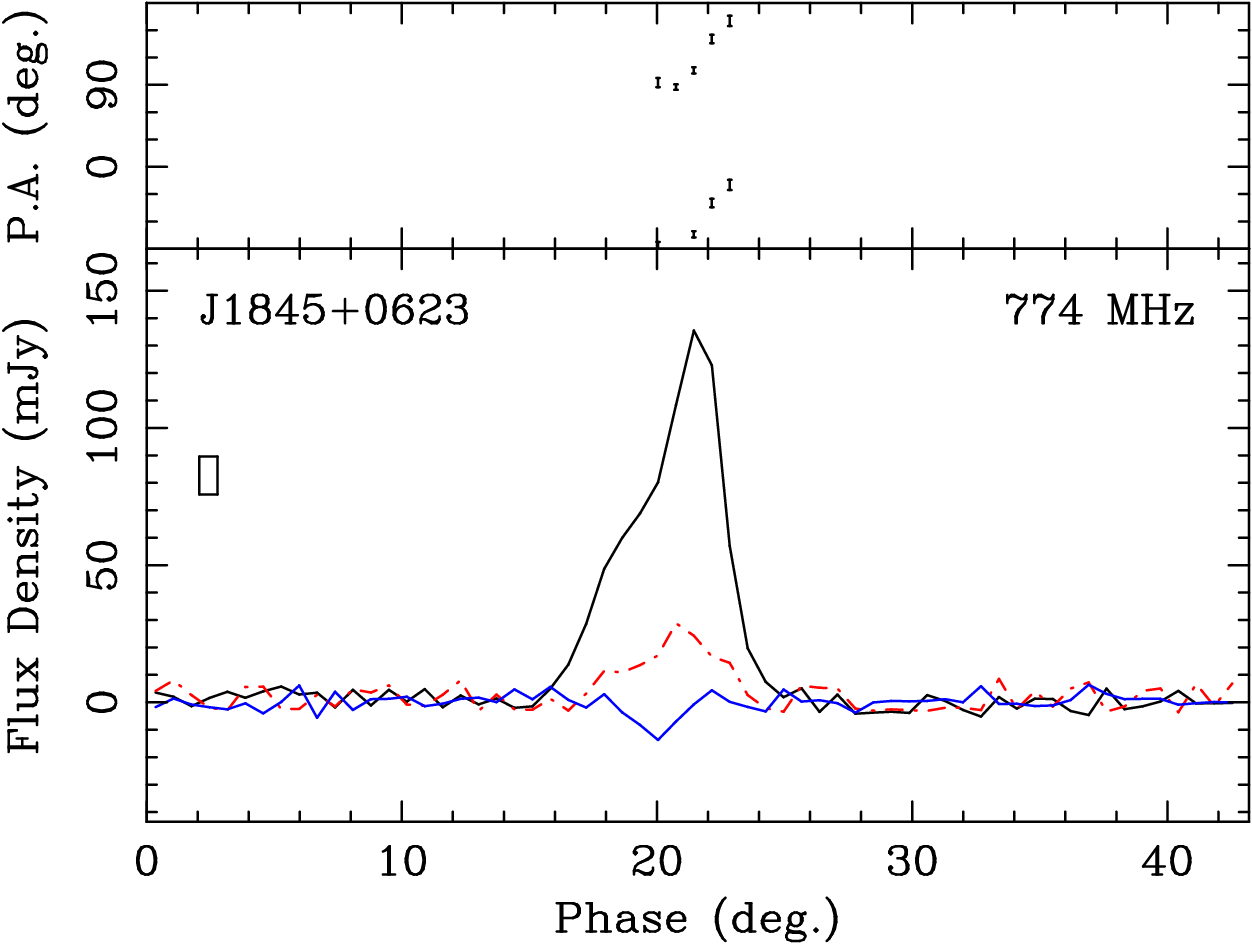,angle=270,width=55mm,height=39mm}}&	 
\mbox{\psfig{file=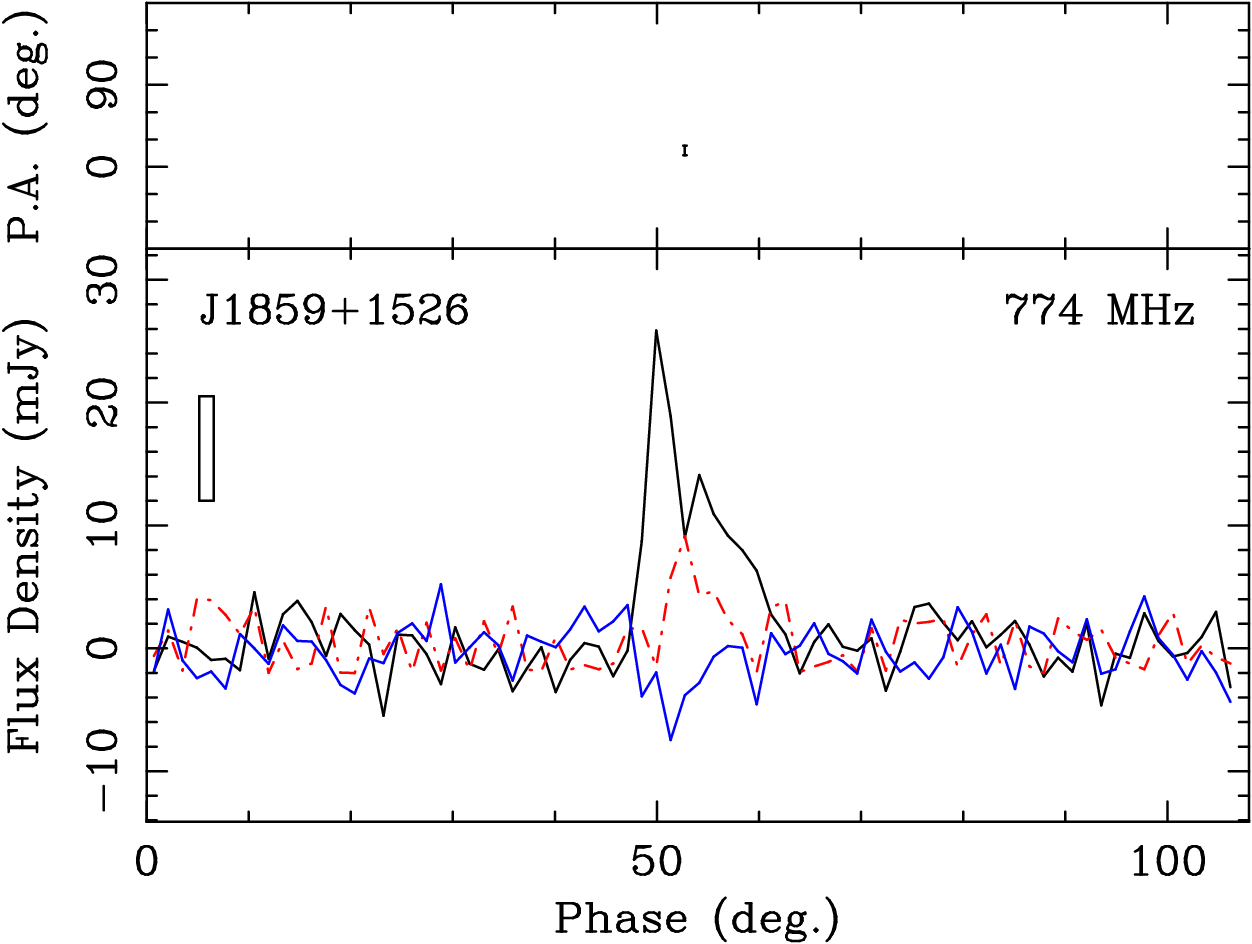,angle=270,width=55mm,height=39mm}}\\
\mbox{\psfig{file=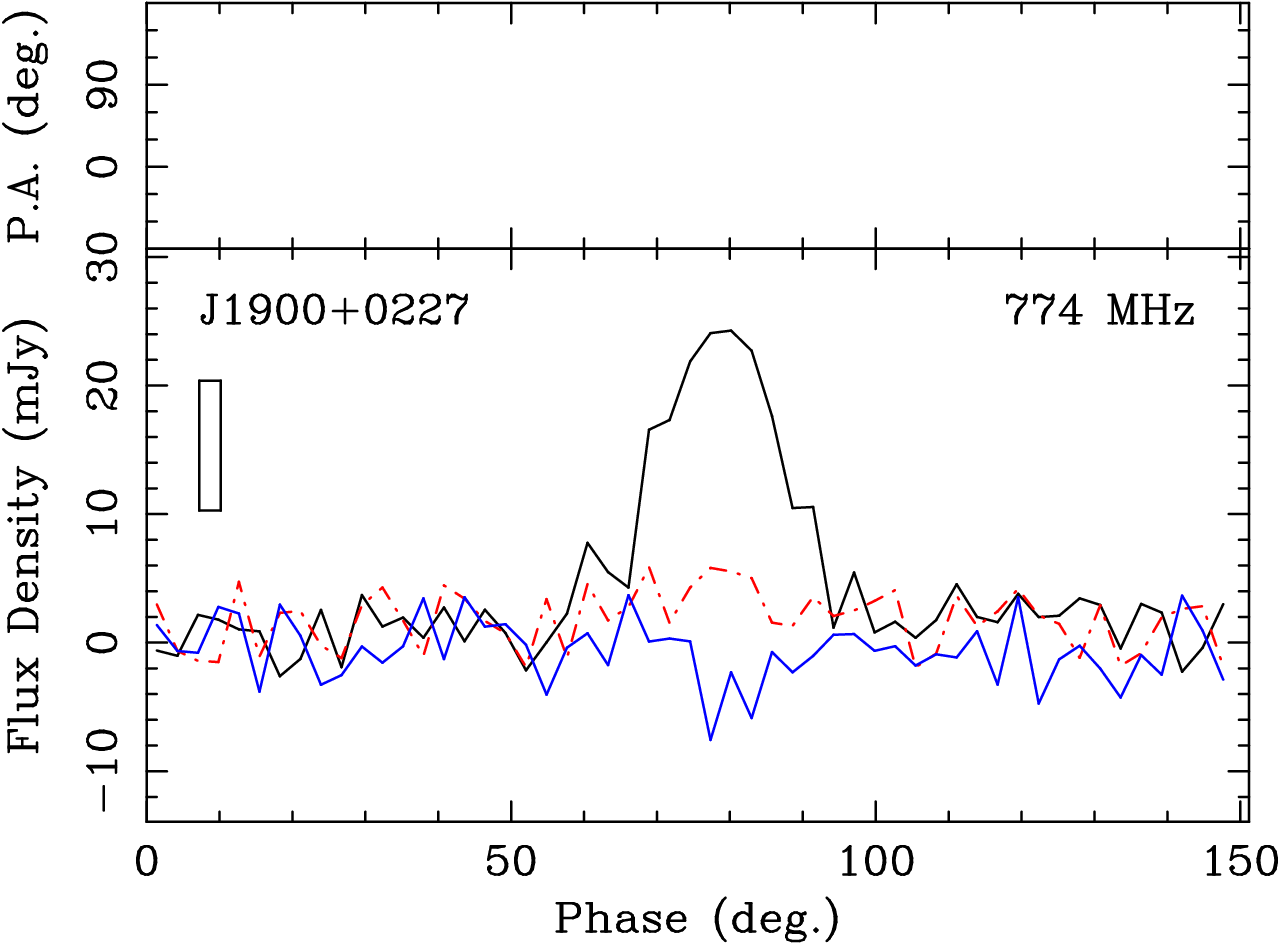,angle=270,width=55mm,height=39mm}}&
\mbox{\psfig{file=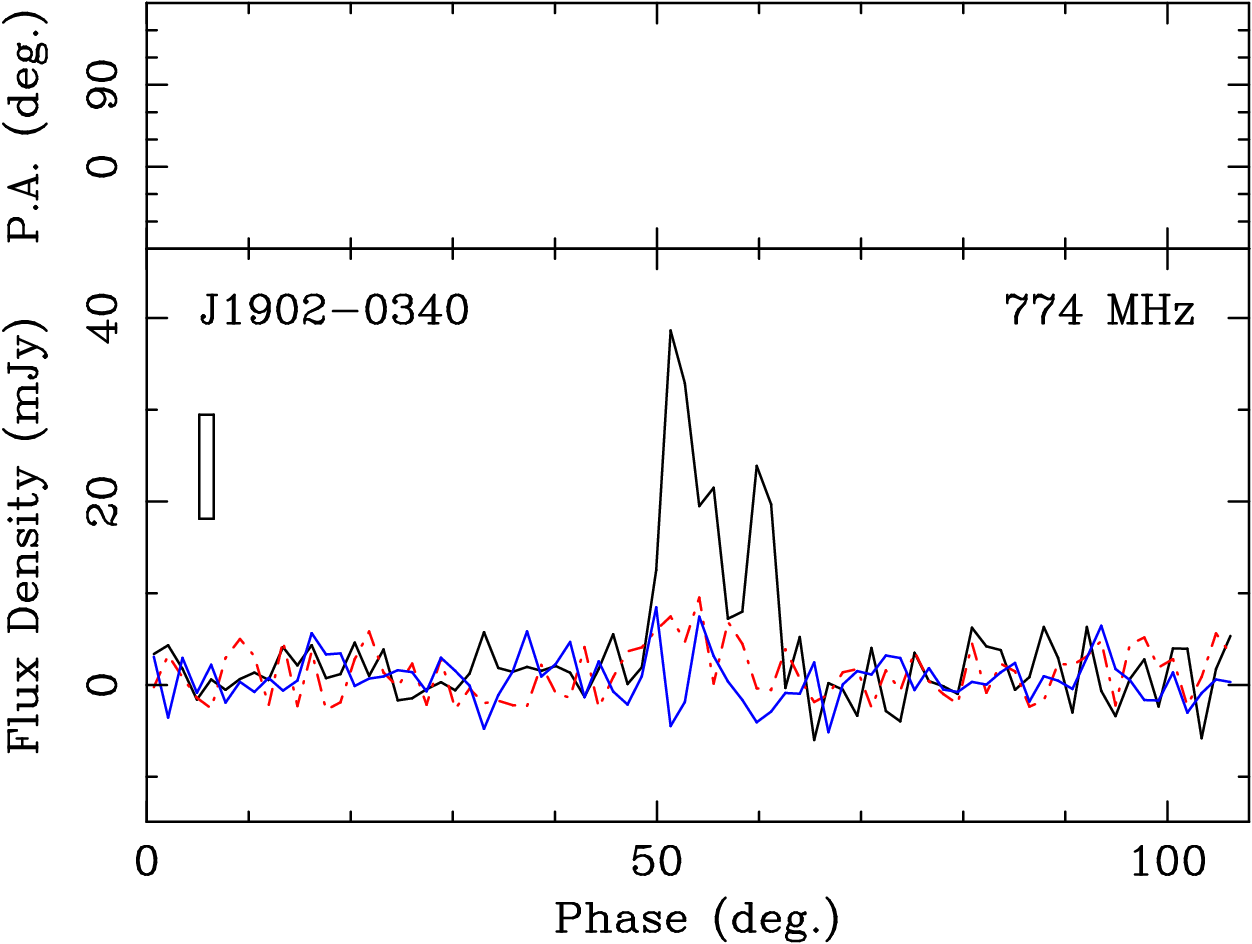,angle=270,width=55mm,height=39mm}}&
\mbox{\psfig{file=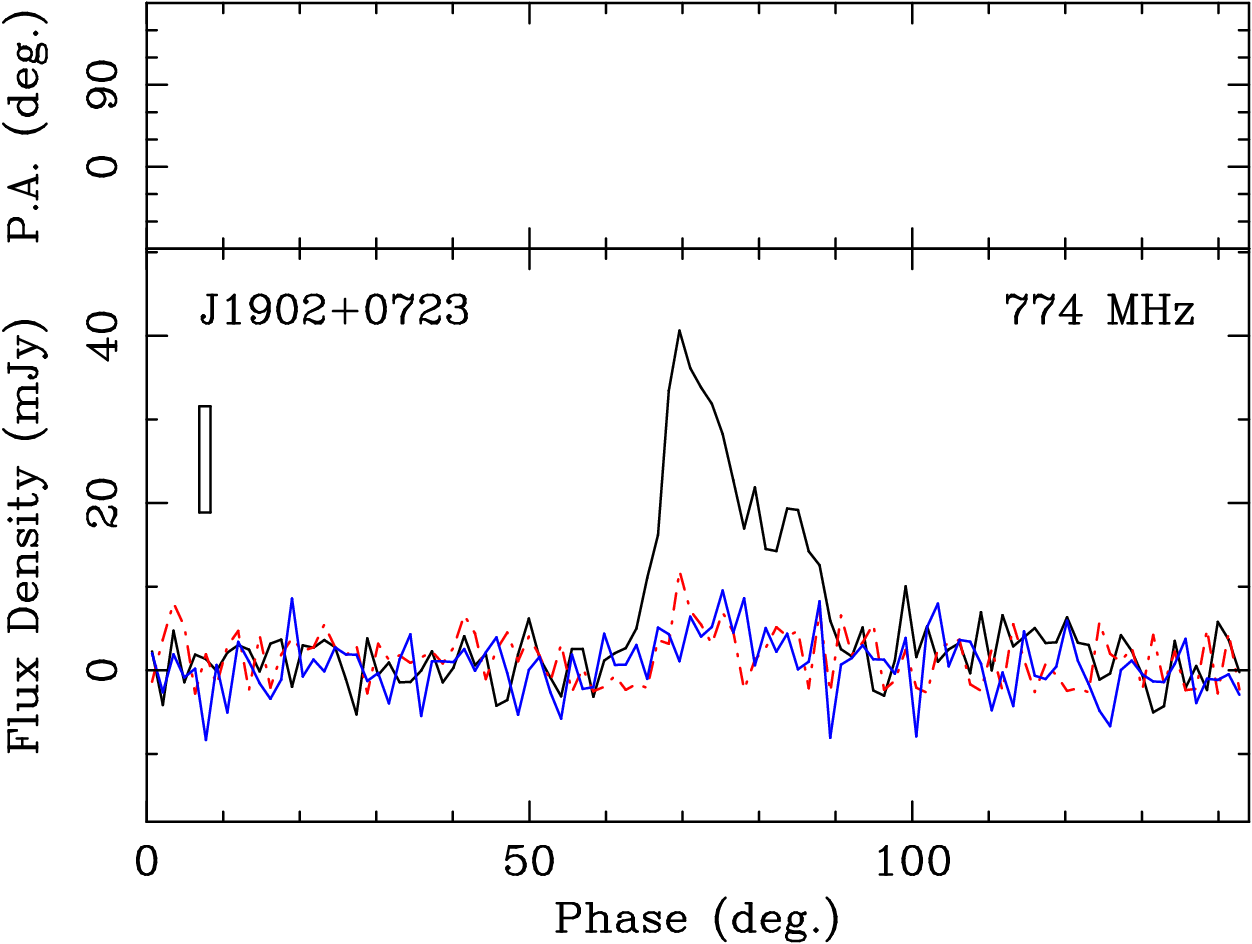,angle=270,width=55mm,height=39mm}}\\	 
\mbox{\psfig{file=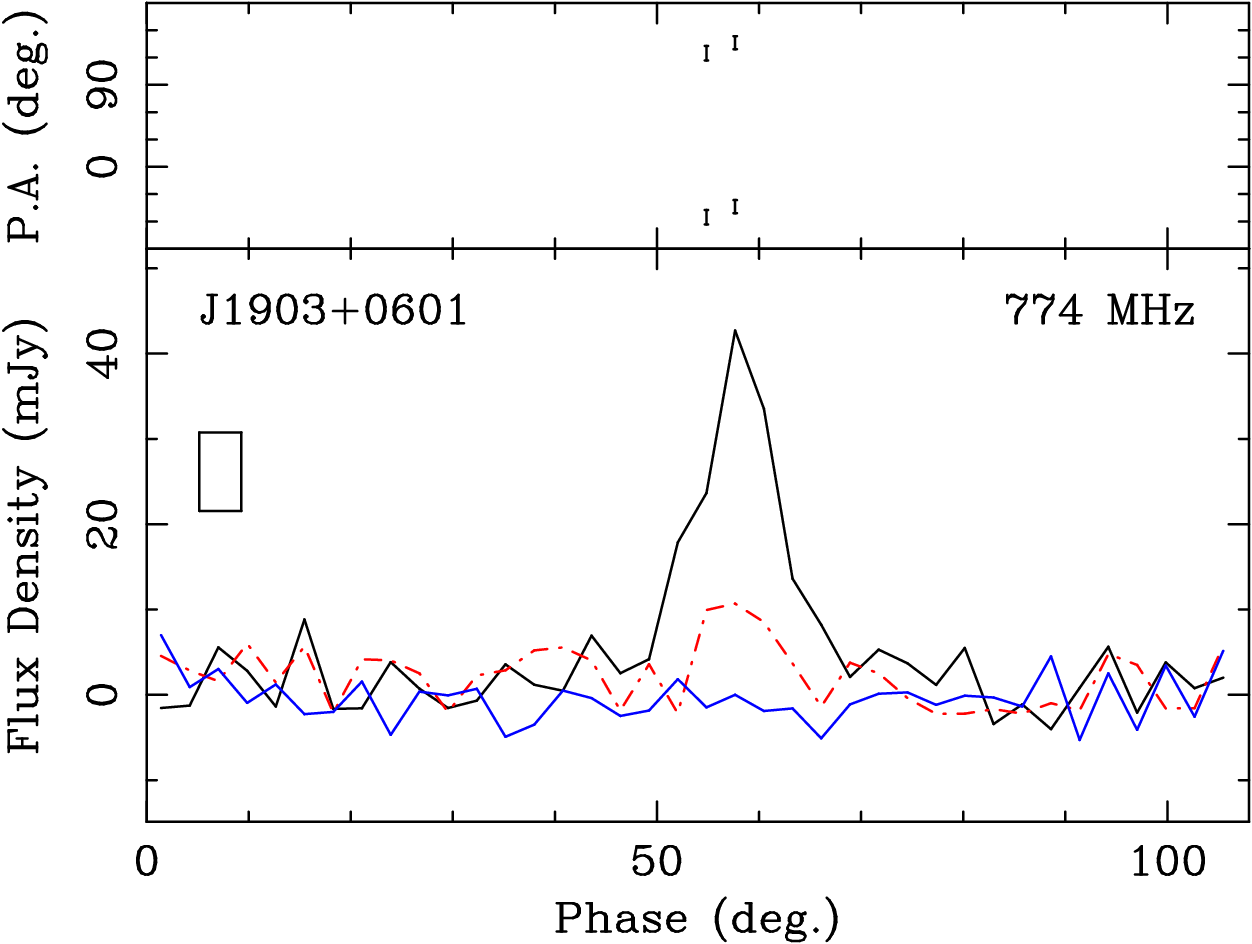,angle=270,width=55mm,height=39mm}}&
\mbox{\psfig{file=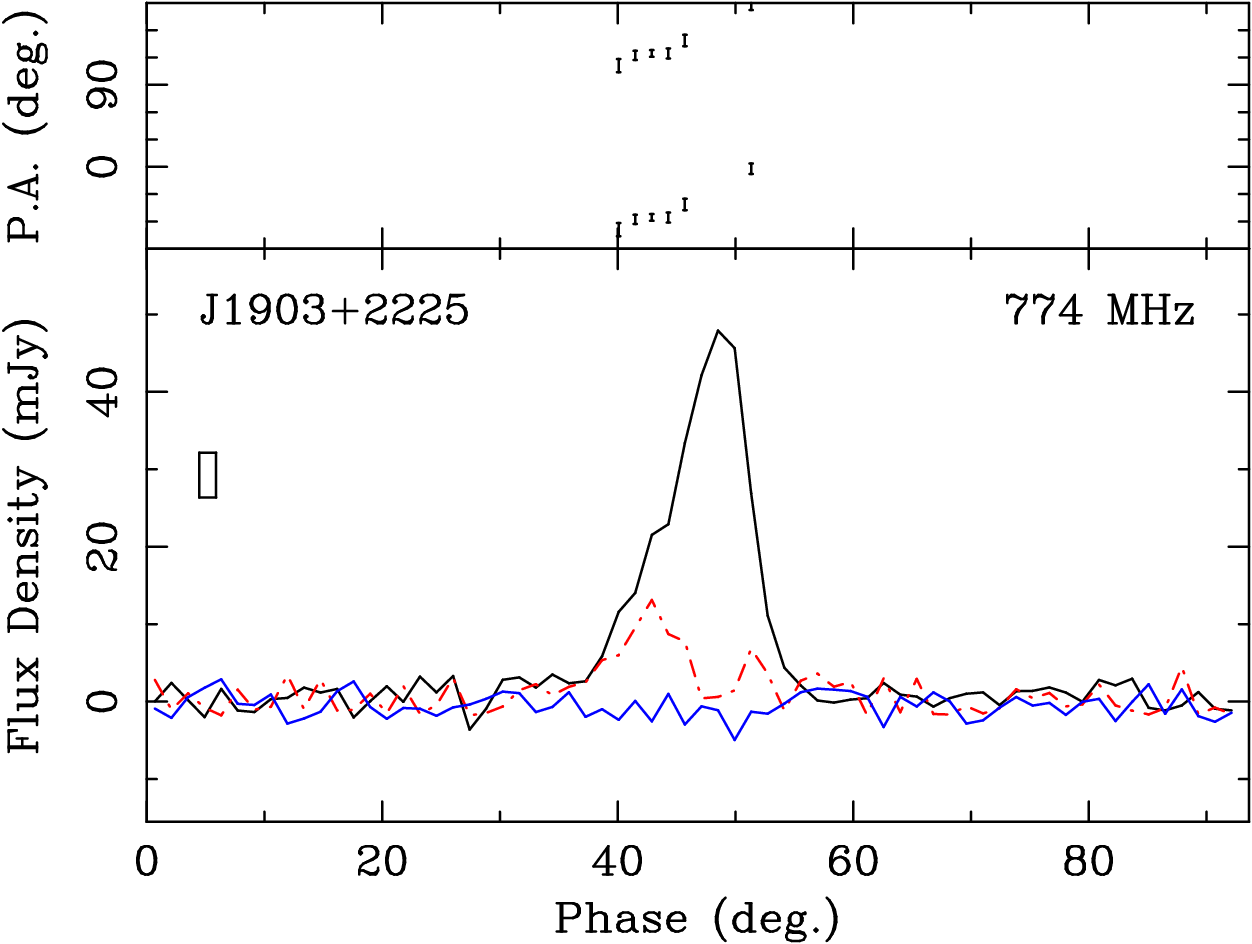,angle=270,width=55mm,height=39mm}}&
\mbox{\psfig{file=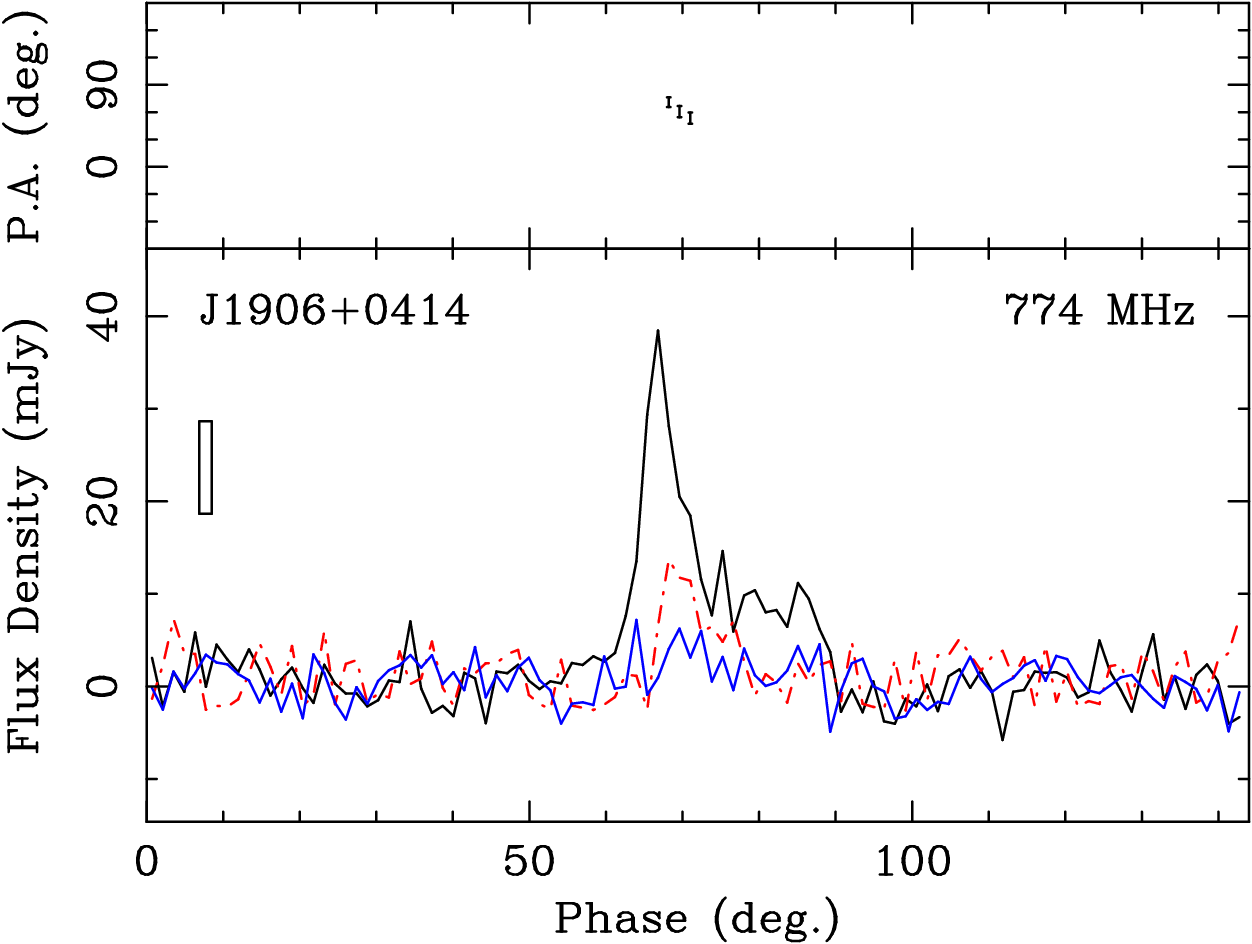,angle=270,width=55mm,height=39mm}}\\
\end{tabular}%
\caption{Polarization profiles of 51 pulsars at 774~MHz -- to be continued.}
\label{psrleft}
\end{figure*}\addtocounter{figure}{-1}
\begin{figure*}
\begin{tabular}{rrrr}
\mbox{\psfig{file=fig9_19.ps,angle=270,width=55mm,height=39mm}}&	 
\mbox{\psfig{file=fig9_20.ps,angle=270,width=55mm,height=39mm}}&	 
\mbox{\psfig{file=fig9_21.ps,angle=270,width=55mm,height=39mm}}\\
\mbox{\psfig{file=fig9_22.ps,angle=270,width=55mm,height=39mm}}&
\mbox{\psfig{file=fig9_23.ps,angle=270,width=55mm,height=39mm}}&	 
\mbox{\psfig{file=fig9_24.ps,angle=270,width=55mm,height=39mm}}\\
\mbox{\psfig{file=fig9_25.ps,angle=270,width=55mm,height=39mm}}&
\mbox{\psfig{file=fig9_26.ps,angle=270,width=55mm,height=39mm}}&
\mbox{\psfig{file=fig9_27.ps,angle=270,width=55mm,height=39mm}}\\	 
\mbox{\psfig{file=fig9_28.ps,angle=270,width=55mm,height=39mm}}&
\mbox{\psfig{file=fig9_29.ps,angle=270,width=55mm,height=39mm}}&
\mbox{\psfig{file=fig9_30.ps,angle=270,width=55mm,height=39mm}}\\	 
\mbox{\psfig{file=fig9_31.ps,angle=270,width=55mm,height=39mm}}&	 
\mbox{\psfig{file=fig9_32.ps,angle=270,width=55mm,height=39mm}}&	 
\mbox{\psfig{file=fig9_33.ps,angle=270,width=55mm,height=39mm}}\\
\mbox{\psfig{file=fig9_34.ps,angle=270,width=55mm,height=39mm}}&
\mbox{\psfig{file=fig9_35.ps,angle=270,width=55mm,height=39mm}}&	 
\mbox{\psfig{file=fig9_36.ps,angle=270,width=55mm,height=39mm}}\\
\end{tabular}%
\caption{Polarization profiles of 51 pulsars at 774~MHz -- to be continued.}
\end{figure*}\addtocounter{figure}{-1}
\begin{figure*}
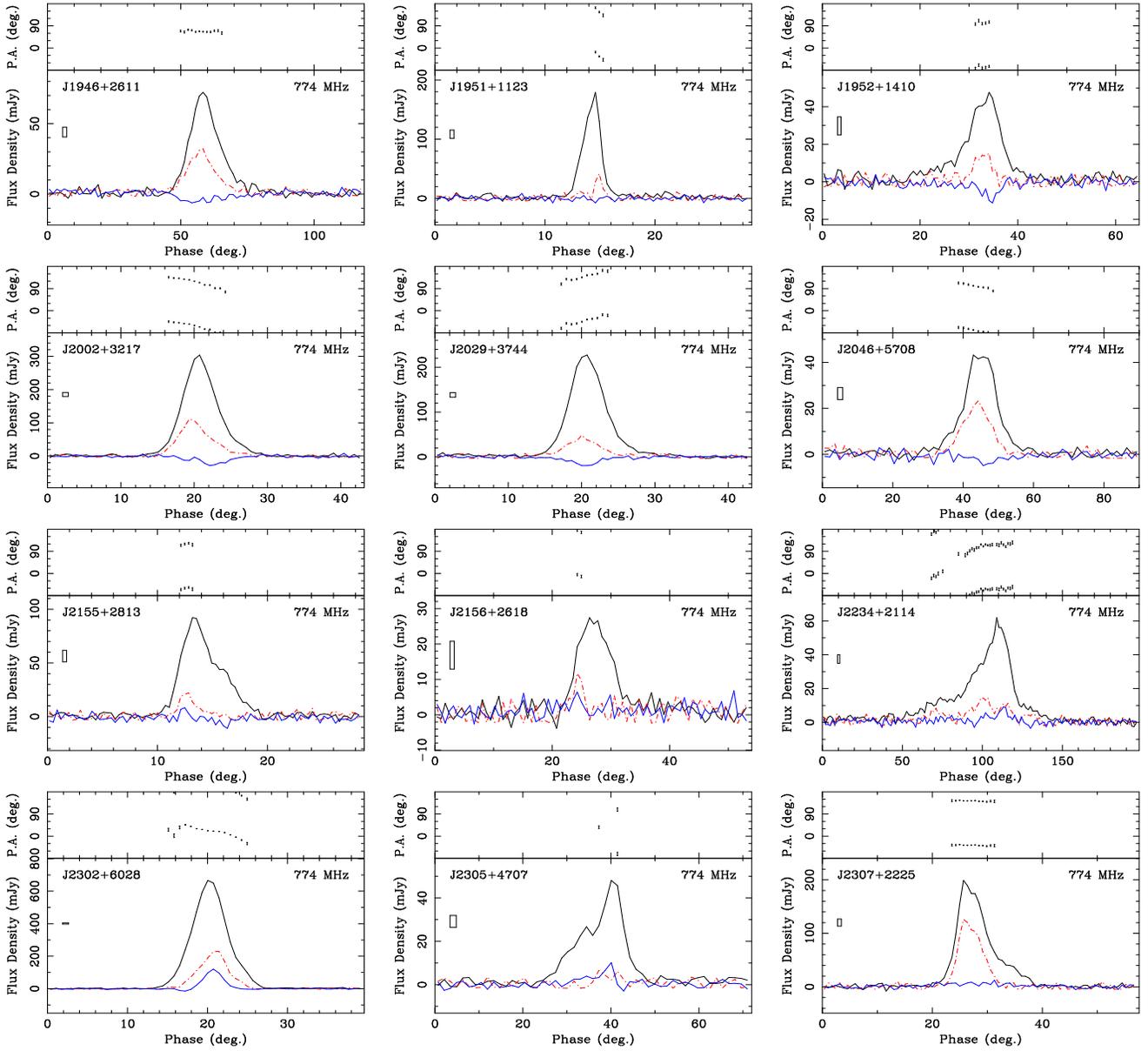

\begin{tabular}{rrrr}
\mbox{\psfig{file=fig9_37.ps,angle=270,width=55mm,height=39mm}}&
\mbox{\psfig{file=fig9_38.ps,angle=270,width=55mm,height=39mm}}&	 
\mbox{\psfig{file=fig9_39.ps,angle=270,width=55mm,height=39mm}}\\
\mbox{\psfig{file=fig9_40.ps,angle=270,width=55mm,height=39mm}}&
\mbox{\psfig{file=fig9_41.ps,angle=270,width=55mm,height=39mm}}&	 
\mbox{\psfig{file=fig9_42.ps,angle=270,width=55mm,height=39mm}}\\	 
\mbox{\psfig{file=fig9_43.ps,angle=270,width=55mm,height=39mm}}&
\mbox{\psfig{file=fig9_44.ps,angle=270,width=55mm,height=39mm}}&
\mbox{\psfig{file=fig9_45.ps,angle=270,width=55mm,height=39mm}}\\	 
\mbox{\psfig{file=fig9_46.ps,angle=270,width=55mm,height=39mm}}&
\mbox{\psfig{file=fig9_47.ps,angle=270,width=55mm,height=39mm}}&
\mbox{\psfig{file=fig9_48.ps,angle=270,width=55mm,height=39mm}}\\
\end{tabular}%
\caption{Polarization profiles of 51 pulsars at 774~MHz -- end.}
\end{figure*}

\subsection{Other Pulsars} 

Our GBT observations of the remaining pulsars in Figure~\ref{psrleft} are
mostly new polarization measurements. Only a few pulsars in this set
have been observed before \citep[GL98]{wcl+99,wck+04}, and our GBT data
generally have better S/N.
The polarization profile of PSRs B0114+58 (J0117+5914) and B1927+13
(J1930+1316) at 430~MHz \citep[GL98]{wck+04} is similar to our 774~MHz
profile.
PSR B2027+37 (J2029+3744) has been observed at 1.4~GHz by \citet{wcl+99} and
410, 610, 1408 and 1642~MHz by GL98. Our polarization profile is consistent
with the 610~MHz data.
Our GBT observations of PSRs B0153+39 (J0156+3949) and B2045+56 (J2046+5708) 
confirm the linear polarization and P.A. at 606~MHz by GL98.

\begin{figure}
\centering
\psfig{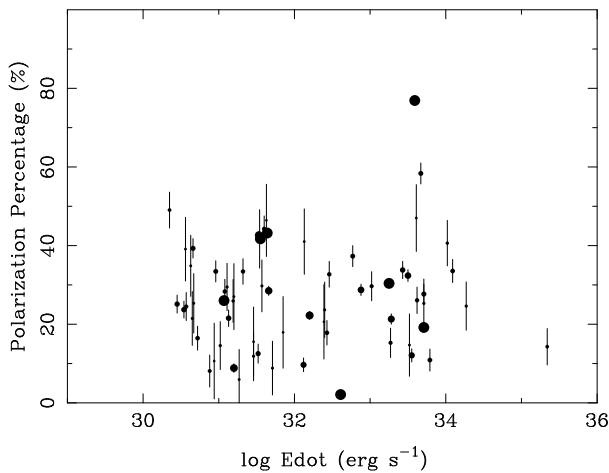}
\caption{Polarization percentage at 774 MHz vs. spin-down luminosity,
$\dot{E}$. The values of $\dot{E}$ were obtained from the ATNF pulsar
catalog \citep{mhth05} and also appear in Table~\ref{tab1}. Here, only
pulsars from our observed sample with an uncertainty of polarization
percentage less than 10\% are plotted. There is no evidence for correlation
between the degree of linear polarization and $\dot{E}$. The size of the
black dots in this plot is inversely proportional to the square root of the
uncertainty in polarization percentage, so that the highest quality data
clearly stand out.}
\label{edotl}
\end{figure}%

\section{Summary and discussions}

We have observed 100 pulsars using the GBT in order to obtain their
polarization profiles at 774~MHz. These are the first published polarization
measurements for 67 of them. Our observations consistently provide better
quality profiles for pulsars which have been previously observed. The
polarization properties of 7 millisecond pulsars in our sample show wide
profiles and flat P.A. curves. We also detected extremely strong linear
polarization from six pulsars. Together with circular polarization, they
are almost 100\% polarized over the whole pulse. About a dozen pulsars have
almost 100\% linear polarization for only a small part of their pulse
phase. Extremely strong circular polarization has been detected from one
component of PSR J1920+2650. Reversals in the sense of circular polarization
have been observed in about 20 pulsars. In only about half of these cases is
it associated with the core emission near the pulse center, and in some it
is definitely associated with a conal emission component. Most of the
observed double-component pulsars are probably conal-double. The correlation
between the direction of the P.A. curve and the sense of circular polarization is not
statistically significant in our data, likely due to the smaller sample size
available here than was used in \citet{hmxq98} and \citet{yh06}. Because
the circular polarization of some pulsars evolves with frequency
\citep[see][]{hmxq98,yh06} and often may have a transit of handedness near
800~MHz, it may be worth further investigating the dependence of the correlation
with frequency in the future.

Another important issue is the relationship between the spin-down luminosity
($\dot{E} = 3.95\times10^{46}$~erg~s$^{-1}\dot{P} P^{-3}$, derived from the
pulsar spin period $P$ and period derivative $\dot{P}$) and the observed
percentage of linear polarization. It has been found that pulsars with high
$\dot{E}$ tend to have stronger linear polarization, at least at 1400~MHz
\citep{hkk98,cmk01}. We have examined our polarization percentage data at
774~MHz, see Figure~\ref{edotl}. Although on average the polarization
percentage for pulsars with $\dot{E}>10^{33}$erg~s$^{-1}$ may be slightly
larger than that for pulsars with $\dot{E}<10^{33}$erg~s$^{-1}$, the
dependence is not significant.  Any relationship here is much less prominent
than that at 1400~MHz \citep{ct07}. Note also that some highly polarized
pulsars in our sample do not have very large $\dot{E}$.

\section*{Acknowledgments}
We thank Professor Don Backer and Professor Ingrid Stairs for use of the GASP pulsar
backend and the referee for a very careful review of the manuscript.
GBT observations were performed under proposal number GBT07A-024.
We sincerely thank Bryan Jacoby for kind contributions to the
original observing proposal.
J.L.H. is supported by the National Natural Science Foundation (NNSF) of China
(10521001, 10773016, and 10833003) and the National Key Basic Research
Science Foundation of China (2007CB815403).
P.B.D. is a Jansky Fellow of the National Radio Astronomy Observatory.
{\it Facility:} \facility{GBT}

\end{document}